\documentstyle[preprint,eqsecnum,aps,epsf]{revtex}
\tightenlines
\newcommand{\be}{\begin{equation}}
\newcommand{\ee}{\end{equation}}
\newcommand{\ba}{\begin{eqnarray}}
\newcommand{\ea}{\end{eqnarray}}
\begin{document}
\title{\Large{\bf{Precession of a Freely Rotating Rigid Body.\\ 
Inelastic Relaxation in the Vicinity of Poles.}}}
\author{\Large{Michael Efroimsky}}
\address{Department of Physics, Harvard University}

\address{~\\}
\address{~\\}
\address{NEW ADDRESS:} 
\address{Institute for Mathematics and Its Applications,
University of Minnesota}
\address{207 Church Street SE, Suite 400, Minneapolis MN 55455 USA\\}
\address{~\\}
\address{e-mail: efroimsk@ima.umn.edu}
\address{telephones: (612) 333 2235 , (612) 625 5532 ; fax: (612) 626 7370}
\author{PUBLISHED IN:} 
\author{\large{\bf{{{Journal of Mathematical Physics}},
~Vol. {\bf{41}}, ~p. 
1854 (2000)}}}
\maketitle
\begin{abstract}
When a solid body is freely rotating at an angular velocity ${\bf \Omega}$, 
the ellipsoid of constant angular momentum, in the space $\Omega_1, \;\Omega_2,
\;\Omega_3$, has poles corresponding to spinning about the minimal-inertia and 
maximal-inertia axes. The first pole may be considered stable if we neglect the
inner dissipation, but becomes unstable if the dissipation is taken into 
account. This happens because the bodies dissipate energy when they rotate 
about any axis different from principal. In the case of an oblate symmetrical 
body, the angular velocity describes a circular cone about the vector of 
(conserved) angular momentum. In the course of relaxation, the angle of this 
cone decreases, so that both the angular velocity and the maximal-inertia axis 
of the body align along the angular momentum. The generic case of an asymmetric
body is far more involved. Even the symmetrical prolate body exhibits a 
sophisticated behaviour, because an infinitesimally small deviation of the 
body's shape from a rotational symmetry (i.e., a small difference between the 
largest and second largest moments of inertia) yields libration: the precession
trajectory is not a circle but an ellipse. In this article we show that often 
the most effective internal dissipation takes place at twice the frequency of 
the body's precession. Applications to precessing asteroids, cosmic-dust 
alignment, and rotating satellites are discussed.

\end{abstract}

\pagebreak
                
\section{Introduction}

A complex rotational motion of a free solid body is an evidence 
of its rotation-axis' wobbling about the angular momentum. Indeed, a 
solid body in a long-established regime of free rotation must have its 
axis of rotation parallel to the angular momentum: this configuration
will minimise the kinetic energy, the angular momentum being fixed. The body
achieves this minimisation by aligning both its axis of rotation and 
axis of maximal inertia parallel to the angular momentum. 
By the end of this relaxation the body comes to steady spinning about its 
maximal-inertia axis. Any deviation from this regime witnesses either 
of the influence of the tidal forces, or (in the case of comets) of the 
result of jetting, or of an impact experienced by the body within its 
characteristic time of relaxation, or of the entire body being a wobbling 
fragment of an asteroid disrupted by a collision (Giblin \& Farinella 1997
)$^1$, (Giblin et al 1998)$^2$, (Asphaug \& Scheeres 1999)$^3$. The contest 
between the impacts (or the tidal forces, or the cometary jetting) on the one 
hand, and the relaxation mechanism(s) on the other hand, determines the 
dynamics of the body rotation. 

A study of the rotation of asteroids and 
comets may thus provide valuable information about their recent history.
Several examples of complex motion have already been registered. Among the 
asteroids, 4179 Toutatis furnishes another example of wobble (Ostro {\textit{
et al}} 1993)$^4$, (Ostro et al 1995)$^5$, (Ostro et al 1999)$^6$; see also $\;
\;$ http://www.eecs.wsu.edu/~hudson/asteroids.html$\;$). Among the comets, 
P/Halley is certainly an example of such a tumbling object (Sagdeev et al 
1989)$^7$, (Peale \& Lissauer 1989)$^8$, (Peale 1991)$^9$, (Belton et al 1991
)$^{10}$, (Samarasinha \& A'Hearn 1995)$^{11}$. Another example is
comet 46/P Wirtanen. ROSETTA mission is supposed to explore this comet soon
(Hubert \& Schwehm 1991)$^{12}$. The spacecraft will be carrying a 
high-resolution OSIRIS imaging system (Thomas et al 1998)$^{13}$ that, 
probably, will enable the astronomers not only to see the precession
but also to observe its relaxation during the 2.5 years of the spacecraft's 
escorting the comet. Our estimates$^{14, 41}$ show that the angular resolution of 
the currently available equipment gives to such sort of experiment a good 
chance of success, provided the experiment includes measurements performed at
least half a year apart from one another. An important factor influencing the
success or failure of such an experiment is the jetting intensity of the
particular comet.

Another field of application for this study is the cosmic-dust alignment: 
some of the alignment mechanisms are very sensitive to coupling between 
the angular velocity and angular momentum of the interstellar grains 
(Lazarian \& Draine 1999)$^{15}$, (Lazarian \& Efroimsky 1999)$^{16}$.

The third possible application of the developed formalism could be spin 
stabilisation of spacecrafts, including spacecrafts with a precession damper
(Chinnery and Hall 1995)$^{17}$, (Hughes 1986)$^{18}$, (Levi 1989)$^{19}$. An 
interest in studies of nonrigid-body dynamics with applications to spacecraft 
motion emerged after launch of "Explorer" satellite in 1958. (I am thankful to 
Vladislav Sidorenko for drawing my attention to this example.) The satellite 
was a very prolate body with 4 small deformable antennas on it. It had been 
supposed that it would rotate about its minimal-inertia axis. Instability of 
this motion was a major surprise for mission experts (Modi 1974)$^{20}$. 

On general grounds, the necessity of relaxation is evident: the system must 
reduce its
\begin{figure}
\centerline{\epsfxsize=3.5in\epsfbox{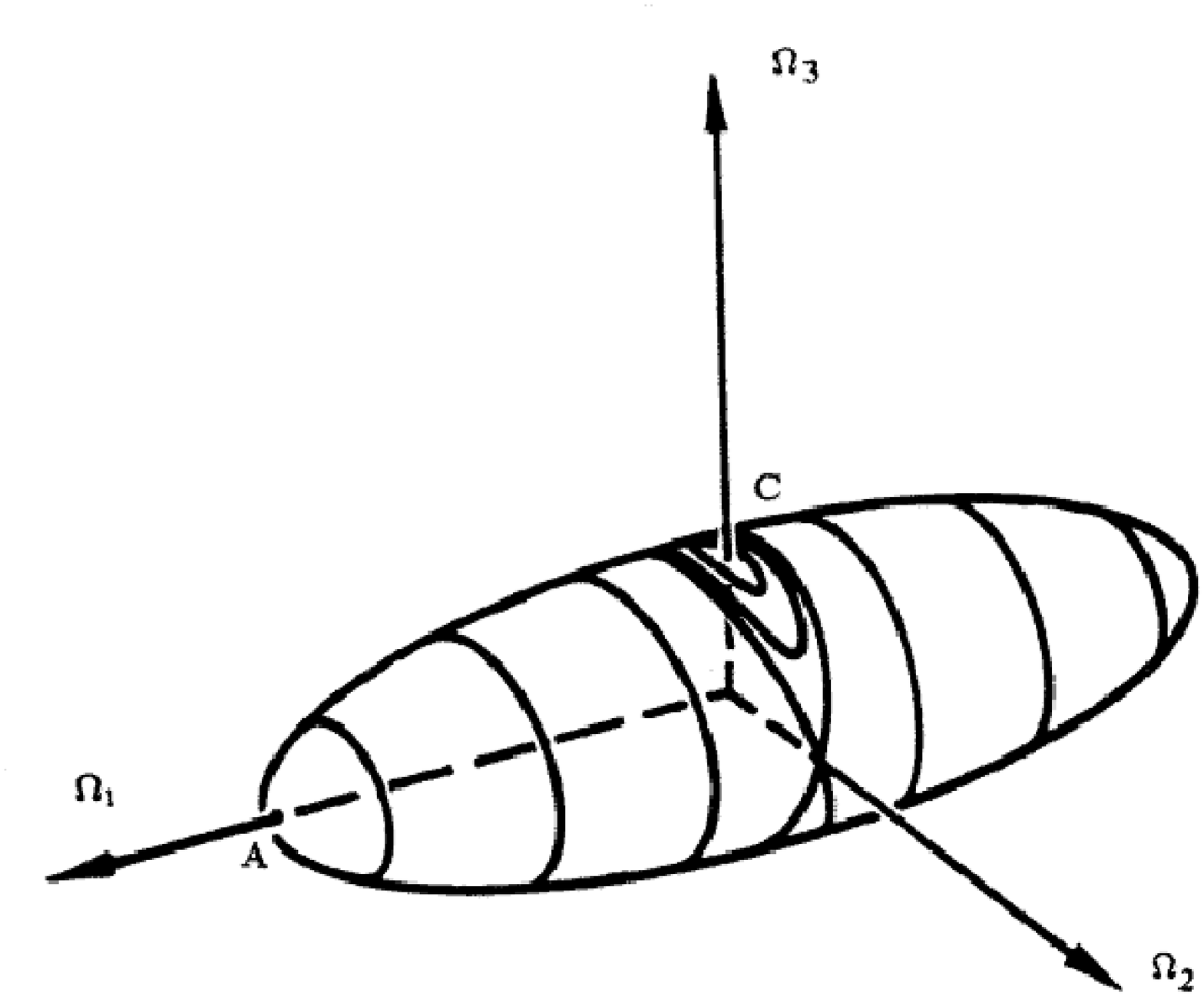}}
\bigskip
\caption{The ellipsoid of constant angular momentum in the angular-velocity space 
($\Omega_1\, , \; \Omega_2\, ,\; \Omega_3$). Lines on its surface denote its
intersections with the ellipsoids appropriate to different values of the 
rotation energy $T_{kin}$. Pole A corresponds to the maximal value  
of $T_{kin}$ and, thereby, to rotation about the minimal-inertia axis. Pole C 
corresponds to a complete relaxation, when $T_{kin}$ is minimal and the body 
is spinning about its major-inertia axis. In the course of precession, the 
angular-velocity vector $\bf \Omega$ moves along the constant-energy lines, and
slowly shifts from one line to another, in the direction from A to C. The above
picture describes the case of an almost prolate body: $I_3 \stackrel{>}{\sim} 
I_2>I_1$. The trajectories are circular near A and, in this case, remain almost
circular all the way from A to the separatrix. Crossing over the separatrix may
yield chaotic flipovers, whereafter the body will begin libration. The 
trajectories will again become circular only in the closemost vicinity of C.
}
\end{figure}
\noindent\\ kinetic energy down to the value that is minimal available for a 
fixed angular momentum. What particular physical effects
provide this relaxation? One phenomenon, relevant to tiny grains (like
those of the interstellar dust) but feeble for large samples, is the
Barnett dissipation called into being by the oscillating (due to
the precession) remagnetisation of the material, caused by the Barnett 
effect. (See, for example, (Lazarian 1994)$^{21}$, (Lazarian \& Draine 
1997)$^{22}$, and references therein.) Another process, relevant in small 
granules, and overwhelmingly leading for large bodies, is the 
inelastic dissipation. It is produced by the precession-caused
alternating stresses and strains.

A pioneering paper on the inelastic dissipation was published by Burns \& 
Safronov$^{23}$ back in 1973. Later the inelastic dissipation in small and in 
large freely-rotating oblate bodies was addressed in (Lazarian \& Efroimsky
1999)$^{16}$ and (Efroimsky \& Lazarian 2000)$^{24}$, correspondingly. In the 
present 
article we shall tackle to the dynamics of a body of arbitrary values of its 
moments of inertia. The issue is nontrivial. For example, a perfectly biaxial 
prolate spheroid 
has two axes of equal maximum moment of inertia and, therefore,
really does not have a stable rotation pole. For a triaxial figure of
a shape slightly deviating from a symmetrical prolateness, the situation is 
that the precession trajectory is not a circle but an elongated ellipse,
with long axis around the "waist" of the "cigar" figure (Fig.1).  
For the tumbling case,
the spin axis circulates all the way around the body, and in fact the
circulation is more or less around the long axis.  This leads to the
curious effect, observed in computer simulations of the rotation of
asteroid (433) Eros, that the body appears much of the time to be spinning 
nearly about its long axis (Black, Nicholson, Bottke, Burns $\&$ Harris 1999)$^{25}$.

In what follows we briefly review the main facts and formulae describing the 
solid-body rotation (section~II), dwelling comprehensively upon the case of an 
almost symmetrical prolate body (section~III). We divide the motion into four 
distinct stages (section~IV). Then we discuss the relaxation rate (section~V) 
and explain the nature of the nonlinearity emerging in this problem 
(section~VI), whereafter we compute the stresses arising in a precessing body, 
and calculate the energy density of the appropriate deformations (section~VII).
In sections~VIII and IX we calculate the rate of internal dissipation.
In section~X we draw conclusions and mention some practical 
applications of the formalism developed. In section ~XI we briefly account of 
the vississitudes of the generic case. 

\section{Notations and Assumptions}

We shall discuss free rotation of a solid body, using two Cartesian 
coordinate systems, each with an origin at the centre of mass of the body. 
The inertial coordinate system ($X$, $Y$, $Z$), with unit vectors 
${\mathbf{e}}_{X}$, ${\mathbf{e}}_{Y}$, ${\mathbf{e}}_{Z}$, will have its $Z$ 
axis parallel to the (conserved) angular momentum $\mathbf{J}$. Coordinates 
with respect to this frame are denoted by the same capital letters: $X$, $Y$, 
and $Z$. We shall also use the body frame defined by the three principal 
axes of inertia: $1$, $2$, and $3$, with coordinates $x$, $y$, $z$ and unit 
vectors $\mathbf{e}$$_{1}$, $\mathbf{e}$$_{2}$, $\mathbf{e}$$_{3}$.  

We denote the angular velocity by $\bf{\Omega}$, while $\bf{\omega}$ 
will stand for the rate of precession. Components of $\bf{\Omega}$ {\it in the 
body frame} will be called $\; \Omega_{1,2,3} \;$. 

Due to the lack of an established convention on notations, we present
a table that hopefully will prevent misunderstandings.
\noindent
\begin{table}
\begin{tabular}{|p{3cm}|p{3cm}|p{3cm}|p{3cm}|}
\hline
          &                  &                             &             \\
          &Principal moments &       Components of angular & Frequency of\\
          &of inertia        &       velocity              & precession  \\
          &                  &                             &             \\
\hline
                     &                             &                    &\\
  (Purcell 1979),    &                             &                    &\\
  (Lazarian \&       &                             &                    &\\ 
  Efroimsky 1999),   &                             &                    &\\
 (Efroimsky 2000),   &                             &                    &\\ 
  (Efroimsky \&      & $I_3\;\geq\;I_2\;\geq\;I_1$  &$\Omega_3\;\;,\;\;\;\Omega_2\;\;,\;\;\;\Omega_1$&
$\omega$\\
  Lazarian 2000),    &                             &                    &\\ 
  present article    &                             &                    &\\
                     &                             &                    &\\
\hline
          &                 &                                           &\\
(Synge \& Griffiths 1959) & $A\geq B\geq C$ & $\omega_1\;\;,\;\;\;\omega_2\;\;,\;\;\;\omega_3$ &$p$\\
 & & &\\
\hline
 & & &\\
(Black et al. 1999) & $C\geq B\geq A$ & $\omega_c\;\;,\;\;\;\omega_b\;\;,\;\;\;\omega_a$ & $\nu$\\
 & & &\\
\hline
\end{tabular}
\end{table}
A free rotation of a body obeys the Euler equations
\be
\frac{d}{dt} \; \left(I_i \; \Omega_i \protect\right) \; = \; \left( 
I_j \; - \; I_k \protect\right) \; \Omega_j \; \Omega_k
\label{2.1}
\ee
where $\; (i, j, k) \; = \; (1, 2, 3) \;$ (with cyclic transpositions), and 
the principal moments of inertia are assumed to obey:
\be
I_1 \; \leq \;   I_2 \; \leq \;   I_3 \; \; \; \; .
\label{2.2}
\ee
In the approximation of an 
absolutely solid body, the equations simplify:
\be
I_i \; {\dot{\Omega}}_i \; = \; \left(I_j \; - \; I_k \protect\right) \; 
\Omega_j \; \Omega_k \; \; \; \; .
\label{2.3}
\ee
This neglect of $\; {\dot{I}}_i \, \Omega_i \;$ against
$\; I_i \; {\dot{\Omega}}_i \;$ does need a justification,  
because the inelastic relaxation we are going to describe is due to
small deformations that yield nonzero $\; \dot{I_i} \;$. To validate the 
neglection, i.e., to prove that $\; \dot{I_i}/{I_i} \; \ll \; 
\dot{\Omega_i}/{\Omega_i} \;$, one must recall that for a rotational
period $\tau$
\begin{eqnarray}
{\dot{\Omega}}_i/{\Omega}_i \; \approx \; {\tau}^{-1} \; \; \; , 
\; \; \; \; \; 
{\dot{I}}_i/{I}_i \; \approx \; {\tau}^{-1} \, \epsilon \; \; \;
\label{2.4}
\end{eqnarray}
$\epsilon$ being a typical value of the relative strain (which in real
life rarely exceeds $10^{-6}$).

Conservation of the angular momentum $\; \bf{J} \;$ and the kinetic energy
$\; T_{kin} \;$ entails:
\be
I_1^2 \, {\Omega}_1^2 \; + \; I_2^2 \, {\Omega}_2^2 \; + \; I_3^2 \, 
{\Omega}_3^2 \; = \; {\bf{J}}^2 \; \; \; , 
\label{2.5}
\ee
\be
I_1 \, {\Omega}_1^2 \; + \; I_2 \, {\Omega}_2^2 \; + \; I_3 \, 
{\Omega}_3^2 \; = \; 2 \; T_{\small{kin}} 
\label{2.6}
\ee
In the context of our study (which is aimed at estimating the rate of 
relaxation), equation (\ref{2.6}) is applicable as long as we accept the 
adiabatic approach, i.e., assume the 
relaxation to be a "slow process", compared to the "fast processes" of 
rotation and precession. A rigorous formulation of this assertion is  
based on formulae (\ref{8.14}) and (\ref{9.8}) obtained below. These formulae 
are to
become the main result of our article. They give the relaxation rate $\;
d{<\sin^2\theta>}/dt\;$, $\;\theta\;$ being the angle between the angular 
momentum and the major-inertia axis of the body, and averaging being performed
over the precession period (Appendix A, formula (A1)). The exact formulation of
the adiabatic approach will read: 
\be
-\;\frac{d\,<\sin^2\theta>}{dt}\,\ll\,\omega\;\;,
\label{adiabaticity}
\ee
$\omega\;$ being the precession rate. This condition will adumbrate the 
applicability realm of the solutions (\ref{8.14}) and (\ref{9.8}) to be 
derived.

Equations (\ref{2.5}) and (\ref{2.6}) may be resolved with respect to
$\; {\Omega}_1^2 \;$ and $\; {\Omega}_2^2 \;$:
\ba
\nonumber
{\Omega}_3^2 \; = \; P \; - \; Q \, {\Omega}_2^2 \;\;\;\;,\;\;\;\;\;\; 
P\;\equiv\;\frac{2\,I_1\,T_{kin}\;-\;{\bf J}^2}{I_3\,(I_1\,-\,I_3)}\;\;\;\;,\;
\;\;\;\;\;\;\;Q\;\equiv\;\frac{I_2}{I_3}\;\frac{I_1\,-\,I_2}{I_1\,-\,I_3}\;\;\;
\;\;\;\;\;\\
\label{2.7}
\\
\nonumber
{\Omega}_1^2 \; = \; R \; - \; S \, {\Omega}_2^2 \; \; \; \; ,\;\;\;\;\;\;
R\;\equiv\;\frac{2\,I_3\,T_{kin}\,-\,{\bf J}^2}{I_1\,(I_3\,-\,I_1)}
\;\;\;\;,\;\;\;\;\;\;\;\;S\,\equiv\,\frac{I_2}{I_1}\,\frac{I_3\,-\,I_2}{I_3\,-\,I_1}\,\;\;\;\;\;\;\;\;
\ea
substitution whereof in  (\ref{2.3}), for $\; {\mathit{i}} = 2 \;$,
gives:
\be
{\dot{\Omega}}_2^2 \; = \; \left(\frac{I_3 \; - \; I_1}{I_2}\protect\right)^2
\;\left( P \; - \; Q \, {\Omega}_2^2 \protect\right) \; 
\left(R \; - \; S \, {\Omega}_2^2 \protect\right) \; \; \; \; .
\label{2.8}
\ee
It is
possible (Synge \& Griffith 1959)$^{26}$ to  pick up such positive functions 
$\; \beta \;$, $\; \omega \;$, $\; k \;$ of the arguments $\; I_{1,2,3}, \; 
T_{kin}, \; {\bf{J}}^2 \;$ that the rescaled time and second component of the 
angular velocity,
\be
t' \; \equiv \; \omega \, t \; \; \; , \; \; \; \; \xi \;\equiv \;\Omega_2/\beta  \; \; \; \; ,
\label{2.9}
\ee
satisfy 
\be
\left(d\xi/dt'\protect\right)^2\;= \;\left(1 \; - \; \xi^2\protect\right) 
\left(1\; -\;k^2\;\xi^2\protect\right)
\label{2.10}
\ee
with $\; k \; < \; 1 \;$. Solution to this equation is the Jacobian
elliptic function $\; {\textit{sn}}{(t')} \;$, so that
\be
\Omega_2 \; = \; \beta \, \; sn\left[\omega (t \, - \, t_0), \; k^2 \protect\right]
\label{2.11}
\ee
$t_o$ being an arbitrary constant. It is known (Ibid.) that
substitution of the latter in (\ref{2.7}) yields, for 
$\;\;{\bf{J}}^2\;>\;2\;I_2\;T_{\small{kin}}\;$ :
\begin{eqnarray}
\Omega_1\;=\;\gamma\;\,{\it{cn}}\left[\omega (t \, - \, t_0), \; k^2 \protect\right]
\; \; \;, \; \; \; \; \; \; 
\Omega_3 \; = \; \alpha\;\,{\it{dn}}\left[\omega (t \, - \, t_0),\;k^2\protect\right]
\label{2.12}
\end{eqnarray}
(Recall that what we call $\;{\Omega_3}\;$, in (Synge \& Griffiths)$^{26}$ is 
called ${\omega_1}$, while (Black et al)$^{25}$ denote it $\;{{\omega}_{c}}$.)
In the above formulae
\begin{eqnarray}
\nonumber\\
\alpha \; = \; \sqrt{
\frac{{\bf{J}}^2\;-\;2\;I_1\;T_{\small{kin}}}{I_3\;(I_3 - I_1)}}\;\;\;,\;
\; \; \; \beta \; = \; \sqrt{\frac{2 \; I_3 \; T_{\small{kin}} \; - \; 
{\bf{J}}^2}{I_2 \; (I_3 - I_2)}} \; \; \; , \; \; \; \; 
\gamma \; = \;\sqrt{\frac{2 \; I_3 \; T_{\small{kin}} \; - \; 
{\bf{J}}^2}{I_1 \; (I_3 - I_1)}} 
\nonumber\\
\omega \; = \; \sqrt{\frac{\left({\bf{J}}^2 \; - \; 2 \; I_1 \; T_{\small{kin}}
\protect\right)\; \left(I_3 \; - \; I_2\protect\right)}{I_1\;I_2\;I_3}}\;\;\;
, \; \; \; \; k \; = \; \sqrt{\frac{I_2 \; - \; I_1}{I_3 \; - \; I_2} \;  \; 
\frac{2 \; I_3 \; T_{\small{kin}} \; - \; {\bf{J}}^2}{{\bf{J}}^2 \; - \; 2\;
I_1 \; T_{\small{kin}} }} \; \; \; ,
\label{2.13}
\end{eqnarray}
while for $ \; {\bf{J}}^2 \; < \; 2\;I_2 \; T_{\small{kin}} \; $ one
arrives to:
\begin{eqnarray}
\Omega_1\;=\;\gamma\;\,{\it{dn}}\left[\omega (t \, - \, t_0), \; k^2 \protect\right]
\; \; \;, \; \; \; \; \; \; 
\Omega_3 \; = \; \alpha \; \, {\it{cn}}\left[\omega (t \, - \, t_0), \; k^2
  \protect\right]
\label{2.14}
\end{eqnarray}
where
\begin{eqnarray}
\nonumber\\
\alpha \; = \; \sqrt{
\frac{{\bf{J}}^2\;-\;2\;I_1\;T_{\small{kin}}}{I_3\;(I_3 - I_1)}}\;\;\;,\;
\; \; \; \beta \; = \; \sqrt{\frac{{\bf{J}}^2 \; - \; 2 \; I_1 \; 
T_{\small{kin}}}{I_2 \; (I_2 - I_1)}} \; \; \; , \; \; \; \; 
\gamma \; = \; - \; \sqrt{\frac{2 \; I_3 \; T_{\small{kin}} \; - \; 
{\bf{J}}^2}{I_1 \; (I_3 - I_1)}} 
\nonumber\\
\omega \; = \; \sqrt{\frac{\left(2 \; I_3 \; T_{\small{kin}} \; - \; {\bf{J}}^2
\protect\right)\; \left(I_2 \; - \; I_1\protect\right)}{I_1\;I_2\;I_3}}\;\;\;
, \; \; \; \; k \; = \; \sqrt{\frac{I_3 \; - \; I_2}{I_2 \; - \; I_1} \;  \; 
\frac{{\bf{J}}^2\;-\; 2 \; I_1 \; T_{\small{kin}}}{{2\;I_3\;T_{\small{kin}} 
\; - \; {\bf{J}}^2 }}} \; \; \; ,
\label{2.15}
\end{eqnarray}
In some books (like for example in Abramovitz \& Stegun
1964)$^{27}$ notation $\, m \, \equiv \, k^2 \,$ is used 

Mind that in ${ Synge\;\&\;Griffith}^{26}$ expression (14.116a) for 
$\gamma$ is given with a wrong sign. Our expression for $\gamma$, as 
given by our formula (\ref{2.13}), makes the expressions (\ref{2.11} - 
\ref{2.12}) coincide, in the limit of oblate symmetry 
($\;I_1\;=\;I_2\;$), with the well-known Eulerian solution: $\;\Omega_1\,=\,
\Omega_{\perp}\,\cos \omega (t-t_0) \;$, $\;\Omega_2\,=\,\Omega_{\perp}\,\sin 
\omega (t-t_0) \;$, $\;\Omega_3\,=\,const \;$, for  $\;\Omega_3\,>\,0 \;$. Our 
choice of signs is correct since it leaves $\;\bf \Omega\;$ parallel to $\;\bf 
J\;$ in the relaxation limit.

Our ultimate goal is to compute the rate of inelastic dissipation
caused by alternating stresses in a wobbling body. To know the picture
of stresses, one should begin with derivation of the acceleration
experienced by a point $\; (x, \; y, \; z) \;$ inside the body. We
mean, of course, the acceleration with respect to 
the inertial coordinate system $\; (X, \; Y, \; Z) \;$, but for
convenience of the further calculations we shall express it in terms
of coordinates   $\; (x, \; y, \; z) \;$ of the body frame  $\; (1, \; 2, \; 
3) \;$. The position, velocity and acceleration in the inertial frame
will be denoted as $\; {\bf{r}}, \; {\bf{v}}, \; {\bf{a}} \;$, while
those relative to the body frame $\; (1, \; 2, \; 3) \;$ will be
called $\; {\bf{r}}', \; {\bf{v}}' \;$ and $\; {\bf{a}}' \;$ (where 
$\; {\bf{r}}' \; = \; {\bf{r}} \;$). The proper acceleration (i.e.,
that relative to the inertial frame) will read:
\be
{\bf{a}} \; \; = \; \; {\bf{a}}' \; + \;{\bf{{\dot{\Omega}}}} \; \times \;
{\bf{r}}' \; + \; 2 \; {\bf{\Omega}} \; \times \; {\bf{v}}' \; + \; 
{\bf{\Omega}} \; \times \; ( {\bf{\Omega}} \times {\bf{r}}' ) \; \; \; \; . 
\label{2.16}
\ee
In the beginning of this section we justified, on the grounds of the strains 
being small, our neglect of $\; {\dot{I_j}}\,{\Omega} \;$ against $\; {I_j}
\, {\dot{\Omega}} \;$. In a similar manner we shall justify the neglection of 
the first and third terms in the right-hand side of the above formula: 
rotation with a period $\; \tau \;$, of a body of size $\; \mathit{l} \;$ will
yield deformations $\; \delta {\mathit{l}} \; \approx \; \epsilon \, 
{\mathit{l}}$ and deformation-caused velocities $\; v' \, \approx \, \delta  
\mathit{l}/\tau \; \approx \; \epsilon \, \mathit{l}/\tau \;$ 
and accelerations $\; a' \, \approx \, \delta \mathit{l}/\tau^{2} \; = 
\; \epsilon \, \mathit{l}/\tau^{2}\;$, $\; \epsilon \;$ being the relative 
strain. We see that $\; v' \;$ and $\; a'\;$ are much less than the velocities 
and accelerations of the body as a whole (that are about $\; {\mathit{l}}/\tau
\;$  and $\; {\mathit{l}}/{\tau}^2\;$, correspondingly). Henceforth,
\ba
{\bf{a}} \; \; \approx \; \; {\bf{{\dot{\Omega}}}} \; \times \;
{\bf{r}}' \; + \; {\bf{\Omega}} \; \times \; ( {\bf{\Omega}} \times 
{\bf{r}}' ) \; = 
\nonumber  \\ 
{{\bf{e}}_1} \; \{ \;
\left[{\dot{\Omega}}_2\;z\; - \;{\dot{\Omega}}_3\;y\protect\right] \; + \; 
{\Omega}_2\; \left(\Omega_1 \; y \; - \; \Omega_2 \; x \protect\right)\; - \;
{\Omega}_3\; \left(\Omega_3 \; x \; - \; \Omega_1 \; z \protect\right)\;
\}  \nonumber  \\
+ \; \;  
{{\bf{e}}_2} \; \{\; 
\left[{\dot{\Omega}}_3\;x\; - \;{\dot{\Omega}}_1\;z\protect\right] \; + \; 
{\Omega}_3\; \left(\Omega_2 \; z \; - \; \Omega_3 \; y \protect\right)\; - \;
{\Omega}_1\; \left(\Omega_1 \; y \; - \; \Omega_2 \; x \protect\right)\;
\}  \nonumber  \\
+ \; \; 
{{\bf{e}}_3} \; \{\; 
\left[{\dot{\Omega}}_1\;y\; - \;{\dot{\Omega}}_2\;x\protect\right] \; + \; 
{\Omega}_1\; \left(\Omega_3 \; x \; - \; \Omega_1 \; z \protect\right)\; - \;
{\Omega}_2\; \left(\Omega_2 \; z \; - \; \Omega_3 \; y \protect\right)\;
\} \; \; =
\nonumber  \\
\nonumber  \\
=\; \; 
{{\bf{e}}_1} \; \{ \;
{\dot{\Omega}}_2\;z\; - \;{\dot{\Omega}}_3\;y \; - \; x \; 
\left(\Omega_2^2 \; + \; \Omega_3^2 \; \protect\right)\; + \;
y \; \Omega_1 \; \Omega_2 \; + \; z \; \Omega_1 \; \Omega_3 \; 
\}  \nonumber  \\
+ \; \;  
{{\bf{e}}_2} \; \{\; 
{\dot{\Omega}}_3\;x\; - \;{\dot{\Omega}}_1\;z \; - \; y \; 
\left(\Omega_3^2 \; + \; \Omega_1^2 \; \protect\right)\; + \;
z \; \Omega_3 \; \Omega_2 \; + \; x \; \Omega_1 \; \Omega_2 \; 
\}  \nonumber  \\
+ \; \; 
{{\bf{e}}_3} \; \{\; 
{\dot{\Omega}}_1\;y\; - \;{\dot{\Omega}}_2\;x \; - \; z \; 
\left(\Omega_1^2 \; + \; \Omega_2^2 \; \protect\right)\; + \;
x \; \Omega_1 \; \Omega_3 \; + \; y \; \Omega_2 \; \Omega_3 \; 
\} \; \; \; .
\label{2.17}
\ea
where, according to (\ref{2.7}) and (\ref{2.11}):
\ba
\Omega_3^2  +  \Omega_2^2 \; = \; P + (1 - Q) \; \beta^2 \; 
{\mathit{sn}}^2\left[\omega (t - t_0), \; k^2\protect\right] \; \;\;,
\label{2.18}
\ea
\ba
\Omega_1^2 +  \Omega_2^2 \; = \; R + (1 - S) \;
\beta^2 \;{\mathit{sn}}^2\left[\omega (t - t_0), \; k^2\protect\right] \;\;\; ,
\label{2.19}
\ea
\ba
\Omega_3^2 \; + \; \Omega_1^2 \; = \; (P \; + \; R) \; - \; (Q \; + \; S) \;
\beta^2\;{\mathit{sn}}^2\left[\omega (t\;-\;t_0),\;k^2\protect\right]\protect\;\; ,
\label{2.20}
\ea
\be
\Omega_3 \; \Omega_1 \; = \; \alpha \; \gamma \; \, {\mathit{dn}} 
\left[\omega (t \; - \; t_0), \;k^2\protect\right] \; \,  {\mathit{cn}} 
\left[\omega (t \; - \; t_0), \;k^2\protect\right] \; \; \; \; ,
\label{2.21}
\ee
\be
{\dot{\Omega}}_2\;=\;\beta \;\omega \; {\mathit{cn}}\left[\omega (t - t_0), \;
k^2\protect\right]\;\,{\mathit{dn}}\left[\omega (t - t_0), \;k^2\protect\right]
\label{2.22}
\ee
the derivative being written in compliance with (Abramovitz \& Stegun 1964)$^{
27}$, equation 16.16.1. The other items emerging in (\ref{2.17}) will read, for
$\;{\textbf{J}}^2\;>\;2\,I_2\,T_{\small{kin}}\;$:   
\be
\Omega_3 \; \Omega_2 \; = \; \alpha \; \beta \; \, {\mathit{dn}} 
\left[\omega (t \; - \; t_0), \;k^2\protect\right] \; \,  {\mathit{sn}} 
\left[\omega (t \; - \; t_0), \;k^2\protect\right] \; \; \; \; ,
\label{2.23}
\ee
\be
\Omega_2 \; \Omega_1 \; = \; \beta \; \gamma \; \, {\mathit{cn}} 
\left[\omega (t \; - \; t_0), \;k^2\protect\right] \; \,  {\mathit{sn}} 
\left[\omega (t \; - \; t_0), \;k^2\protect\right] \; \; \; \; ,
\label{2.24}
\ee
\be
{\dot{\Omega}}_1\; =\;-\;\gamma\;\omega\;{\mathit{sn}}\left[\omega(t - t_0),\; 
k^2\protect\right]\;\,{\mathit{dn}}\left[\omega (t-t_0),\;k^2\protect\right] 
\;\;\;\;,
\label{2.25}
\ee
and
\be
{\dot{\Omega}}_3\;=\;-\;\alpha\;\omega \;k^2 \; {\mathit{cn}}\left[\omega
(t-t_0),\; 
k^2\protect\right]\;\,{\mathit{sn}}\left[\omega (t - t_0), \;k^2\protect\right]
\;\;\;\;,
\label{2.26}
\ee
while for $\; {\textbf{J}}^2 \; < \; 2 \, I_2 \, T_{\small{kin}}\;$:
\be
\Omega_3 \; \Omega_2 \; = \; \alpha \; \beta \; \, {\mathit{cn}} 
\left[\omega (t \; - \; t_0), \;k^2\protect\right] \; \,  {\mathit{sn}} 
\left[\omega (t \; - \; t_0), \;k^2\protect\right] \; \; \; \; ,
\label{2.27}
\ee
\be
\Omega_2 \; \Omega_1 \; = \; \beta \; \gamma \; \, {\mathit{dn}} 
\left[\omega (t \; - \; t_0), \;k^2\protect\right] \; \,  {\mathit{sn}} 
\left[\omega (t \; - \; t_0), \;k^2\protect\right] \; \; \; \; ,
\label{2.28}
\ee
\be
{\dot{\Omega}}_1\;=\;-\;\gamma \; \omega \; k^2 \; {\mathit{sn}}\left[\omega
(t - t_0), 
\;k^2\protect\right]\;\,{\mathit{cn}}\left[\omega (t - t_0), \;k^2
\protect\right]
\;\;\;\;,
\label{2.29}
\ee
and
\be
{\dot{\Omega}}_3\;=\;-\;\alpha \; \omega \; {\mathit{sn}}\left[\omega 
(t - t_0), 
\;k^2\protect\right]\;\,{\mathit{dn}}\left[\omega y(t - t_0), \;k^2\protect\right]
\label{2.30}
\ee

\pagebreak

\section{Almost Prolate Body}

Before pursuing the generic case, let us dwell for a minute on the case of an 
almost 
symmetric prolate top of $\; I_{3} \;$ and  $\; I_{2} \;$ having close values: 
\be
I_{3} \; \stackrel{>}{\sim} \; I_{2} \; > \; I_{1} \; \; \; ,
\label{3.1}
\ee
i.e., 
\be
I_3 \; - \; I_1 \; \stackrel{>}{\sim} \;  I_2 \; - \; I_1  \; \gg
\; I_3 \; - \; I_2 \; \stackrel{>}{\sim} \; 0 \; \; \; \; .
\label{3.2}
\ee
As we saw above, the solution depends upon the sign of $\; ({\textbf{J}}^2 \; 
- \; 2 \, I_2 \,T_{\small{kin}})\;$. According to ({\ref{2.5}} - {\ref{2.6}}), 
\be
{\textbf{J}}^2 \; - \; 2 \, I_2 \, T_{\small{kin}}\;=\;\Omega_3^2\;I_3^2 \;-
\;\Omega_3^2\;I_3\;I_2\;
+\;\Omega_1^2\;I_1^2\;-\;\Omega_1^2\;I_1\;I_2\;=\;
\Omega_3^2\;I_3^2 \; \left[\frac{I_3\;-\;I_2}{I_3}\;-\;
\frac{\Omega_1^2}{\Omega_3^2}\;\frac{I_1}{I_3}\;\frac{I_2\;-\;I_1}{I_3}
\protect\right] 
\label{3.3}
\ee
Based on (\ref{3.1} - \ref{3.2}), one can introduce the
following parameters: 
\be
q\;\equiv\;\frac{I_3\;-\;I_2}{I_3}\;\;\;\;,
\label{3.4}
\ee
and
\be
s\;\equiv\;\frac{I_1}{I_3}\;\;\;\;.
\label{3.5}
\ee
Besides, we shall use a time-dependent quantity $\;
{\Omega_1^2}/{\Omega_3^2}\;$. It is not, of course, a geometrical 
parameter worthy of the name, though {\textit{formally}} one may
consider it as a sort of parameter in (\ref{3.3}). This wanna-be
parameter may be used as a measure of the system's approaching the
steady regime: {\textit{if $\;{\Omega_1^2}/{\Omega_3^2}\;\ll\;1\;$
over the entire time of one wobble then the relaxation is almost over, and the 
angular velocity $\;{\bf{\Omega}}\;$ is precessing about $\;{\bf{J}}\;$, 
with a small amplitude (i.e., describing a narrow cone).}} Below, the meaning 
of the words like {\textit{small}} and {\textit narrow} will become 
understandable. Meanwhile, one can write down 
(\ref{3.3}) as
\be
{\textbf{J}}^2 \; - \; 2 \; I_2 \; T_{\small{kin}}\;=\;\Omega_3^2\;I_3^2 \;
\left[q\;-\;\frac{\Omega_1^2}{\Omega_3^2}\;s\;(1\;-\;s\;-\;q\;)\protect\right] 
\; \; \; \; ,
\label{3.6}
\ee
and easily find that {\textit{for fixed values of}} $\;q\;$ and $\;s\;$ (i.e., 
for a particular prolate body) the narrowness of the precession cone yields: 
\be
{\textbf{J}}^2\;-\;2\;I_2\;T_{\small{kin}}\;\approx\;\Omega_3^2\;I_3^2
\;q \;\;\;.
\label{3.7}
\ee
This approximation becomes true at the late stage of relaxation, when
\be
\frac{\Omega_1^2}{\Omega_3^2}\;\ll\;q\;\frac{1}{s(1-s-q)} 
\label{3.8}
\ee
holds through the duration of one wobble. Assume, following (Black, 
Nicholson, Bottke, Burns $\&$ Harris 1999)$^{25}$, that the moments of inertia of
asteroid (433) Eros relate as $\;\;\;I_1\;:\;I_2\;:\;I_3\;\;=\;\;1\;:\;3\;:\;
3.05\;$. With these numbers plugged in, the above formula will give: 
$\;|\Omega_1/\Omega_3|\;\ll\;0.4\;$. {\textit{Formally}}, (\ref{3.7}) was 
derived from (\ref{3.6}) by keeping $\;q\;$ and 
 $\;s\;$ fixed, and making the "parameter" $\;{\Omega_1^2}/{\Omega_3^2}\;$ 
approach zero. After this is done, one may consider a variety of geometries, 
and make $\;q\;$ approach zero. Then $\;{\textbf{J}}^2\;-\;2\;I_2\;
T_{\small{kin}}\;$ will approach zero, always remaining {\textit{positive}}, 
so that the end of relaxation will be described by the solution (\ref{2.11}, 
\ref{2.12}) with frequency $\;\omega\;$ expressed by (\ref{2.13}).

On the other hand, one might as well perform a different, unphysical trick: 
for a fixed $\;\Omega_1\;$, begin with a limit $\;q\; \rightarrow \;0\;$, and 
only afterwards choose the case of the small-amplitude wobble (i.e., consider 
$\; {\Omega_1^2}/{\Omega_3^2}\; \rightarrow \;0\;$). The first limit 
will give:
\be
{\textbf{J}}^2\;-\;2\;I_2\;T_{\small{kin}}\;\approx\;-\;\Omega_3^2\;I_3^2
\; \frac{\Omega_1^2}{\Omega_3^2}\; s\;(1-s)\;\;\;,
\label{3.9}
\ee  
After the second limit is taken, $\;{\textbf{J}}^2\;-\;2\;I_2\;
T_{\small{kin}}\;$ will approach zero, always remaining {\textit{negative}}, 
so that the final stage of relaxation would obey the solution (\ref{2.11}, 
\ref{2.14}) with frequency $\;\omega\;$ expressed by (\ref{2.15}).

We see that the operations $\;q\;\rightarrow\;0\;$ and 
$\;{\Omega_1^2}/{\Omega_3^2}\;\rightarrow\;0\;$ do not commute. From the 
physical point of view, an observer studying, for a variety of samples, the 
end of relaxation should first fix the shape of the body (i.e., assume, for 
example, that $\;q\;$ is small enough but constant). Only afterwards he may 
state that he is interested only in the final spin, i.e., assume 
$\;{\Omega_1^2}/{\Omega_3^2}\;\ll\;q\;s^{-1}\;(1-s-q)^{-1}\;$. As explained 
above, this observer will see that the end of relaxation takes place at 
frequency $\;\omega\;$ expressed by (\ref{2.13}).

An opposite order of limits would be physically meaningless, in that it would 
not help us to describe the behaviour of a particular body. 

We had to dwell on this issue so comprehensively because it would be very 
important to understand better the following two statements made in 
(Black, Nicholson, Bottke, Burns $\&$ Harris 1999)$^{25}$: 

1). In the limit where $\;I_1\;=\;I_2\;$, the circulation region around the 
maximal-inertia axis vanishes, and 

2). All trajectories circulate around the minimal-inertia axis. 
(This statement is fortified by the following argument: ``the slightest 
perturbation would cause such an object to ``roll'' about its long axis.'') 

To analyse this statement by {{Black, Nicholson, Bottke, Burns $\&$ Harris}}
$^{25}$, let us cast it in a more exact form. First of all, we should 
understand which of the two possible sequencies of limits these authors 
implied. In case their statement implied the limit $\;q\;\rightarrow\;0\;$ 
taken first, and the relaxation limit $\;{\Omega_1^2}/{\Omega_3^2}\;
\rightarrow\;0\;$ taken afterwards, then according to (\ref{2.15}) the 
frequency of precession $\;\omega \;$ will read as
\ba
\omega \;=\;
\sqrt{\frac{\left(2 \; I_3 \; T_{\small{kin}} \; - \; {\bf{J}}^2
\protect\right)\; \left(I_2 \; - \; I_1\protect\right)}{I_1\;I_2\;I_3}}
\;=\;\sqrt{\frac{\left(\Omega_2^2\;I_2\; (I_3\;-\;I_2)\;+\;
\Omega_1^2\;I_1\;(I_3\;-\;I_1)\protect\right)\;\left(I_2\;-\;
I_1\protect\right)}{I_1\;I_2\;I_3}}\;\rightarrow
\nonumber\\
\nonumber\\
\rightarrow\;\sqrt{\frac{\left(I_3\;-\;I_1\protect\right)\;\left(I_2\;-\;
I_1\protect\right)}{I_2\;I_3}}\;|\Omega_1|\;\;\; , \;\;\; for 
\;\;\; q\;\rightarrow\;0\;\;\;and\;\;\;\Omega_1\;\;being\;\;fixed.\;\;\;\;\;\;\;\;\; 
\label{3.10}
\ea
It will then approach zero as $\;|\Omega_1|\;$. However, as explained above, 
such a sequence of limits would be unphysical, while a physical way is to fix 
the body shape first (i.e., to fix the difference $\;(I_3\,-\,I_2)\;$), then 
to take the relaxation limit $\;{\Omega_1^2}/{\Omega_3^2}\;\rightarrow\;0\;$, 
and only after that to let $\;q\;$ approach zero. In this case, (\ref{2.13}) 
will yield $\;\omega\;$ approach zero as $\;\sqrt{q}\;$, i.e., as $\;\sqrt{I_3
\,-\,I_2}\;$:
\ba
\omega\;=\;\sqrt{\frac{\left({\bf{J}}^2 \; - \; 2 \; I_1 \; T_{\small{kin}}
\protect\right)\; \left(I_3 \; - \; I_2\protect\right)}{I_1\;I_2\;I_3}}\;=\;
\sqrt{\frac{\left( \Omega_2^2\;I_2\; (I_2\;-\;I_1) \;+\;
\Omega_3^2\;I_3\; (I_3\;-\;I_1)\protect\right)\;\left(I_3
\;-\;I_2\protect\right)}{I_1\;I_2\;I_3}}\;\rightarrow
\nonumber\\
\nonumber\\
\rightarrow\;\sqrt{\frac{\left(I_3\;-\;I_1\protect\right)\;\left(I_3\;-\;
I_2\protect\right)}{I_2\;I_3}}\;|\Omega_3|\;\;\;,\;\;\; for \;\;\; \Omega_2\;
\rightarrow\;0\;\;\;and\;\;the\;\;shape\;\;being\;\;fixed.\;\;\;\;\;\;\;\;\;
\label{3.11}
\ea  
The above, physically meaningful, expression coincides with formula (2) in 
(Black, Nicholson, Bottke, Burns $\&$ Harris 1999)$^{25}$, which means 
that the 
authors chose the right sequence of limits. We certainly agree with the 
first of the above two statements derived by the authors from this formula: 
in the limit of $\;I_3\;=\;I_2\;$, the circulation region around the 
maximal-inertia axis vanishes. To understand the second of the above two 
statements made in (Black, Nicholson, Bottke, Burns $\&$ Harris 1999)$^{25}$, 
i.e., 
that {\textit{``the slightest perturbation would cause such an object to 
``roll'' about its long axis''}}, note that the relaxation limit was achieved 
in (\ref{3.11}) by letting $\;\Omega_2\;$ vanish, with no assumptions made 
about $\;\Omega_1\;$. This happened because $\;\omega \;$ remains $\;
\Omega_1-$independent as long as (\ref{2.13}) (and therefore (\ref{3.11})) may
be used. 
These formulae may be used when the right-hand side of (\ref{3.3}) (and that 
of (\ref{3.6})) is positive, i.e., when in (\ref{3.8}) we have at least 
``$<$'' (not necessarily ``$\ll$''). In the case of asteroid (433) Eros, for 
example, this will work as long as, approximately, $\;|\Omega_1|\;<\;0.4\,
|\Omega_3|$. 
Thus we must agree with the statement about ``rolling'' around the 
minimal-inertia axis, but we have to add an important comment to it: 

{\textit{As long as this rolling is slow enough, it will leave the spinning 
body within the realm of solution (\ref{2.12}), (\ref{2.13}) appropriate to 
the final stage of relaxation. Too fast rolling will make it obey a different 
solution, (\ref{2.14}), with $\;\omega \;$ expressed by (\ref{2.15}).}} 

As an illustration, let us consider, in the space  
$\;\Omega_1\;,\;\Omega_2\;,\;\Omega_3\;,$ the angular-momentum ellipsoid 
\be
{\bf{J}}^2 \; = \; I_1^2 \; \Omega_1^2 \; +  \; I_2^2 \; \Omega_2^2 \; + 
\; I_3^2 \; \Omega_3^2 \; \; \; ,
\label{3.12}
\ee
and mark on its surface the lines of its intersection with the kinetic-energy 
ellipsoids 
\be
2 \; T_{\small{kin}} \; = \; I_1 \; \Omega_1^2 \; +  \; I_2 \; \Omega_2^2 \; + 
\; I_3 \; \Omega_3^2 \; \; \; ,
\label{3.13}
\ee
at different values of energy. Let, on $\;Fig.\,1\;$, the starting point of 
motion be somewhere close to the pole $\,A\,$: the vector $
\;{\bf{\Omega}}\;$ initially is almost perpendicular to the major-inertia axis 
(3). Precession of the body will correspond to vector $\;{\bf{
\Omega}}\;$ describing a constant-energy ``circle'' in $\;Fig.\,1\;$. The 
word ``circle'' is standing in quotation marks because this trajectory is 
circular as long as we do not approach the separatrix too close. In the cause 
of the body rotation, 
$\;{\bf{\Omega}}\;$ will be gradually changing the ``circles'' it describes, 
and will eventually approach the separatrix, $\;{\it{en \; \, route}}$ whereto
the ``circles'' will be getting more and more distorted. If the dissipation is 
slow, i.e., if the kinetic-energy loss through one precession period is less 
than a typical energy of an occational interaction (like, say, a
 tidal-force-caused perturbation) then chaotic motion is possible when $\;{\bf{
\Omega}}\;$ is crossing the separatrix: the body may perform flip-overs. After 
that the body will embark on the stage of tumbling. As one can see from $\;
Fig.\,1\;$, the tumbling will eventually turn into the final spin, i.e., into an 
almost circular small-amplitude precession of $\;{\bf{\Omega}}\;$ around point 
$\,C\,$. However, this point will never be reached because the alignment of $\;
{\bf{\Omega}}\;$ along $\;{\bf{J}}\;$ has a vanishing rate 
for small residual angles: it is evident from formulae (\ref{8.14}) and 
(\ref{9.8}) below, that at the end of the relaxation process the relaxation 
rate approaches zero, so that small-angle nutations can persist for long times.

Now we understand that if the afore mentioned ``rolling'' becomes too swift, 
this will look as a ``jump'' over the separatrix in {\textit{Fig.1}}. If we 
assume that $\;(I_3\,-\,I_2)/I_3\;$ is infinitesimally small, then the 
separatrix will approach pole $C$ infinitesimally close, and the smallest tidal
interaction will be able to push the vector $\;{\bf{\Omega}}\;$ across the 
separatrix. In other words, {\bf{the body, during the most part of its history,
will be precessing about its minimal-inertia axis.}} For the first time this 
fact was pointed out in ({{Black, Nicholson, Bottke, Burns $\&$ Harris}} 
1999)$^{25}$.

Dependent upon the particular value of $\;(I_3\,-\,I_2)/I_3\;$ 
and upon the intensity of the occational tidal interaction, the vector 
$\;{\bf{\Omega}}\;$ will be either driven from pole $\;C\;$ back to the 
separatrix, without crossing it, or will be forced to ``jump'' over it. 
In the latter case, chaotic flipovers will emerge while $\;{\bf{\Omega}}\;$ 
is crossing the separatrix. 

As already mentioned, $\;{\bf{\Omega}}\;$ will never approach pole $\;C\;$ 
too close because in the vicinity of $\;C\;$ the relaxation rate 
asymptotically vanishes. The behaviour of $\;{\bf{\Omega}}\;$ after its 
crossing the separatrix will be determined by two factors: on the one hand, 
occasional tidal interactions will push
$\;{\bf{\Omega}}\;$ towards or over the separatrix; on the other hand, the 
inelastic-dissipation process will 
always drive $\;{\bf{\Omega}}\;$ in the direction of pole $\;C\;$ (though, 
once again, it will never manage to bring it too close to $\;C\;$, for the 
above mentioned reason). Sometimes this regime will be interrupted by  
collisions which may drive $\;{\bf{\Omega}}\;$ far away from the separatrix, 
in the direction of pole $\;A\;$.

\section{Stages of Motion}

We shall consider motion of a freely rotating body moving through 
four stages of relaxation. 

The first stage will be called ``the initial spin''. It begins when the body 
rotates about some axis (almost) perpendicular to that of major inertia. This 
motion is characterised by negative $\;({\textbf{J}}^2\;-\;2\;I_2\;
T_{\small{kin}})\;$, so that the frequency $\; \omega \;$ and parameter $ \; k \; $
are expressed by (\ref{2.15}). The initial spin starts when $\;\Omega^2_3\;$ 
is small. Namely, according to (\ref{3.3}),
\be
\frac{\Omega^2_1}{\Omega^2_3}\; > \; \frac{I_3-I_2}{I_2-I_1} \; \frac{I_3}{I_1}
\;=\;\frac{q}{s\;(1\,-\,q\,-\,s)} \; \; \; \; \; .
\label{4.1}
\ee
Note that  $\;\Omega^2_2\;$ does not enter this condition at all. 

Now suppose that the body starts its rotation about an axis that is, for 
example, close 
to the minimal-inertia axis, so that  $\;\Omega^2_2\;$ is much less than $\;
\Omega^2_1\,$. Then, according to (\ref{2.5}), (\ref{2.6}) and (\ref{2.15}),
\ba
\nonumber
\omega \; = \protect\sqrt{\left( \frac{I_3-I_1}{I_2}\;\Omega_1^2 \; + \; 
\frac{I_3-I_2}{I_1}\;\Omega_2^2 \protect\right)\frac{I_2-I_1}{I_3}}\;=\\
\nonumber\\
=\;\protect\sqrt{\left( \frac{1-s}{1-q} \; \Omega_1^2 \; + \; 
\frac{q}{s}\;\Omega_2^2 \protect\right)(1-s-q)}\;\;\approx
\; |{\Omega_1}|\;(1\,-\,s)\;\;\;\;\;\;\;\;\;\;
\label{4.2}
\ea
and
\be
k^2 \; = \; \frac{I_3-I_2}{I_2-I_1} \; \; \frac{
I_2\;\left(I_2\;-\;I_1\protect\right)\Omega_2^2\;+\;
I_3\;\left(I_3\;-\;I_1\protect\right)\Omega_3^2}{
I_1\;\left(I_3\;-\;I_1\protect\right)\Omega_1^2\;+\;
I_2\;\left(I_3\;-\;I_2\protect\right)\Omega_2^2} \; < \; 1 \;\;\;\;,
\label{4.3}
\ee
the inequality ensuing from (\ref{3.1}). In particular, for
\be
\Omega^2_2\;\ll\;\Omega^2_1\; \frac{I_1}{I_2} \; \frac{I_3-I_1}{I_3-I_2} \; = 
\; \Omega^2_1\; \frac{s}{1-q} \; \frac{1-s}{q}
\label{4.4}
\ee
and
\be
\Omega^2_3\;\ll\;\Omega^2_1\;\,\frac{I_1}{I_3}\;\,\frac{I_2-I_1}{I_3-I_2} \; = 
\; \Omega^2_1\; s \,\;\frac{1-s-q}{q}
\label{4.5}
\ee
one gets:
\be
k^2\; \ll\;1 \; \; \; \; .
\label{4.6}
\ee
The initial spin comes to its end when the (negative) quantity $\;{\bf{J}}^2\,
-\,2\,I_2\,T_{\small{kin}}\;$ approaches zero, and $\;k\;\rightarrow \; 1 \;$. 
The next, second stage will be precession in the vicinity of separatrix 
(though without crossing it yet). Crossing of the separatrix may result in 
chaotic flipovers. The third stage is that of tumbling. 
It begins when $\;({\textbf{J}
}^2\;-\;2\;I_2\;T_{\small{kin}})\;=\;0\;$ and $\;k\;=\;1 \;$. It ends when $\;
({\textbf{J}}^2\;-\;2\;I_2\;T_{\small{kin}})\;>\;0\;$ and $\;k\;$ is smaller 
than unity, though not small enough to approximate the Jacobi functions by 
trigonometric functions. Mind that the transition from the initial spin to 
tumbling leaves parameters $\,\alpha\,$, $\,\beta\,$, $\,\gamma\,$, $\,
\omega\,$ and $\, k \,$ continuous: none of these undergo a stepwise change. 
The fourth stage will be called ``the final spin'': it takes place when the 
relaxation is almost over and  $\;k\;\rightarrow\;0\;$. At this stage (as well
as during the initial spin) the Jacobi functions may be well approximated by 
trigonometric functions.

As explained in the previous section, the suggested scenario is, of course, 
too idealised: on the one hand, small occasional interactions will easily 
force a nearly-prolate body to rotate around its minimal-inertia axis 
for most time; on the other hand, the relaxation rate will vanish in the 
vicinity of pole $\;C\;$, so that the perfect relaxation will never be 
achieved.  

\section{Dynamics of a Freely Precessing Body. Relaxation Rate.}

We are interested in the rate at which a freely spinning body changes its 
orientation in space, i.e., in the rate of alignment of the maximal-inertia 
axis along the (conserved) angular momentum. 

In the case of an oblate body ($\;I_3\,>\,I_2\,=\,I_1\;$), one could 
start with a trivial formula $\;d \theta /dt\; = \; \left(dT_{kin}/d\theta
\right)^{-1} \;\left(dT_{kin}/dt\right)\;$
where $\;\theta\;$ stands for the angle between the major-inertia axis and the 
angular momentum (Efroimsky \& Lazarian 2000)$^{24}$. This formula would work 
since in the oblate case $\;\theta\;$ remains almost unchanged through a
precession cycle. Unfortunately,
in the general case of a triaxial rotator, even in the absence of dissipation, 
this angle evolves in time. Luckily though, its evolution is periodic (formulae
(A1) - (A4) in Appendix A), so that instead of using  $\;\theta\;$ one 
can use its average over a cycle. In practice, it turns out to be easier 
to operate with the time-average of $\;\sin^2 \theta\;$: 
\be
\frac{d\;<\sin^2 \theta >}{dt}\;=\;\frac{d\;<\sin^2\theta>}{dT_{kin}}\;\;
\frac{dT_{kin}}{dt}\;\;\;.\;\;\;\;
\label{5.1}
\ee
As shown in Appendix A, formula (A10),
\ba
\nonumber
{\bf J}^2\;\,<\sin^2 \theta>\;=\;\;\;\;\;\;\;\;\;\;\;\;\\
\nonumber\\
=\;\frac{I_1}{I_3\,-\,I_1}\,\left(2\,I_3\,T_{kin}\;-\;{\bf J}^2 \right)\;+\;
I_2\;I_3\;\frac{I_2\,-\,I_1}{I_3\,-\,I_1}\;\beta^2\;\frac{1}{2}\;
\left(1\;+\;\frac{k^2}{8}\;+\;\frac{k^4}{16}\;+\;O(k^6)\right)\;\;.\;
\label{5.2}
\ea
A substitution of (\ref{2.13}) for $\;\beta\;$ and $\;k\;$ into the above 
formula entails (see Appendix A, equation (A12)):
\ba
\nonumber
{\bf J}^2\;\left(\frac{d\;<\sin^2 \theta>}{dT_{kin}}\right)_{(near\;C)}
=\;\frac{I_3\,(I_1\,I_3\,+\,I_2
\,I_3\,-\,2\,I_1\,I_2)}{
(I_3\,-\,I_1)\,(I_3\,-\,I_2)}\;+\\
\nonumber\\
+\;\frac{1}{4}\;\frac{2\,I_3\,T_{kin}\;-\;{\bf J}^2}{{\bf J}^2\,-\,2\,I_1\,T_{kin}}\;\left(\frac{I_2\,-\,I_1}{I_3\,-\,I_2}\right)^2\;
\frac{I_3^2}{I_3\,-\,I_1}\;+\;O(k^4)\;\;.\;
\label{5.3}
\ea
while a substitution of (\ref{2.15}) into (\ref{5.2}) yields (Appendix A, 
formula (A15)):
\ba
{\bf J}^2\;\left(\frac{d\;< \sin^2 \theta >}{dT_{kin}}\right)_{(near\;A)}=
\;\frac{I_1\,I_3}{I_3\,-\,I_1}\;-\;\frac{1}{4}\;\frac{I_1\,I_3\,(I_3\,-\,I_2
)}{(I_3\,-\,I_1)\,(I_2\,-\,I_1)}\;\frac{{\bf J}^2\,-\,2\,I_1\,T_{kin}}{2\,I_3
\,T_{kin}\;-\;{\bf J}^2}\;+\;O(k^4)\;.\;
\label{5.4}
\ea
Formulae (\ref{5.3}) and (\ref{5.4}) explain how the losses of 
the kinetic energy of rotation make $\;<\sin^2 \theta>\;$ change.
Since the kinetic energy decreases because of the inelastic dissipation,  
\be
\dot{T}_{kin} \; = \; \dot{W} \;\;\;,\;\;\;
\label{5.5}
\ee
what we have to find is the rate of the elastic-energy losses $\; \dot{W}$, 
quantity $W$
being the time-dependent part of the elastic energy stored in the body due to 
the alternating stresses. Then, with aid of (\ref{5.1}), (\ref{5.3}), 
(\ref{5.4}), we shall compute the rate of alignment:
\be
\frac{d\,<\sin^2\theta>}{dt}\;=\;\frac{d\,<\sin^2\theta>}{dT_{kin}}\;
\frac{dW}{dt}\;
\label{5.6}
\ee
The next four sections will be devoted to the calculation of the 
dissipation rate $\;dW/dt\;$.


\section{Essential Nonlinearity in the Precession-Caused \\
Dissipation. Cases of Hot and Cold Bodies.}

Our goal now is to describe the kinetic-energy dissipation caused by the 
deformations of the body, experienced in the course of its precession. The 
deformation of body is neither purely elastic nor purely plastic, but is a 
superposition of the former and the latter. It is then to be described by the 
tensor 
$\epsilon_{\it{ij}}$ of {\textit {viscoelastic}} strains and by the velocity 
tensor consisting of the time-derivatives $\dot{\epsilon}_{\it{ij}}$. The 
stress tensor will consist of two components: the elastic stress 
and the plastic (viscous) stress. In the simpliest, so-called Maxwell-Voigt 
model, the components are additive (Tschoegl 1989)$^{28}$:
\be
\sigma_{\it{ij}} \; \; = \; \; \sigma_{\it{ij}}^{(e)} \; \; + 
\sigma_{\it{ij}}^{(p)}~~~,
\label{6.1}
\ee
where the components of the elastic stress tensor are interconnected with 
those of the strain tensor (Landau and Lifshitz 1970)$^{36}$:
\be
\epsilon_{ij} \; \; = \; \; \delta_{ij} \; \; \frac{Tr \; 
\sigma^{(e)}}{9 \; K} \; \; + \; \; 
\left( \; \sigma_{\it{ij}}^{(e)} \; \; - \; \; \frac{1}{3} \; \; \delta_{ij} 
\; \; Tr \; \sigma^{(e)} \right) \; \frac{1}{2 \; \mu} \; \; \; ~~~,
\label{6.2}
\ee
\be
\sigma_{ij}^{(e)} \; \; = \; \; K \; \delta_{ij} \; \; Tr \; \epsilon \; \; + 
\; \; 2 \; \mu \; \left(\epsilon_{ij} \; - \; \frac{1}{3} \; \delta_{ij} \; 
\; Tr \; \epsilon \right)~~~,
\label{6.3}
\ee
$\mu$ and $K$ being the {\textit{adiabatic}} shear and bulk moduli, and $Tr$ 
standing for the trace of a tensor. Components of the plastic stress are 
connected with the strain derivatives as 
\be
\dot{\epsilon}_{ij} \; \; = \; \; \delta_{ij} \; 
\frac{Tr \; \sigma^{(p)}}{9 \zeta} \; \; + \; \; \left(\sigma_{ij}^{(p)} \; - 
\; \frac{1}{3} \; \delta_{ij} \;\; Tr \; \sigma^{(p)} \right) \; 
\frac{1}{2 \eta}~~~,
\label{6.4}
\ee
\be
\sigma_{\it{ij}}^{(p)} \; \; = \; \; \zeta \; \delta_{ij} \; \; 
Tr \; {\dot{\epsilon}} \; \; + \; \; 2 \; \eta \; \left(\dot{\epsilon}_{ij}
\; - \; \frac{1}{3} \; \delta_{ij} \; \; Tr \; \dot{\epsilon} \right)~~~, 
\label{6.5}
\ee
where  $\; \eta \;$ and $\; \zeta \;$ are the shear and stretch viscosities.

Dissipation may be taking place at several modes:
\be
\dot{W}\;=\;\sum_{\omega_n} \; \dot{W}{({\omega_n})}\;=\;-\;\sum_{\omega_n}\;
\frac{\omega_n\;W_0({\omega}_n)}{Q({\omega_n})}\;=\;-\;2\;\sum_{\omega_n}\;
\frac{\omega_n\;\,<W({\omega}_n)>}{Q({\omega_n})}\;\;\;,\;\;  
\label{6.6}
\ee
$\;Q(\omega)\;$ being the so-called quality factor of the material, and  
$\;W_0({\omega}_n)\;$ and $\;\,<W({\omega}_n)>\,\;$ being the maximal and 
the average (over a period) values of the appropriate-to-$\omega_n\;$ fraction 
of elastic energy stored in the body. The average (over the precession cycle) 
of the total elastic energy reads 
\be
<W>\;=\;\frac{1}{2}\;\int\;dV\;<\sigma_{ij}\;\epsilon_{ij}> \;\;\;\;,\;\;\;
\label{6.7}
\ee
and it must be decomposed in a sum over the frequencies:
\be
<W>\;=\;\sum_{n}\;\,<W(\omega_n)>\;\;\;\;,\;\;\;
\label{6.8}
\ee
For example, 
in the case of a symmetrical oblate body, studied in (Lazarian \& Efroimsky 
1999)$^{}$ and (Efroimsky \& Lazarian 2000)$^{}$, both the stress tensor 
$\;\sigma\;$ and the strain tensor $\;\epsilon\;$ contain only the precession 
frequency $\;\omega\;$. Therefore their contraction $\;\sigma_{ij}\,
\epsilon_{ij}\;$ contains two frequencies: $\;\omega\;$ and $\;2\omega\;$, and 
hence in this case $\;dW/dV \; = \; dW^{(\omega)}/dV \; + \; dW^{(2\omega)}/dV
\;$.

All in all, the general expression (\ref{6.6}) entails:
\be
\dot{W}\;=\;-\;2\;\sum_{\omega_n}\;\int\;dV\;\left\{\frac{\omega_n}{
Q({\omega_n})}\;\,\frac{d\;<W(\omega_n)>}{dV}\right\}\;\;\;,\;\;
\label{6.9}
\ee
where the integral is taken over the entire volume $V$ of the body. In the 
latter expression we have deliberately put the quality factor under the 
integral, implying its possible coordinate-dependence. The coordinate 
dependence of attenuation should be taken into account whenever one is dealing
with precession of an inhomogeneous body. We mean, for example, the problem of 
rotational stability of a spacecraft. Wobble of a strongly inhomogeneous 
asteroid is another example of relevance of the coordinate dependence of $Q$.

Returning to (\ref{6.6}), it is important to stress that {\bf the dissipation
process is {\underline{essentially}} nonlinear: the generation of the higher 
modes in (\ref{6.6}) is no way to be a higher-order correction. Instead, it is 
the higher-than-$\omega$ frequencies that contribute the overwhelming share of 
the entire effect.} This, crucial circumstance had gone unnoticed in the 
preceding studies (Burns and Safronov 1973)$^{23}$, (Purcell 1979)$^{}$, and 
was studied only this year in (Lazarian \& Efroimsky 1999)$^{16}$ and 
(Efroimsky \& Lazarian 2000)$^{24}$. In the latter two articles we were 
considering a simple case of a symmetrical oblate body ($I_1=I_2$). In that 
case, the second mode was generated due to the quadratic dependence of the 
centripetal acceleration upon the angular velocity $\;\Omega\;$: since the 
angular velocity of an oblate body precesses at a rate $\;\omega\,=\,(h-1)
\Omega_3\;$ (where $\;h\equiv\,I_3/I_1\,=\,I_3/I_2$), the emergence of 
double-frequency terms in the expression for acceleration (and therefore, in 
the expressions for stresses, strains, elastic energy, and finally, in the 
expression for the relaxation rate) is unavoidable. For the first time the 
presence and role of the double-frequency terms was discussed in (Lazarian \& 
Efroimsky 1999)$^{16}$, in the context of cosmic-dust alignment, and in 
(Efroimsky \& Lazarian 2000)$^{24}$, in the context of cometary and asteroid 
wobbling. It turns out that this second mode often gives a leading input into 
the dissipation process. This is an example of a nonlinearity giving birth to a
leading-order effect. It remains a puzzle as to why this, leading, effect had 
not been studied thitherto. It would be though unfair to say that the effect 
had gone completely unnoticed. After our two articles had been published, 
Vladislav Sidorenko drew our attention to the fact that the second mode had 
been mentioned back in fifties by (Prendergast 1958)$^{29}$ and then forgotten.
Prendergast took into account the centripetal acceleration but missed the term 
$\;{\dot{\bf \Omega}}\,\times\,{\bf r}\;$ in his analysis. He also ignored the 
emergence of the higher modes. Anyway, we would credit Prendergast for first 
noticing the nonlinear nature of the process.

In 1973 Peale published an article (Peale 1973) devoted to inelastic relaxation
of nearly spherical bodies, where he did take the second harmonic into account.

Since in the case of an oblate symmetrical body only two modes are 
present, formula (\ref{6.6}) for $\dot{W}$ simplifies a lot: $
{\dot{W}}_{oblate} = \, \dot{W}({\omega})
\, + \, \dot{W}{(2{\omega})}\,=\,-\,2\,[\,\omega\;<{W{({\omega})}}>/{Q{({
\omega})}}\;+\;2\omega\;<{W{({2\omega})}}>/{Q{({2\omega})}}]\;\approx\;-\,
2\,{\omega}\;\,<W{({\omega})} \; + \; 2 W{({2\omega})}{>}\,/Q(
\omega)\;$ where we used the fact that the quality factor depends upon the 
frequency very slowly: $\;Q{({2\omega})}\,\approx\,Q{({\omega})}\,$. This  
neglection of the frequency-dependence of $Q$ is certainly valid when we 
consider inputs from frequencies differing from one another by a factor of 
two. However, in the generic case, when a broader band of frequencies comes 
into play, the frequency-dependence of the quality factor in (\ref{6.6}) must 
be respected. This dependence may be crucial if the body is a composite 
structure with resonant eigenfrequencies of its own: attenuation at these 
may be especially effective. 

As we shall see below, whenever the rotation axis is (almost) parallel either 
to the maximal- or minimal-inertia axis of the body, the dissipation is taking
place on two frequencies solely: $\;\omega\;$ and $\;2\omega\;$. This 
situation will be reminiscent of the above-mentioned case of a symmetrical 
oblate body. Relaxation in the vicinity of the separatrix is a far more 
complicated case; in that case numerous frequencies will be 
generated, and the frequency dependence $\;Q(\omega)\;$ will be relevant.

The quality factor $\;Q\;$ is empirically introduced in acoustics and 
seismology to make up for our inability to describe the total effect of a 
whole variety of the attenuation mechanisms (Nowick and Berry 1972)$^{30}$, 
(Burns 1986)$^{31}$, (Burns 1977)$^{32}$, (Knopoff 1963)$^{33}$. A discourse on
the frequency- and temperature-dependence of the Q-factor is given in Appendix 
B.

Another issue worth touching here is that of elasticity and plasticity 
demonstrated by materials at various temperatures. In order to calculate the 
terms entering (\ref{6.8}), one must know the stress tensor (that can be 
found from knowing the acceleration of an arbitrary point of the body) and the 
strain tensor (that depends upon the stress tensor through the system of 
equations  (\ref{6.1}), (\ref{6.3}) and (\ref{6.5})). In general, it is 
difficult to resolve the system (\ref{6.1}), (\ref{6.3}) and (\ref{6.5}) 
with respect to $\;\varepsilon_{ij}(x,\,y\,,z)\;$. Fortunately, in two simple
practical cases the system solves easily. These are the cases of cold and hot 
(plastic) body. 

As well known, at low 
temperatures materials are fragile: when the deformations exceed some critical
threshold, the body will rather break than flow. At the same time, at these 
temperatures the materials are elastic, provided the deformations are 
{\textit{beneath}} the said threshold: the sound absorption, for example, is 
almost exclusively due to the thermal conductivity rather than to the 
viscosity. These facts may be summarised like this: at low temperatures, the 
viscosity coefficient $\; \eta \;$ has, effectively, two values: one value - 
for small deformations (and this value is almost exactly zero); another value 
- for larger-than-threshold deformations (and that value is  
high. (Effectively it may be put infinity because, as 
explained above, the body will rather crack than demonstrate fluidity.) 
Therefore, within the range from the absolute zero up to at least several 
hundred degrees of Celsius the plastic part of the stress tensor may be well 
neglected.

At high temperatures materials become plastic, which means that the shear 
viscosity $\; \eta \;$ gets its single value, deformation-independent in the 
first approximation. On the one hand, this value will be far from zero (so 
that the scattering of vibrations will now be predominantly due to the 
viscosity, not due to the thermal conductivity). On the other hand, this
value will not be that high: a plastic body will rather yield than break. 
All this is certainly valid for the stretch viscosity $\; \zeta \;$ as well.
As a result, at temperatures higher than about three quarters of the melting 
temperature one may neglect the elastic part of the stress tensor, compared to 
its plastic part. (I am deeply thankful to Shun-Ichiro Karato for a 
consultation on this topic.)

To simplify the stress tensor, we model the body by a rectangular prizm of 
dimensions $\,2\,a\,\times\,2\,b\,\times\,2\,c$. The tensor must obey three 
demands. First, it must satisfy the relation
\be
\partial_{i}\sigma_{ij}\;=\;\rho\;a_j\;\;\;\;
\label{6.10}
\ee
$a_j$ being the time-dependent parts of the acceleration components, and 
$\,\rho\,a_j$ being the time-dependent parts of the components of the 
force acting on a unit volume. Second, tensor $\,\sigma_{ij}$ must be 
symmetrical and, third, it should
obey the boundary conditions, i.e., the product of the stress tensor and the 
normal unit vector, $\,\sigma_{ij}n_j$ should vanish on the boundaries of the
body (this condition was not fulfilled in (Purcell 1979)$^{34}$). 

It would be important to emphasise that the above assumption of the body being 
a prism brings almost no error into claculations performed for real 
irregular-shaped physical objects, like asteroids or cosmic-dust grains. The 
reason for this is that an overwhelming share of dissipation is anyway taking 
place not near the surface but in the depth of the body. This is especially 
evident from formulae (\ref{88.5}) - (\ref{8.10}), and it is just another  
manifestation of Saint-Venant's principle of elasticity. (I am grateful 
to Mark Levi for drawing my attention to this fact.) So, whether the body 
is indeed a rectangular prism or more like an ellipsoid, will not make much 
difference for an estimate of the relaxation time. 
Mind though that for shells Saint-Venant's principle does not work, so that in
the case of spinning spacecrafts the subtleties of their shape may be relevant.

\section{The elastic energy of alternate deformations}

Unless the temperature is too high, the bodies manifest, for small 
deformations, no 
viscosity ($\; \omega \eta \, \sim \, \omega \zeta \; \ll \;  \mu \, \sim \, K
\; $), so that the stress tensor is approximated to a very high accuracy
by its elastic part: instead of the system (\ref{6.1}) - (\ref{6.5}) one may  
write:
\be
\epsilon_{ij} \; \; = \; \; \delta_{ij} \; \; \frac{Tr \; 
\sigma}{9 \; K} \; \; + \; \; 
\left( \; \sigma_{\it{ij}} \; \; - \; \; \frac{1}{3} \; \; \delta_{ij} 
\; \; Tr \; \sigma \right) \; \frac{1}{2 \; \mu} \; \; \; ~~~,
\label{7.1}
\ee
This will enable us to derive an expression for the elastic energy stored in 
a unit volume of the precessing body:
\ba
\frac{d\;\,<W>}{dV}\;=\;\frac{1}{2}\;\,<\epsilon_{ij}\;\sigma_{ij}>\;=
\;\frac{1}{4 \mu}\; 
\left\{ \left(\frac{2 \; \mu}{9\; K} \; - \; \frac{1}{3} \protect\right) \;
\,<\,\left(Tr\;\sigma \protect\right)^2\,>\;+\;<\sigma_{ij}\,\sigma_{ij}> 
\protect\right\} \; = 
\nonumber \\
\nonumber \\
\frac{1}{4\mu}\;\left\{\,-\,\frac{1}{1\,+\,\nu^{-1}}\;\,<\left(Tr\;\sigma
\protect\right)^2>\,+\,<\sigma_{xx}^2>\,+\,<\sigma_{yy}^2>\,+\,<\sigma_{zz}^2
> \, + \, 2 \,<\,\sigma_{xy}^2 \, + \, \sigma_{yz}^2 \, + \, 
\sigma_{zx}^2 \,> \protect\right\}\;\;
\label{7.2}
\ea 
where we have made use of the expressions connecting the shear and bulk moduli
with the Young modulus $E$ and Poisson's ratio $\nu$: since $\; K\, =\,E/[3(1
-2\nu )]\;$ and $\; \mu \,=\,E/[2(1+\nu )] \;$ then $\; 2 \mu/(9K) \, - 1/3 \,
= \, - \, \nu /(1+\nu ) \;$. As Poisson's ratio $\nu$ is, for cold solids,  
typically about $0.25$, one may safely put $\; 2 \mu/(9K) \, - 1/3 \; \approx 
\; - \, 1/5 \;$. 

The above expression (\ref{7.2}) must be decomposed into a sum like 
(\ref{6.8}). In what follows we shall be interested in the energies averaged 
over the precession period. For this reason, in (\ref{7.2}) all $\;\sigma^2_{ij}\;$'s are averaged. (See Appendices C and D for 
details.)

\pagebreak

\section{Relaxation Rate in the Vicinity of Pole C:\\
the Relaxation is Almost Completed and the  
Body is Spinning Almost about its Maximal-Inertia Axis\\}

Near the poles parameter $k$ is close to zero. This justifies   
the following simple asymptotics for elliptic functions  
(Abramovitz \& Stegun 1964, formulae 16.13.1-3)$^{27}$:
\be
cn(u,\;k^2) \; = \; \cos \,u \; + \; \frac{1}{4} \; k^2 \; (u\;-\;\sin u \; 
\cos u) \; \sin u\;+\;O(k^4)\;\;,\;\;\; 
\label{8.1}
\ee
\be
sn(u,\;k^2) \; = \; \sin \,u \; - \; \frac{1}{4} \; k^2 \; (u\;-\;\sin u \; 
\cos u) \; \cos u \;+\;O(k^4)\;\;,\;\;\; 
\label{8.2}
\ee
\be
dn(u,\;k^2)\;=\;1\;-\;\frac{1}{2}\;k^2 \; \sin^2 u \;+\;O(k^4)\;\;,\;\;\;
\label{8.3}
\ee
In the vicinity of pole C, i.e., during the ``final spin'' (when $\;\bf \Omega
\;$ is almost aligned along or opposite $\;\bf J\;$), we substitute 
(\ref{2.11}-\ref{2.13}) and (\ref{2.18}-\ref{2.26}) into (\ref{2.17}). Then we
must use the asymptotics (\ref{8.1}) - (\ref{8.3}) neglecting terms of 
order higher than $\;k^2\;$. Mind that, for $\;{\bf J}^2\,>\,2\,T_{kin}\,I_2
\;$, the parameters $\;\beta\;$ and $\;\gamma\;$ are of same order as $\,k\,$ 
(as evident from (\ref{2.13})), while $R$ is of order $\,k^2\,$ (according to 
(\ref{2.7})). This will give us:
\ba
\nonumber
{\bf a}_{(near\;C)}^{(t)}\;=\;\;\;\;\;\;\;\;\;\;\;\;\;\;\;\;\;\;\;\;\;\;\;\;\;
\;\;\;\;\;\;
\ea
\ba
\nonumber
={\bf e}_1\;\left\{\,-\,x\,\left[P\,+\,(1\,-\,Q)\,\beta^2\;{\it sn}^2[u,\,k^2]
\,\right]\,+\,y\,\left(\alpha\, \omega \, k^2\, +\, \beta\, \gamma \right)\;{\it sn}[u,
\,k^2]
\;{\it cn}[u,\,k^2]\;+ \right.\\
\nonumber
+\;z\,\left. (\beta\, \omega \, +\, \alpha\, \gamma)\;{\it cn}[u,\,k^2]\;{\it dn}[u,\,k^2]
\right\}\;+\\
\nonumber\\
\nonumber
+{\bf e}_2\;\left\{\,x\,\left( \beta \,\gamma\,-\,\alpha\, \omega \, k^2
\right)\;{\it sn}[u,k^2]\;{\it cn}[u,k^2]\,
-\,y\,\left[P\,+\,R\,-\,(Q\,+\,S) \,\beta^2 \;{\it sn}^2[u,k^2]\right]\;+\right.\\
\nonumber
+\,z\,\left.
(\alpha\, \beta\, +\, \gamma\, \omega)\;{\it sn}[u,k^2]\;{\it dn}[u,k^2]\right\}\;+\\
\nonumber\\
\nonumber
+{\bf e}_3\;\left\{\,x\,(\,-\,\beta\, \omega \,+\,\alpha\, \gamma)\;{\it cn}[u,\,k^2]\;{\it dn}[u,\,k^2]\,+\,y\,
(\alpha\, \beta \,-\, \gamma\, \omega)\;{\it sn}[u,\,k^2]\;{\it dn}[u,\,k^2]\; -\right.\\
-\; z\,\left.
\left[R\,+\,(1\,-\,S)\,\beta^2\;{\it sn}^2[u,\,k^2]\right]\right\}=
\;\;\;\;\;\;\;\;\;\;\;
\label{88.4}
\ea
\ba
\nonumber
=\;{\bf e}_1\;\left\{\,\frac{1}{2}\;x\;(1\,-\,Q)\;\beta^2\;\cos 2u\;+\;\frac{1}{2}
\;y\;\left(\alpha\;\omega \;k^2\;+\;\beta\;\gamma \right)\;\sin 2u\;+\;z\;(\beta\,\omega
\,+\,\alpha\,\gamma)\;\cos u\right\}\;+\\
\nonumber\\
\nonumber
+{\bf e}_2\,\left\{\frac{1}{2}x\left(\beta \gamma - \alpha \omega k^2
\right)\,\sin 2u\,- y\,\frac{1}{2}\,(Q + S) 
\beta^2\,\cos 2u\,+\,z(\alpha \beta + \gamma \omega)\,\sin u\right\}+\\
\nonumber\\
\nonumber
+\;{\bf e}_3\;\left\{\,x\;(\,-\,\beta\,\omega \,+\,\alpha\,\gamma)\;\cos u\;+
\;y\;(\alpha\,\beta\,-\,\gamma\,\omega )\;\sin u\;+\;z\,\frac{1}{2}\,\left(1\,-
\,S\right)\;\beta^2\;\cos 2u\,\right\}\;+\\
\nonumber\\
\nonumber
+\;\left\{\;time-dependent\;\;k^2-order\;\;terms\;\;originating\;\;from\;\;the 
\right. \;\;\;\;\;\;\;\;\;\;\;\;\;\;\;\;\\
\nonumber\\
\nonumber
\left. k^2-order\;\;terms\;\;in\;\;expansions\;\;(8.1)-(8.3)\;\right\}\;+\;\;\;
\;\;\;\;\;\;\;\;\;\;\;\;\\
\nonumber\\
+\;\;\left\{\;time-independent\;\;terms\;\right\}\;\;\;\;\;\;\;\;\;\;\;\;\;\;\;
\;\;\;\;\;
\label{8.4}
\ea
where $\;u\;\equiv\;\omega\,(t\,-\,t_0)\;$. In the further calculations we  
shall ignore the time-independent terms emerging in (\ref{8.4}) because, in 
order to calculate the inelastic-dissipation rate, we need only time-dependent 
part of the stress tensor.

Dissipation is taking place in two modes one of which has the frequency of 
precession, while another one is of twice that frequency. (If one 
plugs into (\ref{88.4}) all the high-order terms from (\ref{8.1}) - (\ref{8.3})
they will give an infinite amount of the higher modes in (\ref{88.4}). In the vicinity of poles, we neglect the high-order terms in  (\ref{8.1}) - (\ref{8.3}), and thereby neglect harmonics higher than second.) 
As already mentioned in Section VI, the second mode 
originates from the centripetal term in (\ref{2.16}). This fact is understood 
especially easily if we assume that the body is oblate and symmetric. In this 
case one component of $\;\bf \Omega\;$ (the one parallel to the axis of maximal
inertia) will stay unchanged, while the other two will be proportional to $\;
\sin \omega t\;$ and $\;\cos \omega t\;$ (which simply means that $\;\bf \Omega
\;$ is precessing at rate $\;\omega \;$ about the maximal-inertia axis). Quite 
evidently, squaring of $\;\bf \Omega\;$ in (\ref{2.16}) yields the double 
frequency. Mathematically speaking, in the case of an oblate body the realm of 
applicability of the solution (\ref{2.14}) - (\ref{2.15}) shrinks to a line, 
so that the solution (\ref{2.12}) - (\ref{2.13}) accounts for the entire 
process. As evident from (\ref{2.13}), in the oblate case $\;k\,=\,0\;$ and 
therefore formulae (\ref{8.1}) - (\ref{8.3}) contain only terms of order 
$\,k^0\,$. 

So we shall strip (\ref{88.4} - \ref{8.4}) off its time-independent terms, 
and shall plug the ${\bf time-dependent}$ terms into (\ref{6.10}). Integration 
thereof will then give us expressions for the ${\bf time-dependent}$ components
of the stress tensor: 
\ba
\sigma_{xx}\;=\;
-\;\frac{\rho}{2}\,(1\,-\,Q)\;\beta^2\;\left(x^2\,-\,a^2\right)
\;\left\{{\it sn}^2[\omega (t\,-\,t_0),\,k^2]\;-\;<\,{\it sn}^2[\omega (t\,-
\,t_0),\,k^2]\,>\;\right\}\;=\;\;\\
\label{88.5}
\nonumber\\
=\;-\;\frac{\rho}{2}\,(1\,-\,Q)\;\beta^2\;\left(x^2\,-\,a^2\right)
\;\left\{-\,\frac{1}{2}\;\cos 2[\omega (t\,-\,t_0)]\;+\;O(k^2)\right\}\;\;,\;\;\;\;
\label{8.5}
\ea
\ba
\nonumber\\
\sigma_{yy}\;=\;\frac{\rho}{2}\,(S\,+\,Q)\;\beta^2\;\left(y^2\,-\,b^2
\right)\;\left\{{\it sn}^2[\omega (t\,-\,t_0),\,k^2]\;-\;
<\,{\it sn}^2[\omega (t\,-\,t_0),\,k^2]\,>\;\right\}\;=\;\;\\
\label{88.6}
\nonumber\\
=\;\frac{\rho}{2}\,(S\,+\,Q)\;\beta^2\;\left(y^2\,-\,b^2\right)\;
\left\{-\,\frac{1}{2}\;\cos 2[\omega (t\,-\,t_0)]\;+\;O(k^2)\right\}\;\;,\;\;\;\;\;
\label{8.6}
\ea
\ba
\nonumber\\
\sigma_{zz}\;=\;-\;\frac{\rho}{2}\,(1\,-\,S)\;\beta^2\;\left(z^2\,-\,c^2
\right)\;\left\{{\it sn}^2[\omega (t\,-\,t_0),\,k^2]\;-\;<\,{\it sn}^2[\omega 
(t\,-\,t_0),\,k^2]\,>\;\right\}\;=\;\;\\
\label{88.7}
\nonumber\\
=\;-\;\frac{\rho}{2}\,(1\,-\,S)\;\beta^2\;\left(z^2\,-\,c^2
\right)\;\left\{-\,\frac{1}{2}\cos 2[\omega (t\,-\,t_0)]\;+\;O(k^2)\right\}\;\;
,\;\;\;\;\;
\label{8.7}
\ea
\ba
\nonumber\\
\nonumber
\sigma_{xy}\,=\,\frac{\rho}{2}\,\left\{\left(\beta \gamma + \alpha \omega k^2\right)
\left(y^2 - b^2\right)\,+\right.\;\;\;\;\;\;\;\;\;\;\;\;\;\;\;\;\;\;\;\;\;\;\\
\nonumber\\
\nonumber
\left.+\;\left(\beta \gamma - \alpha \omega  k^2\right)\left(x^2 - a^2\right)\right\}
\;\left\{{\it sn}[\omega (t-t_0),\,k^2]\;{\it cn}[\omega ((t - t_0),\,k^2]\,-
\right.\\
\nonumber\\
\left.-\;<\,{\it sn}[\omega (t-t_0),\,k^2]\;{\it cn}[\omega ((t - t_0),\,k^2]
\,>\;\right\}\;=\;\;\;
\label{88.8}
\ea
\ba
\nonumber\\
\nonumber
=\;\frac{\rho}{2}\,\left\{
\left(\beta\,\gamma\,+\,\alpha\,\omega \,k^2\right)\;
\left(y^2\,-\,b^2\right)\;+\right.\\
\nonumber\\
\left.+\;\left(\beta\,\gamma\,-\,\alpha\,\omega \,k^2\right)\;
\left(x^2\,-\,a^2\right)\right\}\;
\left\{\frac{1}{2}\;\sin 2[\omega (t\,-\,t_0)]\;+\;O(k^2)\right\}\;\;,\;
\label{8.8}
\ea
\ba
\nonumber\\
\nonumber
\sigma_{xz}\,=\,\frac{\rho}{2}\,\left\{
\left(\beta\,\omega \,+\,\alpha\,\gamma\right)\,
\left(z^2\,-\,c^2\right)\,+\right.\;\;\;\;\;\;\;\;\;\;\;\;\;\;\;\;\;\;\;\;\;\\
\nonumber\\
\nonumber
\left.+\;\left(\,-\,\beta\,\omega \,+\,\alpha\,\gamma\right)\,
\left(x^2\,-\,a^2\right)\right\}\,\left\{
{\it dn}[\omega (t\,-\,t_0),\,k^2]\;{\it cn}[\omega (t\,-\,t_0),\,k^2]\;-
\right.\\ 
\nonumber\\
\left.-\;<\,{\it dn}[\omega (t\,-\,t_0),\,k^2]\;{\it cn}[\omega (t\,-\,t_0),\,
k^2] \,>\;\right\}\;=\;\;\;
\label{88.9}
\ea
\ba
\nonumber\\
\nonumber
=\;\frac{\rho}{2}\,\left\{
\left(\beta\,\omega \,+\,\alpha\,\gamma\right)\;
\left(z^2\,-\,c^2\right)\;+\right.\\
\nonumber\\
\left.+\;\left(\,-\,\beta\,\omega \,+\,\alpha\,\gamma\right)\;
\left(x^2\,-\,a^2\right)\right\}\;\left\{\cos [\omega (t\,-\,t_0)]\;+\;O(k^2)
\right\}\;\;,\;\;
\label{8.9}
\ea
\ba
\nonumber\\
\nonumber
\sigma_{yz}\,=\,\frac{\rho}{2}\,\left\{\left(\alpha \beta +\omega \gamma\right)
\left(z^2 - c^2\right)\,+\right.\;\;\;\;\;\;\;\;\;\;\;\;\;\;\;\;\;\;\;\;\;\;\\
\nonumber\\
\nonumber
\left.+\;\left(\alpha \beta - \omega  \gamma\right)
\left(y^2 -\,b^2\right)\right\}\;\left\{
{\it dn}[\omega (t - t_0),\,k^2]\;{\it sn}[\omega (t - t_0),\,k^2]\;-\right.\\
\nonumber\\
-\;\left.<\,{\it dn}[\omega (t - t_0),\,k^2]\;{\it sn}[\omega (t - t_0),\,k^2]\,>\right\}\;=\;\;\;
\label{88.10}
\ea
\ba
\nonumber
=\;\frac{\rho}{2}\,\left\{
\left(\alpha\,\beta\,+\,\omega \,\gamma\right)\;
\left(z^2\,-\,c^2\right)\;+\right.\\
\nonumber\\
\left.+\;\left(\alpha\,\beta\,-\,\omega \,\gamma\right)\;
\left(y^2\,-\,b^2\right)\right\}\;
\left\{\sin [\omega (t\,-\,t_0)]\;+\;O(k^2)\right\}\;\;,\;\;\;\;\;
\label{8.10}
\ea
The symbol $\;<...>\;$ stands for averaging over the mutual period $\;\tau\;$ 
of functions $\;\it sn\;$ and $\;\it cn\;$. (See Appendix A.) In the above 
expressions, it might be better to write $\;\sigma^{(t)}_{ij}\;$ instead of $\;
\sigma_{ij}\;$, in order to stress that we are considering only the 
time-dependent part, but we would rather omit the superscrips for brevity. The 
above expressions (\ref{8.4}), (\ref{8.5}), (\ref{8.6}), (\ref{8.7}), (\ref{8.8}), (\ref{8.9}), (\ref{8.10}) coincide, in the limit of oblate 
symmetry ($\,I_1\,=\,I_2\,$), with formulae (19) - (23) from our previous 
article (Lazarian and Efroimsky 1999)$^{16}$.

The expression (8.5) is exact, while the formulae (8.6) - (8.17) implement the
polynomial approximation to the stress tensor. This approximation keeps the
symmetry and obeys (6.10). The boundary conditions are satisfied only approximately, for the off-diagonal components. This approximation very considerably 
simplifies calculations and yields only minor errors in the numerical factors 
(8.26) - (8.28). A comprehensive analysis of the polynomial approximation will
be presented elsewhere.

To calculate the dissipation rate, we shall need averaged over the 
precession period squares of the above stresses, $\;<\sigma_{ij}^2> \;$, 
as well as $\;<\left( Tr \,\sigma \right)^2>\;$. Moreover, for our goals we
shall need these calculated up to terms of order $k^2$ inclusively. This demand
makes it necessary to have the above expressions (\ref{8.5}), (\ref{8.6}), 
(\ref{8.6}), (\ref{8.7}), (\ref{8.8}), (\ref{8.9}), (\ref{8.10}) with all the 
$k^2-$order terms written explicitly. 
How to get these terms? On the face of it, the answer is trivial and looks like
this. In the above formulae we approximated the elliptic functions using only 
$k^0-$order terms of (\ref{8.1} - \ref{8.3}); now, let us keep also the 
$k^2$-order terms. Surprisingly, this is the case when the simpliest shortcut 
leads to a wrong answer. Plugging of the contained in (\ref{8.1} - \ref{8.3}) 
$k^2-$order terms into (\ref{88.5}), (\ref{88.6}), (\ref{88.6}), (\ref{88.7}),
 (\ref{88.8}), (\ref{88.9}), (\ref{88.10}), with the further squaring thereof, 
will give birth to secular terms in the expressions for $\;<\sigma_{ij}^2>\;$, 
i.e., to terms linear in $\;u\;\equiv\;\omega (t\,-\,t_0)\;$. Averaging of
these terms will entail ambiguities: one will get into an illusion that it does
matter whether to integrate from $\,0\,$ through $\;2\,\pi\;$ or, say, from 
$\;2\,\pi\;$ through $\;4\,\pi\;$. The secular terms have been long known in 
nonlinear mechanics and astronomy where they often tarnish calculations and 
sometimes become a real pain. Luckily, in our case we can sidestep this 
obstacle by employing directly the fundamental definition of the elliptic 
functions: 
\be
{\it sn}(u,\,k^2)\;\equiv\;\sin \phi\;\;,\;\;\;{\it cn}(u,\,k^2)\;\equiv\;
\cos \phi\;\;,\;\;\;{\it dn}(u,\,k^2)\;\equiv\;\left(1\;-\;k^2\;\sin^2 
\phi \right)^{1/2}\;\;
\label{jacobi1}
\ee
the auxiliary quantity $\;\phi\;$ being connected to $\;u\;$ like that:
\be
u\;\equiv\;\int_{0}^{\phi}\frac{d \theta}{\left(1\;-\;k^2\;\sin^2 
\theta \right)^{1/2}}\;\;.\;\;\;
\label{jacobi2}
\ee
This will give us the key to a correct calculation of the averaged-over-period 
quadratic and quartic forms. For example, the average $\;< {\it sn}^2 u\;
{\it dn}^2 u >\;$ will read:
\ba
\nonumber
<{\it sn}^2 (u,\,k^2)\;{\it dn}^2 (u,\,k^2) >\;\equiv\;\frac{1}{\tau}\;
\int_{0}^{\tau}\;{\it sn}^2 (u,\,k^2)\;{\it dn}^2 (u,\,k^2)\;du\;=\\
\nonumber\\
\nonumber
=\;\frac{1}{4\;K}\;\int_{0}^{2 \pi }\;\sin^2 \phi\;\left(1\;-\;k^2\;\sin^2 
\phi \right)\;\frac{du}{d \phi }\;d \phi\;=\;\frac{1}{4\;K}\;\int_{0}^{2 \pi }
\;\sin^2 \phi\;\left(1\;-\;k^2\;\sin^2 \phi \right)^{1/2}\;d \phi\;=\\
\nonumber\\
=\;\frac{2 \pi}{4\;K}\;\frac{1}{2 \pi}\;\int_{0}^{2 \pi}\;\sin^2 \phi\;\left(
1\;-\;\frac{k^2}{2} \sin^2 \phi \right)\;\approx\;\frac{1/2\;-\;k^2/16}{1\;+
\;k^2/4}\;d{\phi}\;\approx\;\frac{1}{2}\;\left(1\;-\;\frac{3}{8}\;k^2\right)\;\;\;\;\;\;
\label{jacobi3}
\ea
where we used (A7). The squared and averaged in the above manner 
stress components are presented in Appendix C, expressions (C2), 
(C4), (C6), (C8), (C10), (C12) and (C14). Substitution thereof into 
(\ref{7.2}) will lead us to the expression for dissipation per unit volume:
\be
\frac{d\;\,<W>}{dV}\;=\;\frac{d\;\,<W^{(2 \omega )}>}{dV}\;
+\;\frac{d\;\,<W^{(\omega )}>}{dV}\;
\label{8.11}
\ee
where the first term stands for the dissipation of oscillations at frequency 
$\;2\,\omega \;$:
\ba
\nonumber
\frac{d\,\;<W^{(2\omega )}>}{dt}\;=\;\;\;\;\;\;\;\;\;\;\;\;\;\;\;\;\;\;\;\;\\
\nonumber\\
\frac{1}{4\,\mu}\;\left\{\,-\,\frac{1}{1\,+\,\nu^{-1}}\;\,<
(Tr\,\sigma)^2>\;+\;<\sigma_{xx}^2>\;+\;<\sigma_{yy}^2>\;+\;<
\sigma_{zz}^2>\;+\;2\;<\sigma_{xy}^2>\;\right\}\;\;\;,
\label{8.12}
\ea
while the second term expresses the dissipation at the principal frequency:
\ba
\frac{d\,\;<W^{(\omega )}>}{dt}\;=\;\frac{1}{4\,\mu}\;\left\{\,2\;\left(
\;<\sigma_{yz}^2>\;+\;<\sigma_{zx}^2>\;\right)\,\right\}\;
\label{8.13}
\ea
Expressions for $\;d\,<W^{(\omega )}>/dV\;$ and $\;d\,<W^{(2\omega )}>/dV\;$ 
in terms of $\;I_{1,2,3}\;$ are presented in the Appendix (formulae (C20) and 
(C21)). 
These expressions should be now multiplied by $\;2\omega /Q(\omega )\;$ and 
$\;4\omega /Q(2\omega )\;$, correspondingly, and integrated over the volume of 
the body, as in (\ref{6.9}). The outcome of this integration will be the total 
dissipation rate $\;\dot{W}\;$ that must be plugged, together with (\ref{5.3})
into (\ref{5.1}). Here follows the result:
\ba
\nonumber
\frac{d\;< \sin^2 \theta>}{dt}\;=\;-\;\frac{2\;I_3 \;\rho^2}{\mu\;Q(\omega)}\;
\frac{\left(2\,I_3\,T_{kin}\,-\,{\bf J}^2\right)}{{\bf J}^2}\;\left\{\,\omega 
\;\left({\bf J}^2\,-\,2\,I_1\,T_{kin}\right)\;H_1\;\left[
\frac{I_1\,I_3\,+\,I_2\,I_3\,-\,2\,I_1\,I_2}{(I_3\,-\,I_1)\,(I_3\,-\,I_2)}\;+
\right. \right.\\
\nonumber\\
\nonumber
\left. +\;\frac{1}{4}\;\frac{I_3\;\left(I_2\,-\,I_1\right)^2}{\left(I_3-I_1 \right)\,\left(I_3-I_2\right)^2}\;
\frac{2\,I_3\,T_{kin}\,-\,{\bf J}^2}{{\bf J}^2\,-\,2\,I_1\,T_{kin}}
\right]\;-\;\omega\;\frac{}{}\;H_0\;\left(2\,I_3\,T_{kin}\;-\;{\bf J}^2\right)\;+\\
\nonumber\\
\left.+\,2\;\omega \;\frac{Q(\omega)}{Q(2\omega )}\;\left(2\,I_3\,T_{kin}\,-\,{\bf J}^2\right)\;H_2\;\frac{I_1\,I_3\,+\,I_2\,I_3\,-\,2\,I_1\,I_2}{(I_3-I_1)
\,(I_2\,-\,I_1)}\;+\,O(k^4)\,\right\}\;\;\;
\label{8.14}
\ea
The ratio $\;Q(\omega )/Q(2\omega )\;$ is typically 
close to unity, unless the structure of the body or the properties of the 
material provide resonances. Terms with $\;H_0\;$ and $\;H_1\;$ are
due to the dissipation of oscillations at frequency $\;\omega\;$, while the 
term with $\;H_2\;$ is due to the vibrations at $\;2\,\omega\;$. 

Numerical coefficients $\;H_0\;$, $\;H_1\;$ and $\;H_2\;$ emerging in (\ref{8.14}) are 
geometrical factors that depend upon the moments of inertia and dimensions of 
the body. General expressions for $\;H_{0,1,2}\;$ are given 
in Appendix C. Obviously, $\;H_0\;$ vanishes in the oblate case. 

Equation (\ref{8.14}), together with (\ref{2.13}) and with equation
\ba
{\bf J}^2\;\,<\sin^2 \theta>_{(near\;C)}\;=\;\frac{1}{2}\;\frac{I_1\,I_3\,+\,I_2
\,I_3\,-\,2\,I_1\,I_2}{(I_3\,-\,I_1)(I_3\,-\,I_2)}\,\left(2\,I_3\,T_{kin}\;-\;
{\bf J}^2 \right)\;+\;O(k^4)\;\;\;\;
\label{8.17}
\ea
connecting $\;T_{kin}\;$ with $\;<\sin^2 \theta>\;$, makes a system of 
equations describing relaxation in the vicinity of pole C. (Equation 
(\ref{8.17}) is a truncated version of (\ref{5.2}). For details see (A11) in
Appendix A.)

Let us elaborate on the factors  $\;H_0\;$, $\;H_1\;$ and $\;H_2\;
$. In the case of a homogeneous body of dimensions $\;2a\,\times\,2b\,\times\,
2c\;$, expressions for the factors read (see the end of Appendix C):
\ba
H_1\;=\;\frac{317}{m^4}\;\;\frac{a\;b\;c^5}{\left(b^2\,+\,c^2\right)\;\left(
a^4\,-\,c^4\right)\;\left(a^2\,+\,b^2\right)}\;\left(\frac{b^4}{b^4\,-\,c^4}\;
+\;\frac{a^4}{a^4\,-\,c^4} \right)\;\;,\;\;
\label{8.18}
\ea
\ba
H_2\;=\;\frac{100}{m^4}\;\;\frac{a^9\,b^9\,c\,-\,a^9\,b^5\,c^5\,-\,a^5\,b^9\,c^5\,+\,0.21\,a^9\,b\,c^9\,+\,0.19 \,a^5\,b^5\,c^9\,+\,0.21\,a\,b^9\,c^9}{\left(a^2\,+\,b^2\right)^2\;\left(a^4\,-\,
c^4 \right)^2\;\left( b^4\,-\,c^4 \right)^2}\;\;
\label{8.19}
\ea
and
\ba
H_0\;=\;\frac{237}{m^4}\;a\;b\;c^5\;(a^2\,-\,b^2)\;\frac{(2.67\,a^4\,b^4\,-\,a^4\,c^4\,-\,1.67\,b^4\,c^4)\,(a^4\,+\,b^4\,-\,2\,c^4)}{(a^2\,+\,b^2)\,(a^2\,-\,c^2)\,(b^2\,-\,c^2)\,(a^4\,-\,c^4)^2\,(b^4\,-\,c^4)^2}\;\;.\;\;
\label{H0}
\ea
The denominators of (C22) - (C24) contain expressions 
$\;(I_3\,-\,I_1)\;$ and $\;(I_3\,-\,I_2)\;$; as a result, the denominators of 
(\ref{8.18}), (\ref{8.19}) and (\ref{H0}) contain $\;\left(
a^4\,-\,c^4\right)^2 \;$ and $\;\left( b^4\,-\,c^4 \right)^2 \;$. It would 
be appropriate to make sure that nothing wrong happens when $\;a\,\rightarrow\,
c\;$ or  $\;a\,\rightarrow\,b\;$. We assumed from the beginning that $\;I_3\,
\geq\,I_2\,\geq\,I_1\;$, i.e., that (for a prism) $\;a\,\geq\,b\,\geq\,c\,$. 
Therefore it would be enough to investigate the case of $\;a\,\rightarrow\,b\;
$. Recall also that the parameter $k$ given by (\ref{2.13}) and (\ref{2.15}) 
never exceeds unity: $\;k\,=\,0\,$ at the poles and  $\;k\,=\,1\,$ at the 
separatrix. From (\ref{2.13}) we see that, for $\;(I_3-I_2)\;\sim\;(b^2-c^2)\;
\rightarrow\;0\,$, the condition $\;k\,\leq\,1\,$ is fulfilled only if $\;
\Omega_1^2/\Omega_3^2\,<\,(I_3(I_3-I_2))/(I_1(I_2-I_1))\;$. In other words, 
making $\,(I_3-I_2)\,$  (and $\,(b-c)\,$) infinitesimally 
small leads to infinitesimal squeezing of the region around pole C between 
the separatrices on Fig.1. Thus the region, where the appropriate solution is 
applicable, shrinks into a point. 

In our analysis it is possible to get rid of the variable $\,T_{kin}\,$ 
completely: one should express it through $\;<\sin^2 \theta>\;$ by means of 
(\ref{8.17}), and plug the result into (\ref{8.14}). This will give us what we 
would call Master Equation, a differential equation for  $\;<\sin^2 \theta>\;$:
\ba
\nonumber
\frac{d\,< \sin^2 \theta>}{dt}\,=\,-\,\frac{4\,{\bf J}^2\,\rho^2}{\mu\;Q(
\omega)}\,\frac{I_3-I_1}{I_1\,I_3\,+\,I_2\,I_3\,-\,2\,I_1\,I_2}\;<\sin^2\theta>
\\
\nonumber\\
\nonumber
\left\{\,\omega \,H_1\,\left[1\,-\,2\;<\sin^2\theta>\;\frac{I_1\;(I_3\,-\,I_2)}{I_1\,I_3\,+\,I_2\,
I_3\,-\,2\,I_1\,I_2}\right]\,\left[I_1\,I_3\,+\,I_2\,I_3\,-\,2\,I_1\,I_2\;+
\right.\right.\\
\nonumber\\
\nonumber
\left.\left.+\;\frac{1}{2}\,\frac{(I_2\,-\,I_1)^2\;I_3^2\;
\;<\sin^2\theta>\;}{(
I_1\,I_3\,+\,I_2\,I_3\,-\,2\,I_1\,I_2)\;-\;2\;<\sin^2\theta>\;I_1\,(I_3\,-\,
I_2)}\right]\;-\right.\\
\nonumber\\
\nonumber
\left.\left.-\;\omega\;H_0\;2\;I_3\;(I_3\;-\;I_2)\;<\sin^2 \theta>\;+\right.\right.\\
\nonumber\\
\left.+\,4\,\omega\,H_2\,\frac{Q(2\omega)}{Q(\omega)}\;< \sin^2 \theta>\;I_3\,
(I_3\,-\,I_2)\right\}\;+\;O(k^4)\;\;\;\;
\label{8.20}
\ea
where, according to (\ref{2.13}) and (\ref{8.17}),
\be
\omega\;=\;\frac{|{\bf J}|}{I_3}\;\sqrt{\frac{(I_3-I_2)(I_3-I_1)}{I_1\;I_2}
\left[   1\;-\;2\;\,<\sin^2 \theta>\,\;\frac{I_1\,(I_3\,-\,I_2)}{I_1 I_3+I_2 I_3 - 2 I_1 I_2}\right]}
\label{omega}
\ee
Equation (\ref{8.20}) is one of the main results of our study. It describes the
relaxation in the vicinity of pole C corresponding to rotation about the 
maximal-inertia axis. Simply from looking at this equation one 
can understand several important features of the relaxation process. To start 
with, it follows from (\ref{8.20}) that $\;d\,< \sin^2 \theta>\,/dt\;$ vanishes
in the limit of $\;(I_3\,-\,I_1)\,\rightarrow\,0\;$, which naturally 
illustrates the absence of relaxation in the case of all moments of inertia 
being equal to one another. Second, the overall factor $\;<\sin^2\theta>\;$ 
standing before the brackets in the right-hand side of (\ref{8.20}) evidences 
of a gradual decrease in the relaxation rate: the major-inertia axis will be 
approaching the angular momentum vector but will never align along it exactly. 

Technically, the Master Equation (\ref{8.20}) becomes a self-consistent 
differential equation, describing the time-evolution of $\;<\sin^2 \theta>\;$,
only after the expression (\ref{omega}) for $\;\omega\;$ is plugged into it. We
did not bother to do this not only for the sake of brevity. In fact, equation 
(\ref{8.20}) as it stands is of more practical interest than the 
self-consistent differential equation. It enables, for example, an astronomer 
to use the measurable quantities $\;\omega\;$ and  $\;<\sin^2 \theta>\;$, to 
predict the relaxation rate in the short run. In the real life ``short run'' 
means: the time span during which the currently available resolution of the 
optical or radio equipment makes it possible to notice the narrowing of the 
precession cone. Nowadays spacecraft-based equipment provides an angular 
precision of $\;0.01^o\;$ and even better. This gives us a chance of observing 
precession damping within a period varying from several months to several 
years, for different objects (Efroimsky 2000)$^{14}$. Soon Rosetta mission will
give it the first try (Hubert \& Schwehm)$^{12}$, (Thomas et al)$^{13}$. 

Now let us briefly dwell on the limit of an oblate body ($\;I_2\,=\,I_1\;$). In
this case, the precession is known to be circular (Efroimsky \& Lazarian 
2000)$^{24}$, so the averaging $\;<...>\;$ may be omitted. The simplified 
Master Equation will then look:
\ba
\nonumber
\left(\frac{d\; (\sin^2 \theta)}{dt}\right)^{(oblate)}\;=\;-\;\frac{4\,{\bf 
J}^2\,\rho^2}{\mu\;Q(\omega)}\,\frac{I_3-I_1}{I_1}\;\sin^2\theta\;\left\{\,
\omega \,H_1\,I_1\;\cos^2 \theta+\right.\\
\nonumber\\
\left.+\;2\;\omega\,H_2\,\frac{Q(2\omega)}{Q(\omega)}\,I_3\;\sin^2 \theta
\right\}\;+\;O(k^4)\;\;\;\;
\label{8.21}
\ea
where 
\be
\omega\;=\;\frac{|{\bf J}|}{I_3}\;\left(\frac{I_3}{I_1}\;-\;1\right)\;\cos \theta\;\;.\;\;\;
\label{omegasimple}
\ee
For an oblate homogeneous rectangular prism, the latter and the former, with 
(\ref{8.18}) and (\ref{8.19}) plugged in, will give:
\ba
\nonumber
\left(\frac{d \theta}{dt}\right)^{(oblate)} \; = \;\;\;\;\;\;\;\;\;\;\;\;\;\;\;\;\;\;\;\;\;\;\;\;\;\;\;\;\;\;\\
\nonumber\\
\nonumber
-\,\frac{3}{2^4} \;\sin^3 \theta  
\;\left[63\left(\frac{c}{a}\right)^4 \, \cot^2 \theta \, + \, 20\,\frac{1-2(c/a)^4+0.61(c/a)^8}{1
\,-\,2\,(c/a)^4\,+\,(c/a)^8}\right]\,\frac{a^2 \, 
\Omega^3_0 \, \rho}{\mu \, Q(\omega)} \,[1+(c/a)^2]^{-4}+\,O(k^4)\;=\\
\nonumber\\
\nonumber\\
=\;- \; \frac{3}{2^4} \; \sin^3 \theta 
\;\;\left[\;\frac{63\;(c/a)^4 \; \cot^2 \theta \; + \; 20}{[1+(c/a)^2]^4}\;
\right]\;\frac{a^2\;\Omega^3_0 \; \rho}{\mu \; Q(\omega)} \;+\;O(k^4\,,\;(c/a)^8)\;\;
\label{8.22}
\ea
where 
\be
\Omega_0\;\equiv\;\frac{J}{I_3} \;\;\;\;
\label{8.23}
\ee
and it is assumed that $\;Q(\omega)\;\approx\;Q(2\omega)\;$. 
This perfectly coincides with the exact formula obtained in (Efroimsky \& 
Lazarian 2000)$^{24}$ by a rigorous treatment possible in the oblate case:
\be
\left(\frac{d \theta}{dt}\right)^{(oblate)}_{(E\&L\;2000)} \; = \; - \; \frac{3}{2^4} \; \sin^3 \theta \;\;\left[\;\frac{63\; 
(c/a)^4 \; \cot^2 \theta \; + \; 20}{[1+(c/a)^2]^4} \; \right]\;\;\frac{a^2 \; 
\Omega^3_0 \; \rho}{\mu \; Q} \;\;\;\;,
\label{8.24}
\ee
Now let us see what happens with the Master Equation (\ref{8.20}) when the 
shape of the body is almost prolate ($\;I_3\,\stackrel{>}{\sim}\,I_2\;$):
\ba
\nonumber
\left(\frac{d\,< \sin^2 \theta>}{dt}\right)^{(prolate)}\;\approx\
\nonumber\\
\nonumber\\
\nonumber
\approx\;-\;\frac{4\;{
\bf J}^2\;\rho^2}{\mu\;Q(\omega)}\;<\sin^2\theta>\;\left\{\,\omega\;H_1\;\left(
I_3\;-\;I_1\right)\,\left(1\,+\,\frac{1}{2}\;<\sin^2\theta>\right)\;-\right.\\
\nonumber\\
\left.-\;\omega\;H_0\;2\;(I_3\;-\;I_2)\;+\;4\;H_2\;\omega\;<\sin^2\theta>\;(I_3\;-\;I_2)\,\right\}
\;+\;O(k^4)\;\;\;\;
\label{8.25}
\ea
Even though the term containing $\;H_2\;$ contains also multiplier $\;(I_3\,-\,
I_2)\;$, it diverges in the limit of prolate symmetry (see (C23)). 
However, there is nothing bad about it, as explained in Section III: one should
not make  $\;(I_3\,-\,I_2)\;$ approach zero for fixed $\;\theta\;$, but rather 
fix some value of  $\;(I_3\,-\,I_2)\;$, small but finite, and then make 
$\;\theta\;$ decrease to zero. As already mentioned (see the comment after 
(\ref{8.19})), as the shape approaches the prolate symmetry, the applicability 
region of the solution shrinks. Still, the fact is that within the 
applicability region (called in Section III the final spin) the second-mode 
term will not necessarily be much less than the 
first one. The ratio of these terms will depend upon $\;\sin^2 \theta\;$, which
means that the typical time of relaxation may be a steep function of the angle.
This typical time must be proportional, for dimensional reasons, to $\;\mu\,Q/
(2\,{\bf J}^2\,\omega\,\rho^2\,(I_3-I_1))\;$, but the numerical factor may be 
quite $\;\theta -$dependent, due to the presence of the second term in (\ref{8.25}). We had to dwell on this subtlety due to its practical relevance. In the 
recent literature they sometimes use the formula for relaxation time, derived 
for oblate bodies, in order to estimate relaxation of a tumbling prolate 
rotator. We did this in (Efroimsky \& Lazarian 2000)$^{24}$ when discussing 
asteroid 4179 Toutatis, while (Black et al 1999)$^{25}$ employed this 
estimation for asteroid 433 Eros. We see now that this was wrong even for the 
final spin about pole C. The more so, it was absolutely unjustified to use this
estimate for a tumbling body (i.e., in the vicinity of the separatrix) as done 
in the said articles. 

\section{Relaxation Rate in the Vicinity of Pole A: \\
the Body is Spinning Almost about its Minimal-Inertia Axis}

In the vicinity of pole A, i.e., during the ``initial spin'' (when the angular 
velocity $\;\bf \Omega \;$ is almost perpendicular to the maximal-inertia axis,
and $\;\Omega_3 \approx 0\;$), we substitute asymptotics (\ref{8.1}-\ref{8.3}) 
into (\ref{2.12}), (\ref{2.14} - \ref{2.15}), (\ref{2.18} - \ref{2.22}) and 
(\ref{2.27} - \ref{2.30}), the results to be plugged into (\ref{2.17}). This 
leads to the following expression for the acceleration:
\ba
\nonumber
{\bf a}_{(near\;A)}^{(t)}\;=\;\;\;\;\;\;\;\;\;\;\;\;\;\;\;\;\;\;\;\;\;\;\;\;\\
\nonumber\\
\nonumber
=\;{\bf e}_1\;\left\{\,x\;\left[P\,+\,(1\,-\,Q)\;\beta^2\;{\mathit{sn}}^2\,u\,
\right]\;+\;y\;\left(\alpha\;\omega \;+\;\beta\;\gamma \right)\;
{\mathit sn}[u,\,k^2]\;{\mathit dn}[u,\,k^2]\;+\right.\\
\nonumber
+\left.\;z\;(\beta\,\omega 
\,+\,\alpha\,\gamma)\;{\mathit cn}[u,\,k^2]\;{\mathit dn}[u,\,k^2]\;
\right\}\;+\\
\nonumber\\
\nonumber
+\,{\bf e}_2\,\left\{\,x\;\left(\beta \,\gamma \, -\, \alpha \,\omega 
\right)\;
{\mathit sn}[u,\,k^2]\;{\mathit dn}[u,\,k^2]\,-\,y\,\left[P\,+\,R\,-\,(Q\,
+\,S)\,\beta^2\;{\mathit sn}^2[u,\,k^2]\right]\;+\right.\\
\nonumber
\left.+\,z\,(\alpha \, \beta \,+ \, \gamma \, \omega \,k^2)\;
{\mathit sn}[u,\,k^2]\;{\mathit cn}[u,\,k^2]\,\right\}+\\
\nonumber\\
\nonumber
+\, {\bf e}_3\,\left\{x\,(\,-\beta \, \omega \,+\, \alpha \, \gamma)\;
{\mathit dn}\,u\;{\mathit cn}[u,\,k^2]\,+\,
y\,(\alpha \,\beta \,- \,\gamma \,\omega \,k^2)\;{\mathit sn}[u,\,k^2]\;
{\mathit cn}[u,\,k^2]\,-\right.\\
\nonumber\\
-\;z \left[R\,
\left.+(1-S)\beta^2\;{\mathit sn}^2[u,\,k^2] \,\right]\right\}\;=\;\;\;\;\;\;\;
\;\;\;\;
\label{9.0}
\ea
\ba
\nonumber
=\;{\bf e}_1\;\left\{\,\frac{1}{2}\;x\;(1\,-\,Q)\;\beta^2\;\cos 2u\;+
\;y\;\left(\alpha\;\omega \;+\;\beta\;\gamma \right)\;\sin u\;+\;z\;(\beta\,\omega 
\,+\,\alpha\,\gamma)\;\cos u\right\}\;+\\
\nonumber\\
\nonumber
+\; {\bf e}_2\;\left\{\,x\;\left(\beta\,\gamma\,-\,\alpha\,\omega \,\right)\,
\sin u\;-\;\frac{1}{2}\;(Q\,+\,S)\;y\;\beta^2\;\cos 2u\;+\;\frac{1}{2}\,z\;
(\alpha\,\beta\,+\,\gamma\,\omega \,k^2)\;\sin 2u\right\}\;+\\
\nonumber\\
\nonumber
+\; {\bf e}_3\;\left\{\,x\;(\,-\,\beta\,\omega \,+\,\alpha\,\gamma)\;\cos u\;+\;
\frac{1}{2}\,y\;(\alpha\,\beta\,-\,\gamma\,\omega \,k^2)\;\sin 2u\;-\;z\;
\frac{1}{2}\;(1\,-\,S)\;\beta^2\;\cos 2u\right\}\;+\;\\
\nonumber\\
+\;\left\{\;time-dependent\;\;\;terms \;\;\;of\;\;\;order\;\;\;k^2\;\right\}\;
\;+\;\;\left\{\;time-independent\;\;terms\;\right\}\;\;.\;\;
\label{9.1}
\ea
$\,\mathit En \; route\,$ from (\ref{9.0}) to (\ref{9.1}) we employed 
asymptotics (\ref{8.1}) - (\ref{8.3}), then separated out the time-independent 
terms (which may be dropped, because they do not influence the 
inner dissipation), and we also neglected terms of order higher than $k^2$ (we
remind that, according to (\ref{2.15}), for $\;{\bf J}^2\,<\,2\,T_{kin}\,
I_2\;$, the parameters $\;\beta\;$ and $\;\alpha\;$ are of order $\,k\,$).
In the above expression, $\;u\;\equiv\;\omega \,(t\,-\,t_0)\;$. 
Parameters 
$\,S\,$ and $\,Q\,$ are expressed by (\ref{2.7}). Parameters $\,\alpha\,$, $\,
\beta\,$, $\,\gamma\,$, $\,\omega \,$ and $\,k\,$ are expressed by (\ref{2.15})
and thus are different from  $\,\alpha\,$, $\,\beta\,$, $\,\gamma\,$, $\,\omega
\,$ and $\,k\,$ used in the preceding section (where they were expressed by 
(\ref{2.13})).

Similarly to the preceding section, we shall use equation (\ref{6.10}) 
and expression (\ref{9.1}) to compute the stress tensor. This will lead us to:
\ba
\nonumber\\
\nonumber
\sigma_{xx}\;=\;
-\;\frac{\rho}{2}\,(1\,-\,Q)\;\beta^2\;\left(x^2\,-\,a^2\right)
\;\left\{{\it sn}^2[\omega ((t\,-\,t_0),\,k^2]\;-\right.\\
\nonumber\\
\left.-\;<{\it sn}^2[\omega ((t\,-\,t_0),\,k^2]>\;\right\}\;=\\
\label{99.2}
\nonumber\\
=\;-\;\frac{1}{2}\;\rho\;(1\,-\,Q)\;\beta^2\;\left(x^2\,-\,a^2\right)
\;\left\{\,-\;\frac{1}{2}\;\cos 2[\omega (t\,-\,t_0)]\;+\;O(k^2)\right\}\;,\;\;
\;\;\;
\label{9.2}
\ea
\ba
\nonumber\\
\nonumber
\sigma_{yy}\;=\;
\frac{\rho}{2}\,(S\,+\,Q)\;\beta^2\;\left(y^2\,-\,b^2
\right)\;\left\{{\it sn}^2[\omega ((t\,-\,t_0),\,k^2]\;-\right.\\
\nonumber\\
-\left.<{\it sn}^2[\omega ((t\,-\,t_0),\,k^2]>\;\right\}\;=\\
\label{99.3}
\nonumber\\
=\;\frac{1}{2}\;\rho\;(S\,+\,Q)\;\beta^2\;\left(y^2\,-\,b^2\
\right)\;\left\{\,-\;\frac{1}{2}\,\cos 2[\omega (t\,-\,t_0)]\;+\;O(k^2)\right\}
\;\;\;,\;\;\;\;\;
\label{9.3}
\ea
\ba
\nonumber\\
\nonumber
\sigma_{zz}\;=\;
-\;\frac{\rho}{2}\,(1\,-\,S)\;\beta^2\;\left(z^2\,-\,c^2
\right)\;\left\{{\it sn}^2[\omega ((t\,-\,t_0),\,k^2]\;-\right.\\
\nonumber\\
\left.-\;<{\it sn}^2[\omega ((t\,-\,t_0),\,k^2]>\;\right\}\;=\\
\label{99.4}
\nonumber\\
=\;-\;\frac{1}{2}\;\rho\;(1\,-\,S)\;\beta^2\;\left(z^2\,-\,c^2\
\right)\;\left\{\,-\;\frac{1}{2}\;\cos 2[\omega (t\,-\,t_0)]\;+\;O(k^2)\right\}\;\;,\;\;\;\;\;
\label{9.4}
\ea
\ba
\nonumber\\
\nonumber\\
\nonumber
\sigma_{xy}\;=\;
\frac{\rho}{2}\,\left\{\left(\beta \gamma \right.\right.\;
\left.\left.+\;\alpha \omega \right)\left(y^2 - b^2\right)\,+\right.\;\;
\;\;\;\;\;\;\;\;\;\;\;\;\;\;\;\;\;\;\;\;\\
\nonumber\\
\nonumber
\left.\left.+\;\left(\beta \gamma - \alpha \omega \right)\left(x^2 - a^2\right)
\right\}\;\left\{{\it sn}[\omega ((t-t_0),\,k^2]\;{\it dn}[\omega ((t - t_0),\,
k^2]\;-\right.\right.\\ 
\nonumber\\
\left.-\;<{\it sn}[\omega ((t-t_0),\,k^2]\;{\it dn}[\omega ((t - t_0),\,k^2]>\right\}\,=\;\;\;\\
\label{99.5}
\nonumber\\
=\;\frac{1}{2}\;\rho\;\left\{\left(\beta\,\gamma\,-\,\alpha\,\omega \right)\;
\left(x^2\,-\,a^2\right)\;+\;
\left(\beta\,\gamma\,+\,\alpha\,\omega \right)\;
\left(y^2\,-\,b^2\right)\right\}\;\left\{
\sin [\omega (t\,-\,t_0)] \;+\;O(k^2)\right\}\;,\;\;\;\;\;
\label{9.5}
\ea
\ba
\nonumber\\
\nonumber
\sigma_{xz}\;=\;
\frac{\rho}{2}\,\left\{
\left(\beta\,\omega \,+\,\alpha\,\gamma\right)\,
\left(z^2\,-\,c^2\right)\,+\right.\;\;\;\;\;\;\;\;\;\;\;\;\;\;\;\;\;\;\;\;\;\\
\nonumber\\
\nonumber
\left.+\;\left(\,-\,\beta\,\omega \,+\,\alpha\,\gamma\right)\,
\left(x^2\,-\,a^2\right)\right\}\;\left\{
{\it dn}[\omega ((t\,-\,t_0),\,k^2]\;{\it cn}[p((t\,-\,t_0),\,k^2]\;-\right.\\
\nonumber\\
\left.-\;<{\it dn}[\omega ((t\,-\,t_0),\,k^2]\;{\it cn}[p((t\,-\,t_0),\,k^2]>
\right\}\;=\;\;\;\\
\label{99.6}
\nonumber\\
=\;\frac{1}{2}\;\rho\;\left\{
\left(\beta\,\omega \,+\,\alpha\,\gamma\right)\;
\left(z^2\,-\,c^2\right)\;+\;
\left(\,-\,\beta\,\omega \,+\,\alpha\,\gamma\right)\;
\left(x^2\,-\,a^2\right)\right\}\;\left\{
\cos [\omega (t\,-\,t_0)]\;+\;O(k^2)\right\}\;\;,\;\;\;\;\;
\label{9.6}
\ea
\ba
\nonumber\\
\nonumber
\sigma_{yz}\;=\;
\frac{\rho}{2}\,\left\{\left(\alpha \beta +\omega \gamma k^2\right)
\left(z^2 - c^2\right)\,+\right.\;\;\;\;\;\;\;\;\;\;\;\;\;\;\;\;\;\;\;\;\;
\;\\
\nonumber\\
\nonumber
\left.+\;\left(\alpha \beta - \omega  \gamma k^2 \right)
\left(y^2 -\,b^2\right)\right\}\;\left\{
{\it cn}[\omega ((t - t_0),\,k^2]\;{\it sn}[\omega ((t - t_0),\,k^2]\;-\right.
\\
\nonumber\\
\left.-\;<{\it cn}[\omega ((t - t_0),\,k^2]\;{\it sn}[\omega ((t - t_0),\,k^2]>
\right\}\;=\;\;\\
\label{99.7}
\nonumber\\
\nonumber
=\;\frac{1}{2}\;\rho\;\left\{
\left(\alpha\,\beta\,+\,\omega \,\gamma\,k^2\right)\;
\left(z^2\,-\,c^2\right)\;+\right.\\
\nonumber\\
\left.+\;\left(\alpha\,\beta\,-\,\omega \,\gamma\,k^2\right)\;
\left(y^2\,-\,b^2\right)\right\}\;\left\{
\frac{1}{2}\;\sin 2[\omega (t\,-\,t_0)]\;+\;O(k^2)\right\}\;\;,\;\;\;\;\;
\label{9.7}
\ea
wherefrom we obtain the (averaged over the precession period) quantities that 
emerge in the expression (\ref{7.2}) for the dissipation rate: 
$\;\sigma_{ij}^2\;$ and $\;\left(Tr \,\sigma \right)^2\;$. These expressions 
are written down in Appendix D. Substitution thereof into (\ref{7.2}), with 
the further integration gives the total energy $\;W\;$ of alternating stresses.

Similarly to (8.6) - (8.17), the above formulae give a polynomial approximation to the stress tensor. (See the comment after formula (8.17).)

All the further scheme of calculation exactly repeats that from the preceding 
section. Energy $\,W\,$ consists of two components, one on the principal 
frequency, another 
on the second mode. These should be multiplied by $\;2\,\omega/Q(\omega)\;$ and
$\;4\,\omega/Q(2\omega)\;$, correspondingly (as in formula (\ref{6.6})). It  
will give us the overall dissipation rate $\;\dot{W}\;$. Plugging this rate,
along with (\ref{5.4}) into (\ref{5.6}) yields:
\ba
\nonumber
\frac{d\,<\sin^2 \theta>}{dt}\,=\,-\,\frac{2\;\rho^2}{\mu\;Q(\omega)\;{\bf 
J}^2}\;\frac{I_1\;I_3}{I_3\,-\,I_1}\left({\bf J}^2\;-\;2\;I_1\;T_{kin}\right)\;
\left\{\,\omega\;\left(2\;I_3\;T_{kin}\;-\;{\bf J}^2\right)\;S_1\;
\left[1\;-\right. \right.\\
\nonumber\\
\left. \left.-\;\frac{1}{4}\;\frac{I_3\,-\,I_2}{I_2\,
-\,I_1}\;\frac{{\bf J}^2\;-\;2\;I_1\;T_{kin}}{2\;I_3\;T_{kin}\;-\;{\bf J}^2}\right]\;-\;\omega\;\left({\bf J}^2\;-\;2\;I_1\;T_{kin}\right)\;S_0
\;+\;2\;\omega\;\frac{Q(\omega)}{Q(2\,\omega)}\;\left({\bf J}^2\;-\;2\;I_1\;T_{kin}\right)\;S_2
\right\} \;\;\;
\label{9.8}
\ea
where the geometrical factors $\;S_0\;$, $\;S_1\;$ and $\;S_2\;$ are given in 
Appendix D.

In the case of a homogeneous rectangular prism of dimensions $\;2a\,\times\,2b
\,\times\,2c\;$, the factors $\;S_0\;$, $\;S_1\;$ and $\;S_2\;$ read (see 
Appendix D):
\ba
S_1\;\equiv\;\frac{86.5}{m^4}\;\frac{a\;b\;c}{\left(a^2\,+\,b^2\right)\;\left(b^2\,+\,c^2  
\right)\;\left(a^4\,-\,c^4\right)}\;\left[ 
\frac{a^8\,+\,1.7\,a^4\,b^4\,+\,b^8}{a^4\,-\,b^4}\,+\,\frac{a^8\,+\,1.7\,a^4\,
c^4\,+\,c^8}{a^4\,-\,c^4}\right]\;\;,\;
\label{9.11}
\ea
\ba
\nonumber
S_2\;\equiv\;\frac{21}{m^4}\;\frac{a\;b\;c}{\left(b^2\,+\,c^2\right)^2\;
\left(a^4\,-\,c^4\right)^2\;\left(a^4\,-\,b^4\right)^2}\;\left[ 
a^8\,b^8\,+\,2.8\,a^8\,b^4\,c^4\,-\,4.8\,a^4\,b^8\,c^4\,+\,a^8\,c^8\,-\right.\\
\left.-\,4.8\,a^4\,b^4\,c^8\,+\,4.8\,b^8\,c^8\right]\;\;\;
\label{9.12}
\ea
and
\ba
S_0\;=\;\frac{32.4}{m^4}\;a\;b\;c\;\frac{\left(a^8\,+\,1.67\,a^4\,c^4\,+\,
c^8\right)\,\left(b^2\,-\,c^2\right)}{(a^4\,-\,b^4)\,(a^4\,-\,c^4)^2\,(b^2\,+
\,c^2)}
\label{S0}
\ea
As explained in the end of the preceding section, multipliers like 
$\;(I_3\,-\,I_1)\;$ and $\;(I_2\,-\,I_1)\;$ in the denominators of the 
expressions for $\;S_0,\;S_1\;,S_2\;$ presented in Apendix D, as well as 
multipliers $\;\left(a^4\,-\,c^4\right)^2 \;$ and $\;\left(a^4\,-\,b^4\right)^2
\;$ in the denominators of (\ref{9.11}), (\ref{9.12}) and (\ref{S0}) are 
harmless. 

Similarly to pole C, in (\ref{9.8}) we have two contributions: the one with 
$S_0$ and $S_1$ originates 
from the dissipation of oscillations at frequency $\;\omega\;$, while the 
one with $S_2$ comes from $\;2\,\omega\;$. Often it is the second mode that 
dominates the dissipation. For the case of an oblate body ($\;I_3\;>\;I_1\;=\;
I_2\;$) this fact was proven in (Efroimsky and Lazarian 2000)$^{24}$. In the 
case of an almost
prolate rotator, the importance of the second mode can be easily understood 
simply from looking at Fig. 1. We see that the trajectories described by the 
vector $\;\Omega\;$ remain more or less circular up to a close vicinity of the
separatrix, i.e., that the trigonometric approximation of the Jacobi functions 
is valid through a fairly large region. In this region, therefore, our 
formalism does work. Let us estimate the input of the second mode at points $D$
and $\,F\,$ on Fig. 1. Point $\,D\,$ depicts the situation when $\,\Omega_3\,=
\,0\,$ and $\;\Omega_2/ \Omega_1\,=\,1/2\;$, while point $\,F\,$ corresponds to
$\;\Omega_2/ \Omega_1\,=\,1/7\,$. Following (Black {\it et al} 1999)$^{25}$ we 
have 
prepared the picture so that it corresponds to an example from real life, 
asteroid (433) Eros. To that end we assumed $\;I_2\,=\,3\,I_1\;, \;\;I_3\,=\,
3.05\,I_1\;$, which is the same as $\;a\,=\,2.19\,c\;,\;\;b\,=\,1.05\,c\,$. 
A simple calculation using (\ref{9.11} - \ref{9.12}) and (\ref{2.5} - 
\ref{2.6}) shows that at point $D$ the second-mode term in (\ref{9.8}) is less 
than one tenth of the principal-mode term with $S_1$ ($S_0$ being negligibly 
small):
\be
\frac{2\,S_2}{S_1}\;\left(\frac{{\bf J}^2\;-\;2\;I_1\;T_{kin}}{2\;I_3\;T_{kin}
\;-\;{\bf J}^2}\right)_D\;=\;0.08
\label{9.13}
\ee
At point $F$ though, the second-mode contribution slightly dominates: 
\be
\frac{2\,S_2}{S_1}\;\left(\frac{{\bf J}^2\;-\;2\;I_1\;T_{kin}}{2\;I_3\;T_{kin}
\;-\;{\bf J}^2}\right)_F\;=\;1.02\;\;,\;\;\;
\label{9.14}
\ee
though in reality the nonlinear input at $F$ is higher; first, because of the 
less-than-unity multiplier accompanying $\;\omega\,S_1\;$ in (\ref{9.8}) and, 
second, because of the higher-than-second harmonics. It is perhaps pointless to
approach the separatrix closer than $F$, because the higher-frequency terms 
omitted in (\ref{8.1} - \ref{8.3}) will become relevant. Anyway, their 
relevance will only add to the nonlinearity.

Equation (\ref{9.8}), along with (\ref{2.15}) and equation 
\ba
{\bf J}^2\;<\sin^2 \theta>_{(near\;A)}\;=\;\frac{I_1}{I_3\,-\,I_1}\;\left( 
2\,I_3\,T_{kin}\;-\;{\bf J}^2\right)\;+\;\frac{1}{2}\;\frac{I_3}{I_3\,-
\,I_1}\;\left({\bf J}^2\,-\,2\,I_1\,T_{kin}\right)\;+\;O(k^4)\;\;
\label{9.15}
\ea
that follows from (\ref{5.2}) (see also formula (A14) in Appendix A),
constitute a self-consistent system of equations describing relaxation in the 
vicinity of pole A. By means of the latter equation, one may express $\;T_{kin}
\;$ through $\;<\sin^2 \theta>\;$ and plug the result into (\ref{9.8}). This 
would yield a differential Master Equation for  $\;<\sin^2 \theta>\;$:
\ba
\nonumber
\frac{d\,<\sin^2 \theta>}{dt}\,=\\
\nonumber\\
\nonumber
-\,\frac{4\;\rho^2\;{\bf J}^2}{\mu\;Q(\omega)}\;\left(I_3\,-\,I_1\right)\;
\left(1\,-\;<\sin^2 \theta > \,\right)\;\left\{\,\omega\,S_1\,\left[2\;<\sin^2 
\theta>\;-\;1\;-\right.\right.\\
\nonumber\\
\nonumber
\left.\left.-\;\frac{1}{2}\;\frac{I_3\,-\,I_2}{I_2\,-\,I_1}\;\frac{I_1}{I_3}\;
\left(1\,-\;<\sin^2 \theta > \,\right)\right]\;-\right.\\
\nonumber\\
\left.-\;\omega\;S_0\;\frac{2\,I_1}{I_3}\;\left(1\,-\;<\sin^2 \theta > \,
\right)\;+\;2\,\omega\,S_2\;\frac{Q(2\omega)}{Q(\omega)}\;\frac{2\,I_1}{I_3}\;
\left(1\,-\;<\sin^2 \theta>\,\right)\right\}\;.\;\;
\label{9.16}
\ea
where, according to (\ref{2.15}) and (\ref{9.15}),
\be
\omega\;=\;\frac{{|{\bf J}|}}{I_1}\;\sqrt{\frac{(I_3-I_1)\,(I_3-I_2)}{I_2\,I_3}}\;
\sqrt{2\;<\sin^2 \theta>\;-\,1}
\label{9.17}
\ee
For $\;\sin^2 \theta\,=\,0\;$, i.e., for $\;\bf \Omega\;$ parallel to axis (1),
the relaxation rate $\;d\,<\sin^2 \theta>/{dt}\;$ vanishes. This means that 
vector $\;\bf \Omega\;$ will be leaving the unstable-equilibrium point (pole A
on Fig.1) infinitesimally slowly.

When $\;(I_3\,-\,I_1)\,\rightarrow\,0\;$, the expressions (D18) - (D20) for $\,
S_{0,1,2}\,$ diverge as $\;(I_3\,-\,I_1)^{-2}\;$. Therefore the right-hand side
of (\ref{9.16}) will diverge as $\;(I_3\,-\,I_1)^{-1}\;$, instead of 
approaching zero as one might expect on physical grounds. The reason for this 
would-be divergence is that our analysis is applicable, as explained in Section
 II, only in the adiabatic approximation. Therefore (\ref{9.16}), as well as 
(\ref{8.20}), works for as long as inequality (\ref{adiabaticity}) holds.

Similarly to the Master Equation (\ref{8.20}), the Master Equation (\ref{9.16})
becomes a self-consistent differential equation for $\;<\sin^2 \theta>\;$ only 
after (\ref{9.17}) is substituted into it. Similarly to (\ref{8.20}), here we 
have deliberately abstended from plugging (\ref{9.17}) into (\ref{9.16}), 
because without this substitution (\ref{9.16}) is of more practical use. (See 
the explanation in the end of the preceding section, between equations 
(\ref{omega}) and (\ref{21})).

\section{Conclusions and Practical Applications.}

Formulae (\ref{9.16}), for pole A, and (\ref{8.20}), for pole 
C, constitute the main result of this article. These are differential equations
for the relaxation rate of a precessing homogeneous body of arbitrary moments 
of inertia. The relaxation rate is defined as the rate at which the 
major-inertia axis approaches the angular momentum about which it is 
precessing.

Formula  (\ref{8.20}) describes the relaxation rate of a body spinning 
approximately about its maximal-inertia axis (pole C), while (\ref{9.16}) 
describes the relaxation of a body rotating almost about its minimal-inertia 
axis (pole A).

In these formulae the contributions of the modes $\;\omega\;$ and $\;2\,\omega
\;$ are manifestly separated ($\omega$ being the precession rate). Often the 
dissipation at $\;2\,\omega \;$ gives a considerable input (as shown in the 
example (\ref{9.14}) in the end of section IX), or even dominates,
as in the case of oblate body, when $\;I_1\,=\,I_2\;$ - see (Efroimsky 
\& Lazarian 2000)$^{24}$.

Our formulae (\ref{9.16}) and (\ref{8.20}) were derived in assumption of 
parameter $\;k\;$ being small. When does this assumption work? To get a 
simple answer to this question, let us look again at $\;Fig.\;1\,$. The 
approximation is valid on the part of the ellipsoid surface, that is covered
with almost circular trajectories; a divergence of trajectory shape from a 
circle signals of inacceptability of the approximation (\ref{8.1} - \ref{8.3}).
We see from the picture that, for example, for an almost prolate body our 
approximation remains valid not only in the vicinity of pole A but almost all 
way up to the separatrix. However, after the separatrix is crossed and the 
body begins librations, our approximation will regain its validity only in the
closemost vicinity of pole C. Our formula (\ref{8.20}) coincides, in the limit
of oblate symmetry, with the exact formula (\ref{8.24}) derived for oblate 
bodies in (Efroimsky \& Lazarian 2000)$^{24}$.

The developed formalism has two immediate applications we know of. First of 
all, it is the study of wobbling asteroids (Harris 1998)$^{35}$. Wobbling may 
provide valuable 
information on the composition and structure of asteroids and on their recent 
history of external impacts. Astronomers observing precessing asteroids often 
ask: ``Why do we have so few excited rotators in the Solar System?'' (Steven 
Ostro, private communication). One of the reasons for this deficiency is that, 
due to the dissipation in the second mode (Efroimsky and Lazarian 2000)$^{24}$ 
and higher modes, the effectiveness of the inelastic relaxation turns to be 
much higher that thought previously.
 
Another obvious application of our formalism is the problem of cosmic-dust 
alignment. When they talk in astrophysics about the cosmic-dust alignment, they
imply not the alignment of $\;\bf \Omega\;$ or the major-inerta axis along 
the angular momentum $\;\bf J\;$ but the alignment of the major-inerta axis 
relative to the 
interstellar magnetic field. This is a well-known phenomenon that manifests 
itself through the observable polarisation of starlight. There exists a whole 
bunch of physical mechanisms that make the cosmic dust align with respect to 
the magnetic field. Different types of these mechanisms dominate under 
different physical conditions, but all of them are based on 
the fact that cosmic-dust granules are swiftly precessing about the magnetic 
field lines. This precession is called into being by the interaction of the 
magnetic moment of the granule with the field. The magnetic moment is created 
by the Barnett effect and is thus parallel to the angular velocity $\;\bf 
\Omega\;$. Some of the known mechanisms of cosmic-dust alignment are very 
sensitive to the coupling between $\;\bf \Omega\;$ and $\;\bf J\;$, and this 
is when the inelastic relaxation comes into play. Until recently the Barnett 
relaxation was believed to be the leading relaxation mechanism. This viewpoint
was expressed in the long-standing article by (Purcell 1979)$^{34}$. 
Remarkably, Purcell underestimated the effectiveness of the inelastic 
relaxation in the same manner as (Burns and Safronov 1973))$^{23}$ did it for 
asteroids: he missed the input 
provided by the second mode. Besides, he failed to satisfy the boundary 
conditions on the stresses. As a result, he underevaluated the effectiveness 
of the inelastic dissipation by several orders (see (Lazarian and Efroimsky 
2000))$^{24}$). A study of the role of the inelastic dissipation in various 
mechanisms of cosmic-dust alignment is now on the way, and some results have 
already been published by (Lazarian and Draine 1999))$^{15}$.

There may be a possibility that the developed formalism finds its third 
application in the research on spacecraft-rotation damping. 

\section{The generic case}

In the article thus far we have studied the dissipation in the vicinity of 
poles, i.e., the case when the body rotates about an axis that is close either 
to that of minimal or maximal inertia. All our formulae were derived up to 
$\;O(k^4)\;$, parameter $\;k\;$ being small near poles A and C (Fig. 1). 

In the general case, $\;k\;$ is not small. For example, near the separatrix 
(see Fig. 1) it approaches unity. We have solved the general case in terms of 
the elliptic integrals. In particular, in the vicinity of the separatrix the 
solution may be expanded over the small parameter $\;k'\,=\,\sqrt{1\,-\,k^2}
\;$. These results will be published elsewhere.\\
\\

{\bf Acknowledgements}  

My profoundest gratitude goes to Vladislav Sidorenko who read
the manuscript and came up with extremely helpful criticisms. I wish to thank 
Alex Lazarian, Mark Levi, Steven Ostro and Daniel Scheeres for stimulating 
discussions. I am also thankful to Brian Marsden and Irwin Shapiro for the 
encouragement they provided.

\pagebreak

\appendix

\section{\\
Derivation of equations (5.2) - (5.4) \label{A}}

In this Appendix we shall compute the derivative $\;d{\,<\sin^2 \theta>}/dT_{
kin}\;$. Angle $\;\theta\;$ is the one between the angular momentum vector $\;
\bf J\;$ and the plane determined by the body's minor- and middle-inertia axes,
1 and 2. In the case of an oblate body, this angle also remains virtually 
unchanged through a period of precession, and changes slowly through many 
cycles (Efroimsky and Lazarian 2000)$^{24}$. In the general case of a triaxial 
body, this angle is not preserved through a precession period, though after one
period of wobble it returns (adiabatically) to its initial value. The word 
``adiabatically'' is in order here because in the course of many cycles angle 
$\;\theta\;$ slowly decreases and thus deviates from the exact periodicity.

In the body frame ($\,{\bf e}_1, \;{\bf e}_2,\;{\bf e}_3\,$) associated with 
the principal axes (1, 2, 3), the angular momentum components look: $\;J_i\;=\;
\Omega_i\;I_i\;$. Hence, according to (\ref{2.7}),
\be
|{\bf J}|\;\sin \theta\;=\;\sqrt{I_1^2\,\Omega_1^2\;+\;I_2^2\,\Omega_2^2}\;=\;
\sqrt{I_1^2\;R\;+\;(I_2^2\;-\;I_1^2\;S)\;\Omega_2^2}\;\;\;.
\label{A1}
\ee
This shows that the angle $\;\theta\;$ evolves with the same period as 
$\;\Omega_2^2\;=\;\beta^2\;{\it sn}^2[\omega (t\,-\,t_0);\,k^2 ]\;$. We remind 
that
the periods of $\;{\it sn}[\omega (t\,-\,t_0);\,k^2 ]\;$ and $\;{\it cn}[\omega
(t\,-\,t_0);\,k^2 ]\;$ are equal to
\be
\tau\;=\;4\;K(k^2)\;\equiv\;4\;\int_{0}^{\pi/2}\;
(1\,-\,k^2\,\sin^2 \theta )^{-1/2}\;d\,\theta\;=\;2\,\pi\;\left(1\;+\;
\frac{k^2}{4}\right)\;+\;O(k^4)\;\;
\label{A2}
\ee
which is twice the period of $\;{\it dn}[\omega (t\,-\,t_0);\,k^2 ]\;$. 
(See formula (16.1.1) in (Abramovitz and Stegun 1964)$^{27}$.) The 
$k^2-$approximation in the right-hand side of (A2) follows from the expansion 
(A7) below. The periods of the squares of $\;sn\;$ and $\;cn\;$ are equal to 
\be
\frac{\tau}{2}\;=\;2\;K(k^2)\;\equiv\;2\;\int_{0}^{\pi/2}\;(1\,-\,k^2\,\sin^2 
\theta )^{-1/2}\;d\,\theta
\label{A3}
\ee
which is easy to understand from Fig. 16.1 in (Abramovitz and Stegun 
1964)$^{27}$. 

Squaring and averaging of (A1) yields:
\ba
\nonumber
{\bf J}^2\;<\sin^2 \theta>\;=\;I_1^2\,R\;+\;(I_2^2\,-\,I_1^2\,S)\;<\Omega_2^2>
\;=\\
\nonumber\\
=\;\frac{I_1}{I_3\,-\,I_1}\,\left(2\,I_3\,T_{kin}\;-\;{\bf J}^2 \right)\;+\;
I_2\;I_3\;\frac{I_2\,-\,I_1}{I_3\,-\,I_1}\;\beta^2\;<{\it sn}^2 u>\;
\;.
\label{A4}
\ea
where $\;u\,\equiv\,\omega \,(t-t_0)\;$. The trajectories described by the 
angular velocity vector $\;\bf \Omega\;$ on the surface of the constant-$\bf J$
ellipsoid (Fig.1) are cyclic. The averaging may be performed over one such 
cycle, i.e., over the period $4K$ given by (A2). We shall though average over 
quarter of the total period, i.e., over $K$. This will be sufficient since in 
the above expressions (A1) and (A4) we have only squared components of $\;
\bf \Omega\;$. Thus, 
\ba
<{\it sn}^2 u>\;=\;\frac{{\it Sn}(K)}{K}\;\;\;
\label{A5}
\ea
where the function $\;\it Sn\;$ is, by definition (formula (16.25.1) in 
Abramovitz and Stegun 1964)$^{27}$, squared $\;\it sn\;$ integrated from zero 
to $K$. According to formula (16.26.1) in (Abramovitz and Stegun 1964)$^{27}$, 
$\;{\it Sn}(K)\,=\,(-\,E(K)\,+\,K)/k^2\;$ where $\;E(K(k^2))\,\equiv\,
\int_{0}^{\pi/2}\;(1\,-\,k^2\,\sin^2 \theta)^{1/2}\,d \theta\,$. Hence,
\be
<{\it sn}^2 u>\;=\;\frac{1}{k^2}\;\left\{1\;-\;\frac{E(K)}{K} \right\}\;\;\;.
\label{A6}
\ee
Series expansions (formulae (17.3.11) and (17.3.12) in (Abramovitz and Stegun 
1964)$^{27}$
\be
K(k^2)\;=\;\frac{\pi}{2}\;\left[1\;+\;\left(\frac{1}{2}\right)^2\,k^2\;+\;
\left(\frac{1\,\cdot\,3}{2\,\cdot\,4}\right)^2\,k^4\;+\;
\left(\frac{1\,\cdot\,3\,\cdot\,5}{2\,\cdot\,4\,\cdot\,6}\right)^2\,k^6\;+\;
...\right]
\label{A7}
\ee
and
\be
E(k^2)\;=\;\frac{\pi}{2}\;\left[1\;-\;\left(\frac{1}{2}\right)^2\,k^2\;-\;
\left(\frac{1\,\cdot\,3}{2\,\cdot\,4}\right)^2\,\frac{k^4}{3}\;-\;
\left(\frac{1\,\cdot\,3\,\cdot\,5}{2\,\cdot\,4\,\cdot\,6}\right)^2\,\frac{k^6}{5}\;-\;...\right]
\label{A8}
\ee
entail:
\be
<{\it sn}^2 u>\;=\;\frac{1}{2}\;+\;\frac{1}{16}\,k^2\;+\;\frac{1}{32}\,k^4\;
+\;O(k^6)\;\;
\label{A9}
\ee
so that
\ba
\nonumber
{\bf J}^2\;<\sin^2 \theta>\;=\;I_1^2\,R\;+\;(I_2^2\,-\,I_1^2\,S)\;<\Omega_2^2>
\;=\\
\nonumber\\
=\;\frac{I_1}{I_3\,-\,I_1}\,\left(2\,I_3\,T_{kin}\;-\;{\bf J}^2 \right)\;+\;
I_2\;I_3\;\frac{I_2\,-\,I_1}{I_3\,-\,I_1}\;\beta^2\;\frac{1}{2}\;
\left(1\;+\;\frac{k^2}{8}\;+\;\frac{k^4}{16}\;+\;O(k^6)\right)\;\;.\;
\label{A10}
\ea
It follows from (A10) and (\ref{2.13}) that in the vicinity of pole C
\ba
\nonumber
{\bf J}^2\;<\sin^2 \theta>_{(near\;C)}\;=\;\frac{1}{2}\;\frac{I_1\,I_3\,+\,I_2
\,I_3\,-\,2\,I_1\,I_2}{(I_3
\,-\,I_1)(I_3\,-\,I_2)}\,\left(2\,I_3\,T_{kin}\;-\;{\bf J}^2 \right)\;+\\
\nonumber\\
\nonumber
+\;\frac{1}{16}\;\frac{I_3\;(I_2\,-\,I_1)^2}{(I_3\,-\,I_1)\,(I_3\,-\,I_2)^2}\;
\frac{(2\,I_3\,T_{kin}\,-\,{\bf J}^2)^2}{{\bf J}^2\,-\,2\,I_1\,T_{kin}}\;+\;
\frac{1}{32}\;\frac{I_3\,(I_2\,-\,I_1)^3}{(I_3\,-\,I_1)\,(I_3\,-\,I_2)^3}\;
\frac{(2\,I_3\,T_{kin}\,-\,{\bf J}^2)^3}{({\bf J}^2\,-\,2\,I_1\,T_{kin})^2}\\
\nonumber\\
+\;O(k^8)\;\;\;\;\;\;\;\;\;\;\;\;\;\;\;\;\;\;\;\;\;\;\;\;\;\;\;\;\;\;\;
\label{A11}
\ea
wherefrom
\ba
\nonumber
{\bf J}^2\;\left(\frac{d\,<\sin^2 \theta>}{dT_{kin}}\right)_{(near\;C)}
=\;\frac{I_3\,(I_1\,I_3\,+\,I_2
\,I_3\,-\,2\,I_1\,I_2)}{
(I_3\,-\,I_1)\,(I_3\,-\,I_2)}\;+\\
\nonumber\\
+\;\frac{1}{4}\;\frac{2\,I_3\,T_{kin}\;-\;{\bf J}^2}{{\bf J}^2\,-\,2\,I_1\,T_{kin}}\;\left(\frac{I_2\,-\,I_1}{I_3\,-\,I_2}\right)^2\;
\frac{I_3^2}{I_3\,-\,I_1}\;+\;O(k^4)\;\;.\;
\label{A12}
\ea
Together, the former and the latter yield:
\ba
\nonumber
{\bf J}^2\;\left(\frac{d\,<\sin^2 \theta>}{dT_{kin}}\right)_{(near\;C)}\;
=\;\frac{I_3\,(I_1\,I_3\,+\,I_2
\,I_3\,-\,2\,I_1\,I_2)}{
(I_3\,-\,I_1)\,(I_3\,-\,I_2)}\;+\\
\nonumber\\
+\;\frac{1}{2}\;\frac{(I_2\,-\,
I_1)^2}{(I_3\,-\,I_2)(I_3\,-\,I_1)}\;\frac{I_3^2\;<\sin^2 \theta>}{I_2\,I_3\,-
\,I_2\,I_1\,-\,2\,(I_3-I_2)\,I_1\,<\sin^2 \theta>}\;+\;O(k^4)\;\;\;\;.\;\;\;\;\;\;\;
\;\;\;\;
\label{A13}
\ea
Now we shall derive similar formulae for the vicinity of pole
A. Plugging (\ref{2.15}) into (A10) we get:
\ba
\nonumber
{\bf J}^2\;<\sin^2 \theta>_{(near\;A)}\;=\;\frac{I_1}{I_3\,-\,I_1}\;\left( 
2\,I_3\,T_{kin}\;-\;{\bf J}^2\right)\;+\;\frac{1}{2}\;\frac{I_3}{I_3\,-
\,I_1}\;\left({\bf J}^2\,-\,2\,I_1\,T_{kin}\right)\;+\\
\nonumber\\
+\;\frac{1}{16}\;
\frac{\left({\bf J}^2-2\,I_1 T_{kin}\right)^2}{\left( 
2\,I_3 T_{kin}-{\bf J}^2\right)}\;\frac{(I_3\,-\,I_2)\,I_3}{(I_3-I_1)(
I_2-I_1)}\;+\;\frac{1}{32}\;\frac{I_3\,(I_3\,-\,I_2)^2}{(I_3-I_1)(
I_2-I_1)^2}\;\frac{\left({\bf J}^2-2\,I_1 T_{kin}\right)^3}{\left( 
2\,I_3 T_{kin}-{\bf J}^2\right)^2}\;+\;O(k^8)\;\;.\;\;\;
\label{A14}
\ea
This enables us to write down the derivative we need:
\ba
{\bf J}^2\;\left(\frac{d\,<\,\sin^2 \theta\,>}{dT_{kin}}\right)_{(near\;A)}=
\;\frac{I_1\,I_3}{I_3\,-\,I_1}\;-\;\frac{1}{4}\;\frac{I_1\,I_3\,(I_3\,-\,I_2
)}{(I_3\,-\,I_1)\,(I_2\,-\,I_1)}\;\frac{{\bf J}^2\,-\,2\,I_1\,T_{kin}}{2\,I_3
\,T_{kin}\;-\;{\bf J}^2}\;+\;O(k^4)\;.\;
\label{A15}
\ea
In the limit of oblate ($\;I_3\,>\,I_2\,=\,I_1\;$) or prolate ($\;I_3\,=\,I_2\,>\,I_1\;$) symmetry, expressions (A13) and (A15) simplify a lot:
\ba
{\bf J}^2\;\left(\frac{d\,<\,\sin^2 \theta\,>}{dT_{kin}}\right)_{(near\;C)}^{
oblate}=\;\frac{2\,I_3\,I_1}{
I_3\,-\,I_1}\;\;\;, \;\;\;\;\;\;\;\;
{\bf J}^2\;\left(\frac{d\,<\,\sin^2 \theta\,>}{dT_{kin}}\right)_{(near\;A)}^{
prolate}=\;\frac{I_1\,I_3}{I_3\,-\,I_1}\;\;\;\;\;\;\;\;\;
\label{A16}
\ea
In the general case of a triaxial rotator, it ensues from (A14) and (A15) that
\ba
\nonumber
{\bf J}^2\;\left(\frac{d\,<\,\sin^2 \theta\,>}{dT_{kin}}\right)_{(near\;A)}\;=
\;\;\;\;\;\;\;\;\;\;\;\;\;\;\;\;\;\;\;\;\;\;\;\\
\nonumber\\
=\;\frac{I_1\,I_3}{I_3\,-\,I_1}\;-\;\frac{1}{2}\;\frac{I_1^2}{I_2\,-
\,I_1}\;\frac{I_3\,-\,I_2}{I_3\,-\,I_1}\;\frac{1\;-\;<\sin^2\theta>}{2\;
<\sin^2\theta>\,-\,1}\;+\;O(k^4)\;\;.\;\;\;
\label{A17}
\ea

\section{\\
Several words on the quality factor\label{B}}

In some situations it is possible to calculate the quality factor $Q(\omega)$ 
exactly. These are situations when one particular mechanism of attenuation 
dominates the others. For example, $Q$ may be derived analytically for sound 
dissipation in a viscous liquid. It is said in (Landau and Lifshitz 
1970)$^{36}$ that the calculation of the quality factor
in a solid body would basically follow the same steps as in the case of a 
viscous liquid, in faith whereof the authors even present some laborious 
thermodynamical calculations. Unfortunately, in many cases this is not true, 
and the viscousity of solids contributes almost nothing to the 
attenuation (unless the body is warmed up to a plastic state). A much larger  
contribution to the attenuation is brought, in many materials, by phonon 
scattering on defects, and by a whole variety of related quantum effects 
(Nowick and Berry 1972)$^{30}$. In rocks, the attenuation is determined 
predominantly by displacement of defects. The numerous phenomena participating 
in the attenuation are subtle and bear a complicated dependence upon the 
temperature and frequency. Also mind a dramatic dependence of $Q$ upon the 
humidity, as well as upon the presence of some other saturants. In many 
minerals, including for example silicate rocks, several monolayers of water may
decrease $Q$ by a factor of about 55 (Tittman, Ahlberg, and Curnow 
1976)$^{37}$. It is for this reason that the moonquakes cause an echo that 
keeps propagating and reflecting for long, almost without any attenuation. The 
echo would be dumped much faster, should the lunar lithosphere contain even a 
tiny fraction of water. Presumably, the moisture affects the inter-grain 
interactions in minerals. Another factor influencing $Q$ is the confining 
pressure, but the pressure dependence is very weak within a broad (several 
orders) interval of pressures, and may be neglected. 

Returning to the frequency dependence, we would say that, fortunately, the 
overall frequency-dependence of $Q$ is normally very smooth and slow, like for 
example, in the case of geological materials. Here follows the empirical 
temperature- and frequency-dependence of the quality factor, well supported by
a vast experimental evidence (Karato 1998)$^{38}$:
\be
\; Q \; \sim \; \left[ \omega \; \exp(A^*/RT) \protect\right]^{\alpha} \;, 
\label{B1}
\ee
where $\, A^* \, $ may vary from 150 - 200 $\, kJ/mol $ (for dunite and 
polycristalline forsterite) up to 450 $kJ/mol$ (for olivine). 
This interconnection between the frequency- and temperature-dependences 
tells us that whenever we lack a pronounced frequency-dependence, the 
temperature-dependence is absent too. At room temperature and pressure, at low 
frequencies ($10^{-3}  - \, 1 \, Hz$) the shear $Q-$factor is 
frequency-independent for granites, and almost frequency-independent (except 
some specific peak of attenuation, that makes $Q$ increase twice) for basalts
(Brennan 1981)$^{39}$. It means that within this range of frequencies $\,\alpha
\,$ is close to zero, and $Q$ may be assumed also temperature-independent. 
For higher frequencies ($10 Hz  - 1 \, MHz$), the power $\alpha $ is
remarkably frequency-insensitive (and equals approximately $0.25$ for most 
silicate rocks). For a recent discussion and references on the $Q$-factor of
asteroids see (Efroimsky and Lazarian 2000)$^{24}$. As for the $Q$-factor of 
the cometary nuclei, its value is unknown. Presumably, it should be close to 
the values of the $Q$-factor that are typical for snow and firn: between 0.5 
and 5.

\section{\\
Averaged over the precession period squares of the components of 
the stress tensor, in the vicinity of pole C \label{C}}

Formulae (\ref{8.5}) - (\ref{8.10}) trivially yield:
\ba
<\sigma_{xx}^2>\;=\;\frac{\rho^2}{4}\;\left(x^2\;-\;a^2\right)^2\;\left(1\,-\,Q
\right)^2\;\beta^4\;\;\Xi_1\;\;=\;\;\;\;\;\;\;\;\;\\
\label{C1}
\nonumber\\
=\;\frac{\rho^2}{32}\;\left(2\;I_3\;T_{kin}\;-\;
{\bf J}^2\right)^2
\;\left\{\frac{I_3\;\left(I_1\;-\;I_3\right)\;-\;I_2\;\left(I_1\;-\;
I_2\right)}{I_2\;I_3\;\left(I_1\;-\;I_3\right)\;\left(I_3\;-\;I_2\right)\;}
\right\}^2\;\left(x^2\;-\;a^2\right)^2\;\left\{1\;+\;O(k^4)\right\}\;\;,\;\;\;
\label{C2}
\ea
\ba
\nonumber\\
<\sigma_{yy}^2>\;=\;\frac{\rho^2}{4}\;\left(y^2\;-\;b^2\right)^2\;\left(S\,+\,Q
\right)^2\;\beta^4\;\;\Xi_1\;\;=\;\;\;\;\;\;\;\;\;\;\;\;\;\\
\label{C3}
\nonumber\\
= \;\frac{\rho^2}{32}\;\left(2\;I_3\;T_{kin}\;-\;
{\bf J}^2\right)^2
\;\left\{\frac{I_3\;\left(I_3\;-\;I_2\right)\;-\;I_1\;\left(I_1\;-\;
I_2\right)}{I_1\;I_3\;\left(I_3\;-\;I_1\right)\;\left(I_3\;-\;I_2\right)\;}
\right\}^2\;\left(y^2\;-\;b^2\right)^2\;\left\{1\;+\;O(k^4)\right\}\;\;,\;\;\;
\label{C4}
\ea
\ba
\nonumber\\
<\sigma_{zz}^2>\;=\;\frac{\rho^2}{4}\;\left(z^2\;-\;c^2\right)^2\;\left(1\,-\,S
\right)^2\;\beta^4\;\;\Xi_1\;\;=\;\;\;\;\;\;\;\;\;\;\;\;\;\\
\label{C5}
\nonumber\\
= \;\frac{\rho^2}{32}\;\left(2\;I_3\;T_{kin}\;-\;
{\bf J}^2\right)^2
\;\left\{\frac{I_1\;\left(I_3\;-\;I_1\right)\;-\;I_2\;\left(I_3\;-\;
I_2\right)}{I_1\;I_2\;\left(I_3\;-\;I_1\right)\;\left(I_3\;-\;I_2\right)\;}
\right\}^2\;\left(z^2\;-\;c^2\right)^2\;\left\{1\;+\;O(k^4)\right\}\;\;,\;\;\;
\label{C6}
\ea
\ba
\nonumber\\
<\sigma_{xy}^2>\;=\;\frac{\rho^2}{4}\;\left\{\left(\beta\,\gamma\,+\,\alpha\,
\omega\,k^2\right)\,\left(y^2\,-\,b^2\right)\,+\,\left(\beta\,\gamma\,-\,
\alpha\,\omega\,k^2\right)\,\left(x^2\,-\,a^2\right)\right\}^2\;\;\Xi_2\;\;
=\;\;\;\;\;\;\;\;\;\;\;\;\;\;\\
\label{C7}
\nonumber\\
= \;\frac{\rho^2}{32}\;\frac{\left(2\;I_3\;T_{kin}\;-\;{\bf J}^2
\right)^2}{I_1\;I_2\;I_3^2\;(I_3\;-\;I_1)\;(I_3\;-\;I_2)}\;\left\{\left(I_3\,
+\,I_1\,-\,I_2\right)\,\left(x^2\;-\;a^2\right)\;+\right.\;\;\;
\;\;\;\;\;\;\;\;\;\;\;\;\;\;\;\;\;\;\;\\
\label{C8}
\nonumber\\
\nonumber
+\;\left(I_3\,+\,I_2\,-\,I_1\right)\left.\;\left(y^2\;-\;b^2\right)\protect\LARGE\right\}^2\;\left\{1\;+\;O(k^4)\right\}\;\;\;,
\ea
\ba
\nonumber\\
<\sigma_{xz}^2>\;=\;\frac{\rho^2}{4}\;\left\{
\left(\beta\,\omega\,+\,\alpha\,\gamma\right)\,
\left(z^2\,-\,c^2\right)\,+\,
\left(\,-\,\beta\,\omega\,+\,\alpha\,\gamma\right)\,
\left(x^2\,-\,a^2\right)\right\}^2\;\;\Xi_3\;\;=\;\;\;\;\;\;\;\;\;\;\;\\
\label{C9}
\nonumber\\
\nonumber
= \left\{\frac{\rho^2}{8}\;\frac{\left(2I_3T_{kin}-
{\bf J}^2\right) \left({\bf J}^2-2I_1T_{kin} \right)}{
I_1\;I_2^2\;I_3\;(I_3\;-\;I_1)^2}\left\{\left(z^2-c^2\right)\left(
I_3+I_2-I_1\right)+\left(x^2-a^2\right)\left(I_1+I_2-I_3\right)\protect\LARGE\right\}^2\;-\right.\\
\nonumber\\
\nonumber
\left.-\;\frac{3}{8}\;\frac{\rho^2}{8}\;\frac{\left(2I_3T_{kin}-
{\bf J}^2\right)^2 \left(I_2\,-\,I_1\right)}{
I_1\;I_2^2\;I_3\;(I_3\;-\;I_1)^2\,(I_3\;-\;I_2)}\left\{\left(z^2-c^2\right)
\left(I_3+I_2-I_1\right)\;+\right.\right.\\
\nonumber\\
\left.\left.+\;\left(x^2-a^2\right)\left(I_1+I_2-I_3\right)\protect\LARGE\right\}^2\right\}\;\left\{1\,+\,O(k^4)\right\}\;\;\;\;,\;\;
\label{C10}
\ea
\ba
\nonumber\\
<\sigma_{yz}^2>\;=\;\frac{\rho^2}{4}\;  \left\{
\left(\alpha\,\beta\,+\,\omega\,\gamma\right)\,
\left(z^2\,-\,c^2\right)\;+\;
\left(\alpha\,\beta\,-\,\omega\,\gamma\right)\,
\left(y^2\,-\,b^2\right)\right\}^2\;\;\Xi_4\;\;=\;\;\;\;\;\;\;\;\;\;\;\\
\label{C11}
\nonumber\\
\nonumber
= \left\{\frac{\rho^2}{8}\;\frac{\left(2I_3T_{kin}-
{\bf J}^2\right)\left({\bf J}^2-2I_1T_{kin}\right)}{I_1^2\;
I_2\;I_3\;(I_3\;-\;I_1)\;(I_3\;-\;I_2)}\,\left\{(I_1+I_3-I_2)
(z^2-c^2)+(I_1-I_3+I_2)(y^2-b^2)\protect\LARGE\right\}^2\;-\right.\\
\nonumber\\
\nonumber
\left.-\;\frac{5}{8}\; \frac{\rho^2}{8}\;\frac{\left(2I_3T_{kin}-
{\bf J}^2\right)^2\left(I_2\,-\,I_1\right)}{I_1^2\;
I_2\;I_3\;(I_3\;-\;I_1)\;(I_3\;-\;I_2)^2}\,\left\{(I_1+I_3-I_2)
(z^2-c^2)\;+\right.\right.\\
\nonumber\\
\left.\left.+\;(I_1-I_3+I_2)(y^2-b^2)\protect\LARGE\right\}^2\right\}\;\left\{1\;+\;O(k^4)\right\}\;\;,\;\;
\label{C12}
\ea
\ba
\nonumber\\
<(Tr\,\sigma)^2>\,=\,\frac{\rho^2}{4}\;\beta^4\;\left\{
\left(x^2-a^2\right)\,\left(Q-1\right)\,+\,
\left(y^2-b^2\right)\,\left(Q+S\right)\,+\,
\left(z^2-c^2\right)\,\left(1-S\right)
\right\}^2\;\;\Xi_1\;\;=\;\;\\
\label{C13}
\nonumber\\
\nonumber
=\;\frac{\rho^2}{32}\;\frac{\left(2\;I_3\;T_{kin}\;-\;{\bf J}^2\right)^2}{I_1^2\;I_2^2\;I_3^2\;(I_3\;-\;I_2)^2\;(I_3\;-\;I_1)^2}\;\left\{
\left[
I_1I_2\left(I_1-I_2\right)-I_1I_3\left(I_1-I_3\right)
\right]\left(x^2-a^2\right)+\;\right.\;\\
\nonumber\\
\nonumber
+\,\left[I_2\,I_3\,\left(I_2\,-\,I_3\right)-I_2\,I_1\,\left(I_2\,-\,I_1
\right)\right]
\left(y^2-b^2\right)+\left.\left[
I_3\,I_1\,\left(I_3\,-\,I_1\right)\;-\right.\right.\\
\nonumber\\
\left.\left.-\;I_3\,I_2\,\left(I_3\,-\,I_2\right)
\right]\left(z^2-c^2\right)\protect\LARGE\right\}^2\;\left\{1\;+\;O(k^4)
\right\}\;\;.\;\;
\label{C14}
\ea
In the above formulae, factors $\;\Xi_1\;$, $\;\Xi_2\;$, $\;\Xi_3\;$, and $\;
\Xi_4\;$ stand for averaged powers of the elliptic functions:
\ba
\nonumber
\Xi_1\;\equiv\;<\,\left(\,{\it sn}^2(u,\,k^2)\;-\;<\,{\it sn}^2(u,\,k^2)\,>\,
\right)^2\,>\;=\\
\nonumber\\
=\;<\,{\it sn}^4(u,\,k^2)\,>\;-\;<\,{\it sn}^2(u,\,k^2)\,>^2\;=\;\frac{1}{8}\;
+\;O(k^4)\;\;,\;\;
\label{C15}
\ea
\ba
\nonumber
\Xi_2\;\equiv\;<\,\left(\,{\it sn}(u,\,k^2)\;{\it cn}(u,\,k^2)\;-\;<\,{\it sn}(u,\,k^2)\;{\it cn}(u,\,k^2)\,>\,\right)^2\,>\;=\\
\nonumber\\
=\;<\,{\it sn}^2(u,\,k^2)\;{\it cn}^2(u,\,k^2)\,
>\;-\;<\,{\it sn}(u,\,k^2)\;{\it cn}(u,\,k^2)\,>^2\;=\;\frac{1}{8}\;
+\;O(k^4)\;\;,\;\;
\label{C16}
\ea
\ba
\nonumber
\Xi_3\;\equiv\;<\,\left(\;{\it cn}(u,\,k^2)\;{\it dn}(u,\,k^2)\;-\;<\,{\it cn}(u,\,k^2)\;{\it dn}(u,\,k^2)\,>\,\right)^2\,>\;=\\
\nonumber\\
=\;<\,{\it cn}^2(u,\,k^2)\;{\it dn}^2(u,\,k^2)\,
>\;-\;<\,{\it cn}(u,\,k^2)\;{\it dn}(u,\,k^2)\,>^2\;=\;\;\;\\
\nonumber\\
\nonumber
=\;\frac{1}{2}\;\left(1\;-\;\frac{3}{8}\;k^2\right)\;+\;O(k^4)\;,\;\;\;\;\;\;\;\;\;\;\;\;
\label{C17}
\ea
\ba
\nonumber\\
\nonumber
\Xi_4\;\equiv\;<\,\left(\,{\it sn}(u,\,k^2)\;{\it dn}(u,\,k^2)\;-\;<\,{\it sn}(u,\,k^2)\;{\it dn}(u,\,k^2)\,>\,\right)^2\,>\;=\\
\nonumber\\
=\;<\,{\it sn}^2(u,\,k^2)\;{\it dn}^2(u,\,k^2)\,
>\;-\;<\,{\it sn}(u,\,k^2)\;{\it dn}(u,\,k^2)\,>^2\;=\;\;\;\\
\nonumber\\
\nonumber
=\;\frac{1}{2}\;\left(1\;-\;\frac{5}{8}\;k^2\right)\;+\;O(k^4)\;\;\;\;\;\;\;\;\;\;\;\;\;
\label{C18}
\ea
where the averaging implies:
\ba
\nonumber
<...>\;\equiv\;\frac{1}{\tau}\;\int_{0}^{\tau}\;\,.\,.\,.\,\;du\;\;\;,\;\;\;
\ea
$\tau\;$ being the period expressed by (A2). The approximations were obtained 
by the trick (\ref{jacobi1}) - (\ref{jacobi3}) explained in Section VIII. 
(Expansions of $\;\Xi_{i}\;$ over $\,k^n\,$ cannot be obtained by 
plugging (\ref{8.1}) - (\ref{8.3}) into (C15) - (C18) because this would 
produce secular terms.) 

Plugging of  (C2), (C4), (C6), (C8), (C10), (C12) and (C14) into (\ref{7.2}) 
will lead us to the following expression for dissipation per unit volume:
\be
\frac{d\,<W>}{dV}\;=\;\frac{d\,<W^{(2\omega )}>}{dV}\;+\;\frac{d\,<W^{(\omega )
}>}{dV}\;\;
\label{C19}
\ee
where the first term stands for the dissipation associated with oscillations at
the second mode:
\ba
\nonumber
\frac{d\,<W^{(2\omega )}>}{dV}\;=\;\;\;\;\;\;\;\;\;\;\;\;\;\;\;\;\;\;\;\;\\
\nonumber\\
\nonumber
=\;\frac{1}{4\,\mu}\;\left\{\,-\,\frac{1}{1\,+\,\nu^{-1}}
\,<(Tr\,\sigma)^2>\;+\;<\sigma_{xx}^2>\;+\;<\sigma_{yy}^2>\;+\;<\sigma_{zz}^2>
\;+\;2\;<\sigma_{xy}^2>\;\right\}\;=\\
\nonumber\\
\nonumber
=\;\frac{1}{4\,\mu}\;\frac{\rho^2}{32}\;\left(2\,I_3\,T_{kin}\;-\;{\bf J}^2 
\right)^2 \;\frac{1}{I_1^2\,I_2^2\,I_3^2\,(I_3\,-\,I_2)^2\,(I_3\,-\,I_1)^2}\,
\left\{\,-\,\frac{1}{1\,+\,\nu^{-1}}\,\left[\left(x^2\,-\right.\right.\right.\\
\nonumber\\
\nonumber
-\left.\,a^2\right)\,\left(I_1\, I_2\,(I_1\,-\,I_2)\,-\,I_1\,I_3\,(I_1\,-\,I_3)
\,\right)\;+\;
\left(y^2\,-\,b^2\right)\,\left(I_2\, I_3\,(I_2\,-\,I_3)\,-\,I_2\,I_1\,(I_2\,-
\,I_1)\,\right)\;+\\
\nonumber\\
\nonumber
+\left.\left(z^2-c^2\right)\,\left(I_3I_1(I_3-I_1)-I_3I_2
(I_3-I_2)\right)\right]^2\,+\,
\left(x^2-a^2\right)^2\,\left[I_1I_2(I_1-I_2)-I_1I_3(I_1
-I_3) \right]^2\,+\\
\nonumber\\
\nonumber
+\left(y^2-b^2\right)^2\,\left[I_2I_3(I_2-I_3)-I_2I_1
(I_2-I_1) \right]^2\,+\,
\left(z^2-c^2\right)^2\,\left[I_3I_1(I_3-I_1)-I_3I_2(I_3
-I_2) \right]^2 \,+\\
\nonumber\\
+\,\left.I_1\,I_2\,(I_3-\,I_1)\,(I_3-\,I_2)\,\left[
(I_3+\,I_1-\,I_2)\left(x^2-\,a^2\right)\;+\;(I_3+\,I_2-\,I_1)\left(
y^2-\,b^2\right)\right]^2\right\}\;+\;O(k^6)\;\;\;\;\;\;
\label{C20}
\ea
while the second term stands for the dissipation at the frequency of precession:
\ba
\nonumber
\frac{d\,<W^{(\omega )}>}{dV}\;=\;\;\;\;\;\;\;\;\;\;\;\;\;\;\;\;\;\;\;\;\;\;\\
\nonumber\\
\nonumber
=\;\frac{1}{4\,\mu}\;\left\{\,2\;\left(<\sigma_{xz}^2>\;+\;
<\sigma_{yz}^2>\;\right)\,\right\}\;=\;\frac{1}{2\,\mu}\;\frac{\rho^2}{8}\;\frac{
1}{I_1\,I_2\,I_3\,(I_3\,-\,
I_1)}\;\left(2\,I_3\,T_{kin}\;-\;{\bf J}^2\right)\,\left({\bf J}^2\;-\right.\\
\nonumber\\
\nonumber
\left.-\;2\,I_1\,T_{kin}\right)\;\left\{
\frac{1}{I_2\,(I_3\,-\,I_1)}\;\left[(I_3\,-\,I_1)(z^2\,-\,x^2\,+\,a^2\,-\,c^2)
\;+\;I_2\,(z^2\,+\,x^2\,-\,a^2\,-\,c^2)\right]^2\;+\right.\\
\nonumber\\
\nonumber
+\;\left.\frac{1}{I_1\,(I_3\,-\,I_2)}\;\left[(I_3\,-\,I_2)(z^2\,-\,y^2\,+\,
b^2\,-\,c^2)\;+\;I_1\,(z^2\,+\,y^2\,-\,b^2\,-\,c^2)\right]^2\right\}\;-\\
\nonumber\\
\nonumber
-\;\frac{1}{2\,\mu}\;\frac{\rho^2}{8}\;\frac{
1}{I_1\,I_2\,I_3\,(I_3\,-\,
I_1)}\;\left(2\,I_3\,T_{kin}\;-\right.\\
\nonumber\\
\nonumber
\left.-\;{\bf J}^2\right)^2\,\frac{I_2\;-\;I_1}{I_3\;-
\;I_2}\;\left\{
\frac{1}{I_2\,(I_3\,-\,I_1)}\;\frac{3}{8}\;\left[(I_3\,-\,I_1)(z^2\,-\,x^2\,+\,a^2\,-\,c^2)
\;+\;I_2\,(z^2\,+\,x^2\,-\,a^2\,-\,c^2)\right]^2\;+\right.\\
\nonumber\\
+\;\left.\frac{1}{I_1\,(I_3\,-\,I_2)}\;\frac{5}{8}\;\left[(I_3\,-\,I_2)(z^2\,-\,y^2\,+\,
b^2\,-\,c^2)\;+\;I_1\,(z^2\,+\,y^2\,-\,b^2\,-\,c^2)\right]^2\right\}\;
\;+\;O(k^6)\;\;.\;\;\;
\label{C21}
\ea
As expected, we have obtained the result, (C20 - C22), in the spectral form 
(\ref{6.8}). After integration (like (\ref{6.9})) this result must be plugged 
into (\ref{5.4}), which will lead to (\ref{8.14}). It follows from (C20) and 
(C21) that the geometrical factors emerging in (\ref{8.14}) will read:
\ba
\nonumber
H_1\;=\;\frac{1}{8}\;\frac{1}{I_1\,I_2\,I_3\,(I_3\,-\,I_1)}\;\int_{-a}^{a}dx
\int_{-b}^{b}dy\int_{-c}^{c}dz\,\left\{\frac{1}{I_2\,(I_3\,-\,I_1)}\,
\left[(I_3\,-\,I_1)(z^2\,-\,x^2\,+\,a^2\,-\,c^2)\,+\right. \right. \\
\nonumber\\
\nonumber
\left.+\;I_2\,(z^2\,+\,x^2\,-\,a^2\,-\,c^2)\right]^2\;+\;\;\;\;\;\;\\
\nonumber\\
+\;\left.\frac{1}{I_1\,(I_3\,-\,I_2)}\;\left[(I_3\,-\,I_2)(z^2\,-\,y^2\,+\,
b^2\,-\,c^2)\;+\;I_1\,(z^2\,+\,y^2\,-\,b^2\,-\,c^2)\right]^2\right\}\;\;,\;\;
\label{C22}
\ea
\ba
\nonumber
H_2\;=\;\frac{1}{64}\;\frac{1}{I_1^2\,I_2^2\,I_3^2\,(I_3\,-\,I_1)^2\,(I_3\,-\,
I_2)^2}\;\int_{-a}^{a}\;dx\;\int_{-b}^{b}\;dy\;\int_{-c}^{c}\;dz\;
\left\{\,-\,\frac{1}{1\,+\,\nu^{-1}}\,\left[\left(x^2\,-\right.\right.\right.\\
\nonumber\\
\nonumber
-\left.\,a^2\right)\,\left(I_1\, I_2\,(I_1\,-\,I_2)\,-\,I_1\,I_3\,(I_1\,-\,I_3)
\,\right)\;+\;
\left(y^2\,-\,b^2\right)\,\left(I_2\, I_3\,(I_2\,-\,I_3)\,-\,I_2\,I_1\,(I_2\,-
\,I_1)\,\right)\;+\\
\nonumber\\
\nonumber
+\left.\left(z^2-c^2\right)\,\left(I_3I_1(I_3-I_1)-I_3I_2
(I_3-I_2)\right)\right]^2\,+\,
\left(x^2-a^2\right)^2\,\left[I_1I_2(I_1-I_2)-I_1I_3(I_1
-I_3) \right]^2\,+\\
\nonumber\\
\nonumber
+\left(y^2-b^2\right)^2\,\left[I_2I_3(I_2-I_3)-I_2I_1
(I_2-I_1) \right]^2\,+\,
\left(z^2-c^2\right)^2\,\left[I_3I_1(I_3-I_1)-I_3I_2(I_3
-I_2) \right]^2 \,+\\
\nonumber\\
+\;\left.I_1\,I_2\,(I_3\,-\,I_1)\,(I_3\,-\,I_2)\,\left[
(I_3\,+\,I_1\,-\,I_2)\left(x^2\,-\,a^2\right)\;+\;(I_3\,+\,I_2\,-\,I_1)\left(
y^2\,-\,b^2\right)\right]^2\right\}\;\;.\;\;\;
\label{C23}
\ea
and
\ba
\nonumber
H_0\;=\;\frac{1}{64}\;\frac{I_1\,I_3\,+\,I_2\,I_3\,-\,2\,I_1\,I_2}{(I_3\,-\,I_1
)^2\,(I_3\,-\,I_2)^2}\;\frac{I_2\,-\,I_1}{I_1\,I_2\,I_3}\;\int_{-a}^{a}dx
\int_{-b}^{b}dy\int_{-c}^{c}dz\,\left\{\frac{3}{I_2\,(I_3\,-\,I_1)}\,\left[(z^2\,-\right.\right.\\
\nonumber\\
\nonumber
\left.\left.-\,c^2)(I_3\,+\,I_2\,-\,I_1) \;+\;(x^2\,-\,a^2)(I_1\,+\,I_2\,-\,I_3)\right]^2\;+\;\frac{5}{I_1\,(I_3\,-\,I_2)}\,\left[(z^2\,-\right.\right.\\
\nonumber\\
\left.\left.-\,c^2)(I_1\,+\,I_3\,-\,I_2)\;+\;(y^2\,
-\,b^2)(I_1\,+\,I_2\,-\,I_3)\right]^2\right\}\;\;.\;\;\;\;\;\;\;
\label{C24}
\ea
For a homogeneous prism of dimensions $\;2a\,\times\,2b\,\times\,2c\;$:
\ba
H_1\;=\;\frac{317}{m^4}\;\;\frac{a\;b\;c^5}{\left(b^2\,+\,c^2\right)\;\left(
a^4\,-\,c^4\right)\;\left(a^2\,+\,b^2\right)}\;\left(\frac{b^4}{b^4\,-\,c^4}\;
+\;\frac{a^4}{a^4\,-\,c^4} \right)\;\;,\;\;
\label{C25}
\ea
\ba
\nonumber\\
H_2\;=\;\frac{100}{m^4}\;\;\frac{a^9\,b^9\,c\,-\,a^9\,b^5\,c^5\,-\,a^5\,b^9\,c^5\,+\,0.21\,a^9\,b\,c^9\,+\,0.19 \,a^5\,b^5\,c^9\,+\,0.21\,a\,b^9\,c^9}{\left(a^2\,+\,b^2\right)^2\;\left(a^4\,-\,
c^4 \right)^2\;\left( b^4\,-\,c^4 \right)^2}\;\;
\label{C26}
\ea
and
\ba
H_0\;=\;\frac{237}{m^4}\;a\;b\;c^5\;(a^2\,-\,b^2)\;\frac{(2.67\,a^4\,b^4\,-\,a^4\,c^4\,-\,1.67\,b^4\,c^4)\,(a^4\,+\,b^4\,-\,2\,c^4)}{(a^2\,+\,b^2)\,(a^2\,-\,c^2)\,(b^2\,-\,c^2)\,(a^4\,-\,c^4)^2\,(b^4\,-\,c^4)^2}\;\;.\;\;
\label{C27}
\ea
Calculating $\,H_1\,$ and $\,H_2\,$ we assumed that the quantity 
$\;1/(1+\nu^{-1})\;$ 
emerging in (\ref{7.2}) and (\ref{8.12}) is $\;1/5\;$ (as it normally is 
for solid materials). We also used the standard formulae for the moments of 
inertia:
\be
I_1\;=\;\frac{m}{3}\;\left(b^2\;+\;c^2 \right)\;\;\;, \;\;\;\;etc...
\label{C28}
\ee
$\;m\;$ being the mass of the homogeneous body.

\section{\\ 
Averaged over the precession period squares of the components of the
stress tensor, in the vicinity of pole A \label{D}}

By squaring each of the expressions (\ref{9.2}) - (\ref{9.7}), and averaging 
the result, one will easily arrive to the following formulae:

\ba
<\sigma_{xx}^2>\;=\;
\frac{\rho^2}{4}\,(1\,-\,Q)^2\;\beta^4\;\left(x^2\,-\,a^2\right)^2\;\;\Xi_1
\;\;=\;\;\;\;\;\;\;\;\;\;\;\;\;\;\;\\
\label{D1}
\nonumber\\
\nonumber\\
\frac{\rho^2}{32}\;\frac{\left({\bf J}^2\;-\;2\;I_1\;T_{kin}
\right)^2}{I_2^2\,I_3^2\,\left(I_1\;-\;I_3\right)^2\,\left(I_2\;-\;I_1
\right)^2}\;\left\{I_3\;\left(I_1\;-\;I_3\right)\;-\;I_2\;\left(I_1\;-\;I_2
\right)\right\}^2\;\left(x^2\;-\;a^2\right)^2\;\left\{1\,+\,O(k^4)\right\}\;\;,\;\;\;
\label{D2}
\ea
\ba
\nonumber\\
<\sigma_{yy}^2>\;=\;\frac{\rho^2}{4}\,(S\,+\,Q)^2\;\beta^4\;\left(y^2\,-\,b^2
\right)^2\;\;\Xi_1\;\;=\;\;\;\;\;\;\;\;\;\;\;\;\;\;\;\;\;\\
\label{D3}
\nonumber\\
\nonumber\\
\frac{\rho^2}{32}\;\frac{\left({\bf J}^2\;-\;2\;I_1\;T_{kin}
\right)^2}{I_1^2\,I_3^2\,\left(I_1\;-\;I_3\right)^2\,\left(I_2\;-\;I_1
\right)^2}\;\left\{I_1\;\left(I_2\;-\;I_1\right)\;-\;I_3\;\left(I_2\;-\;I_3
\right)\right\}^2\;\left(y^2\;-\;b^2\right)^2\;\left\{1\,+\,O(k^4)\right\}\;\;,\;\;\;
\label{D4}
\ea
\ba
\nonumber\\
<\sigma_{zz}^2>\;=\;\frac{\rho^2}{4}\,(1\,-\,S)^2\;\beta^4\;\left(z^2\,-\,c^2
\right)^2\;\;\Xi_1\;\;=\;\;\;\;\;\;\;\;\;\;\;\;\;\;\;\\
\label{D5}
\nonumber\\
\nonumber\\
\frac{\rho^2}{32}\;\frac{\left({\bf J}^2\;-\;2\;I_1\;T_{kin}
\right)^2}{I_1^2\,I_2^2\,\left(I_1\;-\;I_3\right)^2\,\left(I_2\;-\;I_1
\right)^2}\;\left\{I_2\;\left(I_3\;-\;I_2\right)\;-\;I_1\;\left(I_3\;-\;I_1
\right)\right\}^2\;\left(z^2\;-\;c^2\right)^2\;\left\{1\,+\,O(k^4)\right\}\;\;,\;\;\;
\label{D6}
\ea
\ba
\nonumber\\
<\sigma_{xy}^2>\;=\;
\frac{\rho^2}{4}\,\left\{\left(\beta \gamma + \alpha \omega k^2\right)
\left(y^2 - b^2\right)\,+
\;\left(\beta \gamma - \alpha \omega  k^2\right)\left(x^2 - a^2\right)\right\}^2\;\;\Xi_2\;\;=\;\;\;\;\;\;\;\;\;\;\;\;\;\;\;\\
\label{D7}
\nonumber\\
\nonumber\\
\frac{\rho^2}{8}\;\frac{\left({\bf J}^2\;-
\;2\;I_1\;T_{kin}\right)\;\left(2\;I_3\;T_{kin}\;-\;{\bf J}^2\right)}{I_1
\;I_2\;I_3^2\;(I_3\;-\;I_1)\;(I_2\;-\;I_1)}\;\left\{I_3\,\left(x^2\;+\;y^2\;
-\;a^2\;-\;b^2\right)\;+\right.\;\;\;\;\;\;\;\;\;\;\;\;\;\;\;\;\;\;\;\;\;\;\\
\label{D8}
\nonumber\\
\nonumber
+\;(I_2\,-\,I_1)\left.\;\left(x^2\;
-\;y^2\;+\;b^2\;-\;a^2\right)\protect\LARGE\right\}^2\;\left\{1\,+\,O(k^4)\right\}\;\;\;,\;\;\;
\ea
\ba
\nonumber\\
<\sigma_{xz}^2>\;=\;\frac{\rho^2}{4}\,\left\{
\left(\beta\,\omega \,+\,\alpha\,\gamma\right)\,
\left(z^2\,-\,c^2\right)\,+
\;\left(\,-\,\beta\,\omega \,+\,\alpha\,\gamma\right)\,
\left(x^2\,-\,a^2\right)\right\}^2\;\;\Xi_3\;\;=\;\;\;\;\;\;\;\;\;\;\;\;\;\;\;\\
\label{D9}
\nonumber\\
\nonumber
=\;\left\{\frac{\rho^2}{8}\;
\frac{\left(2\;I_3\,T_{kin}\,-\;
{\bf J}^2\right)\,\left({\bf J}^2\,-\;2\;I_1\;T_{kin} \right)}{
I_1\;I_2^2\;I_3\;(I_3\,-\,I_1)^2}\;\left\{(I_3\,-\,I_1)\,(x^2-z^2+c^2-a^2)\;+
\right.\right.\;\;\;\;\;\;\;\;\;\;\;\;\;\;\;\;\;\;\;\;\;\\
\nonumber\\
\nonumber
+\;\left.\left.I_2\,(z^2+x^2-a^2-c^2)
\protect\LARGE\right\}^2\;-\right.\\
\nonumber\\
\left.-\;\frac{3}{8}\;
\frac{\rho^2}{8}\;
\frac{\left(I_3\;-\;I_2\right)\,\left({\bf J}^2\,-\;2\;I_1\;T_{kin} \right)^2}{
I_1\;I_2^2\;I_3\;(I_3\,-\,I_1)^2\,(I_2\,-\,I_1)}\;\left\{(I_3\,-\,I_1)\,(x^2-
z^2+c^2-a^2)\;+\right.\right.\;\;\;\;\;\;\;\;\;\;\;\;\;\;\;\;\;\;\;\;\;\\
\nonumber\\
\nonumber
+\;\left.\left.I_2\,(z^2+x^2-a^2-c^2)
\protect\LARGE\right\}^2\;\right\}\;\left\{1\;+\;O(k^4) \right\}\;,\;\;
\label{D10}
\ea
\ba
\nonumber\\
<\sigma_{yz}^2>\,=\,
\frac{\rho^2}{4}\,\left\{\left(\alpha \,\beta +\omega\, \gamma\,k^2\right)
\left(z^2 - c^2\right)\,+
\;\left(\alpha \, \beta - \omega \, \gamma \, k^2\right)
\left(y^2 -\,b^2\right)\right\}^2
\;\;\Xi_4\;\;=\;\;\;\;\;\;\;\;\;\;\;\;\;\;\;\\
\label{D11}
\nonumber\\
\nonumber\\
=\;\frac{\rho^2}{32}\;\frac{\left({\bf J}^2\,-\,2\;I_1\;T_{kin}
\right)^2}{I_1^2\;I_2\;I_3\;(I_3\;-\;I_1)\;(I_2\;-\;I_1)}\,
\left\{(I_1+I_3-I_2)\,(z^2-c^2)\,+\,(I_1-I_3+I_2)\,(y^2-b^2)
\protect\LARGE\right\}^2\;\left\{1\;+\;O(k^2)\right\}\;,
\label{D12}
\ea
\ba
\nonumber\\
<(Tr\,\sigma)^2>\;=\;
\frac{\rho^2}{4}\;\beta^4\;\left\{
\left(x^2-a^2\right)\,\left(Q-1\right)\,+\,
\left(y^2-b^2\right)\,\left(Q+S\right)\,+\,
\left(z^2-c^2\right)\,\left(1-S\right)
\right\}^2\;\;\Xi_1\;\;=\;\;\;\;\;\\
\label{D13}
\nonumber\\
\nonumber\\
\nonumber
=\;\frac{\rho^2}{32}\;\frac{\left({\bf J}^2\,-\,2\;I_1\;T_{kin}\right)^2}{I_1^2\;I_2^2\;I_3^2\;(I_1\;-\;I_2)^2\;(I_3\;-\;I_1)^2}\;\left\{
\left[
I_1I_2\left(I_1-I_2\right)-I_1I_3\left(I_1-I_3\right)
\right]\left(x^2-a^2\right)+\;\right.\;\\
\nonumber\\
\nonumber
+\,\left[I_2\,I_3\,\left(I_2\,-\,I_3\right)-I_2\,I_1\,\left(I_2\,-\,I_1
\right)\right]
\left(y^2-b^2\right)+\left.\left[
I_3\,I_1\,\left(I_3\,-\,I_1\right)\;-\right.\right.\\
\nonumber\\
-\;\left.\left.I_3\,I_2\,\left(I_3\,-\,I_2\right)
\right]\left(z^2-c^2\right)\protect\LARGE\right\}^2\;\left\{1\;+\;O(k^4)
\right\}.
\label{D14}
\ea
where $\;\Xi_{1,2,3,4}\;$ are given by (C15) - (C18).  
Just as in the preceding case of pole C, the above expressions (D2), (D4), 
(D6), (D8), (D10), (D12), (D14) form pole A are to be plugged in (\ref{7.2}). It will entail:
\be
\frac{d\,<W>}{dV}\;=\;\frac{d\,<W^{(\omega )}>}{dV}\;+\;\frac{d\,<W^{(2 \omega )}>}{dV}\;\;
\label{D15}
\ee
where 
\ba
\nonumber
\frac{d\,<W^{(\omega )}>}{dV}\;=\;
\frac{1}{4\,\mu}\;\left\{\,2\;<\sigma_{xy}^2>\;+\;2\;<\sigma_{zx}^2>\,\right\}
\;= \;\;\;\;\;\;\;\;\;\;\;\;\;\;\;\;\;\;\;\;\;\;\;\;\;\;\;\;\;\;\;\;\;\;\\
\nonumber\\
\nonumber
=\;\frac{1}{2\,\mu}\;\frac{\rho^2}{8}\;\frac{
\left( {\bf J}^2\;-\;2\;I_1\;T_{kin} \right) \;
\left( 2\;I_3\;T_{kin}\;-\;{\bf J}^2 \right) \;
}{I_1\,I_2\,I_3\,(I_3\,-\,I_1)}\;
\left\{\frac{1}{I_2\,(I_3\,-\,I_1)} \left[ \left(I_3\,-\,I_1\right)\,\left(x^2
\,-\,a^2\,-\,z^2\,+\,c^2\right)\,+\right.\right.\\
\nonumber\\
\nonumber
\left.\left.+\,I_2\,\left(x^2\,-\,a^2\,+\,z^2\,-\,c^2\right) \right]^2 \;+\right.\\
\nonumber\\
\nonumber
+\;\left.\frac{1}{I_3\,(I_2\,-\,I_1)}\left[I_3\,\left(x^2\,-\,a^2\,+\,y^2\,-\,b^2
\right)\,+\,(I_2\,-\,I_1)\,\left(x^2\,-\,a^2\,-\,y^2\,+\,b^2\right) 
\right]^2 \right\}\;-\\
\nonumber\\
\nonumber
-\;\frac{1}{2\,\mu}\;\frac{\rho^2}{8}\;\frac{3}{8}\;\frac{
\left( {\bf J}^2\;-\;2\;I_1\;T_{kin} \right)^2 \;
\left( I_3\;-\;I_2 \right) \;
}{I_1\,I_2^2\,I_3\,(I_3\,-\,I_1)^2\;(I_2\;-\;I_1)}\;\left\{
 \left(I_3\,-\,I_1\right)\,\left(x^2
\,-\,a^2\,-\,z^2\,+\,c^2\right)\;+\right.\\
\nonumber\\
\left.+\;I_2\,\left(x^2\,-\,a^2\,+\,z^2\,-\,c^2  
\right)\right\}\;+\;O(k^6)\;\;\;\;\;\;\;\;\;
\label{D16}
\ea
and
\ba
\nonumber
\frac{d\,<W^{(2 \omega )}>}{dV}\;=\;\;\;\;\;\;\;\;\;\;\;\;\;\;\;\;\;\;\;\;\;\\
\nonumber\\
\nonumber
=\;\frac{1}{4\,\mu}\;\left\{\,-\,\frac{1}{1+
\nu^{-1}}\,<\left( Tr\, \sigma
\right)^2>\;+\;<\sigma_{xx}^2>\;+\;<\sigma_{yy}^2>\;+\;<\sigma_{zz}^2>\;+\;2\;
<\sigma_{yz}^2> \,\right\}\;=\;\;\;\;\;\;\;\;\\
\nonumber\\
\nonumber
=\frac{1}{4 \mu}\;\frac{\rho^2}{32}\;\frac{\left( {\bf J}^2\;-\;2\;I_1\;T_{
kin} \right)^2}{I_1^2I_2^2I_3^2(I_3-I_1)^2(I_2-I_1)^2}\;\left\{
\left[I_1 I_3(I_1-I_3)-I_1 I_2(I_1-I_2)
\right]^2\left(x^2-a^2\right)^2+\right.\\
\nonumber\\
\nonumber
+\;\left[I_2 I_1\,(I_2\,-\,I_1)\,-\,I_2 I_3\,(I_2\,-\,I_3)
\right]^2\,\left(y^2\,-\,b^2\right)^2\;+\\
\nonumber\\
\nonumber
+\;\left[I_3 I_2\,(I_3\,-\,I_2)\,-\,I_3 I_1\,(I_3\,-\,I_1)
\right]^2\,\left(z^2\,-\,c^2\right)^2\;+\\
\nonumber\\
\nonumber
+\;2\,I_2\,I_3\,(I_3-I_1)\,(I_2-I_1)\,\left[
(I_1+I_2+I_3)\,(z^2-c^2)\,+\,(I_1-I_3+I_2)\,(y^2-b^2)
\right]^2 \,-\\
\nonumber\\
\nonumber
-\;\frac{1}{1+{\nu}^{-1}}\;\left\{
\left[
I_1\,I_2\left(I_1-\,I_2\right)\,-\,I_1\,I_3\left(I_1-\,I_3\right)
\right]\left(x^2-\,a^2\right)\;+\;\left[I_2\,I_3\,\left(I_2\,-\,I_3\right)\;-
\right.\right.\\
\nonumber\\
\left.\left.-\;I_2\,I_1\,\left(I_2\,-\,I_1
\right)\right]
\left(y^2-b^2\right)+\left.\left[
I_3\,I_1\,\left(I_3\,-\,I_1\right)-I_3\,I_2\,\left(I_3\,-\,I_2\right)
\right]\left(z^2-c^2\right)\protect\LARGE\right\}^2\right\}\;+\;O(k^6)\;\;.\;\;
\label{D17}
\ea
Integration of the two above expressions over the volume of the body gives  
expressions for the two components of the time-dependent elastic energy 
deposited in the body: $\;W^{(\omega)}\;$ and $\;W^{(2 \omega)}\;$, plugging 
whereof into (\ref{6.6}) will yield (\ref{9.8}). Expressions for the 
geometrical factors emerging in (\ref{9.8}) are:
\ba
S_1\;=\;\;\;\;\;\;\;\;\;\;\;\;\;\;\;\;\;\;\;\;\;\;\;\;\;\;\;\;\;\;\;\;\;\;\;\\
\label{D18}
\nonumber\\
\nonumber
\frac{1}{16}\;\frac{1}{I_1\,I_2\,I_3\,(I_3\,-\,I_1)}\;
\int_{-a}^{a}dx\,\int_{-b}^{b}dy\,\int_{-c}^{c}dz\;\left\{\frac{1}{I_2\,(I_3\,-\,I_1)} \left[ \left(I_3\,-\,I_1\right)\,\left(x^2
\,-\,a^2\,-\,z^2\,+\,c^2\right)\,+\right.\right.\\
\nonumber\\
\nonumber
\left.\left.+\,I_2\,\left(x^2\,-\,a^2\,+\,z^2\,-\,c^2\right) \right]^2 \;+\right.\\
\nonumber\\
\nonumber
+\;\left.\frac{1}{I_3\,(I_2\,-\,I_1)}\left[I_3\,\left(x^2\,-\,a^2\,+\,y^2\,-\,b^2
\right)\,+\,(I_2\,-\,I_1)\,\left(x^2\,-\,a^2\,-\,y^2\,+\,b^2\right) 
\right]^2 \right\}
\ea
\ba
S_2\;=\;\;\;\;\;\;\;\;\;\;\;\;\;\;\;\;\;\;\;\;\;\;\;\;\;\;\;\;\;\;\;\;\;\;\;\\
\label{D19}
\nonumber\\
\nonumber
\frac{1}{128}\;\frac{1}{I_1^2\,I_2^2\,I_3^2\,(I_3-I_1)^2\,(I_2-I_1)^2}
\;\int_{-a}^{a}dx\,\int_{-b}^{b}dy\,\int_{-c}^{c}dz\;\left\{
\left[I_1 I_3(I_1-I_3)-I_1 I_2(I_1-I_2)
\right]^2\left(x^2-a^2\right)^2+\right.\\
\nonumber\\
\nonumber
+\;\left[I_2 I_1\,(I_2\,-\,I_1)\,-\,I_2 I_3\,(I_2\,-\,I_3)
\right]^2\,\left(y^2\,-\,b^2\right)^2\;+\\
\nonumber\\
\nonumber
+\;\left[I_3 I_2\,(I_3\,-\,I_2)\,-\,I_3 I_1\,(I_3\,-\,I_1)
\right]^2\,\left(z^2\,-\,c^2\right)^2\;+\\
\nonumber\\
\nonumber
+\;2\,I_2\,I_3\,(I_3-I_1)\,(I_2-I_1)\,\left[
(I_1+I_2+I_3)\,(z^2-c^2)\,+\,(I_1-I_3+I_2)\,(y^2-b^2)
\right]^2 \,-\\
\nonumber\\
\nonumber
-\;\frac{1}{1+{\nu}^{-1}}\;\left\{
\left[
I_1I_2\left(I_1-I_2\right)-I_1I_3\left(I_1-I_3\right)
\right]\left(x^2-a^2\right)+\;\right.\;\\
\nonumber\\
\nonumber
+\,\left.\left[I_2\,I_3\,\left(I_2\,-\,I_3\right)-I_2\,I_1\,\left(I_2\,-\,I_1
\right)\right]
\left(y^2-b^2\right)+\left.\left[
I_3\,I_1\,\left(I_3\,-\,I_1\right)-I_3\,I_2\,\left(I_3\,-\,I_2\right)
\right]\left(z^2-c^2\right)\protect\LARGE\right\}^2\right\}
\ea
The factor $S_0$ is equal to
\ba
\nonumber
S_0\;=\;\;\;\;\;\;\;\;\;\;\;\;\;\;\;\;\;\;\;\;\;\;\;\;\;\;\;\;\;\;\;\;\;\;\;\\
\nonumber\\
\nonumber
\frac{3}{128}\;\frac{I_3\;-\;I_2}{I_1\,I_2^2\,I_3\,(I_3\,-\,I_1)^2\,(I_2\,-\,I_1)}\;
\int_{-a}^{a}dx\,\int_{-b}^{b}dy\,\int_{-c}^{c}dz\;\left\{\left(I_3\,-\,I_1\right)\,\left(x^2
\,-\,a^2\,-\,z^2\,+\,c^2\right)\,+\right.\\
\nonumber\\
\left.+\,I_2\,\left(x^2\,-\,a^2\,+\,z^2\,-\,c^2\right) 
\right\}^2\;\;\;\;
\label{D20}
\ea
and becomes negligibly small in the case of the body approaching the prolate 
symmetry ($\;I_3\,-\,I_2\,\rightarrow\,0\;$).

For a homogeneous prism of sizes
$\;2a\,\times\,2b\,\times\,2c\;$, the factors much simplify:
\ba
S_1\;=\;\frac{173}{m^4}\;\frac{a\;b\;c}{\left(a^2\,+\,b^2\right)\;\left(b^2\,+\,c^2  
\right)\;\left(a^4\,-\,c^4\right)}\;\left[ 
\frac{a^8\,+\,1.7\,a^4\,b^4\,+\,b^8}{a^4\,-\,b^4}\,+\,\frac{a^8\,+\,1.7\,a^4\,
c^4\,+\,c^8}{a^4\,-\,c^4}\right]\;\;,\;
\label{D21}
\ea
\ba
\nonumber
S_2\;=\;\frac{42}{m^4}\;\frac{a\;b\;c}{\left(b^2\,+\,c^2\right)^2\;
\left(a^4\,-\,c^4\right)^2\;\left(a^4\,-\,b^4\right)^2}\;\left[ 
a^8\,b^8\,+\,2.8\,a^8\,b^4\,c^4\,-\,4.8\,a^4\,b^8\,c^4\,+\,a^8\,c^8\,-\right.\\
\left.-\,4.8\,a^4\,b^4\,c^8\,+\,4.8\,b^8\,c^8\right]\;\;\;
\label{D22}
\ea
and
\ba
S_0\;=\;\frac{32.4}{m^4}\;a\;b\;c\;\frac{\left(a^8\,+\,1.67\,a^4\,c^4\,+\,
c^8\right)\,\left(b^2\,-\,c^2\right)}{(a^4\,-\,b^4)\,(a^4\,-\,c^4)^2\,(b^2\,+
\,c^2)}
\label{D23}
\ea

\pagebreak

\end{document}